\documentclass[
 superscriptaddress, reprint,
 amsmath,amssymb,
 aps, floatfix
]{revtex4-2}
\usepackage{graphicx}
\usepackage{dcolumn}
\usepackage{bm}
\usepackage[%
]{hyperref}

\usepackage{physics}

\graphicspath{figures/}

\begin{document}

\title{Quantum theory for nonlinear optical effects in the ultra-strong light-matter coupling regime}

\author{Thomas Krieguer} 
\email{thomas.krieguer@espci.fr}
\author{Yanko Todorov}
\email{yanko.todorov@espci.fr}
\affiliation{\small{Laboratoire de Physique et d'Étude des Matériaux, LPEM, UMR 8213, ESPCI Paris, Université PSL, CNRS, Sorbonne Université, F-75005 Paris, France}}

\date{\today}

\begin{abstract}
We present a microscopic quantum theory for nonlinear optical phenomena in semiconductor quantum well heterostructures operating in the regime of ultra-strong light matter coupling regime. This work extends the Power-Zienau-Wooley (PZW) formulation of quantum electrodynamics to account for nonlinear interactions based on a fully fermionic approach, 
without resorting to any  bosonization approximation. It provides a unified description of the microcavity  and the local field enhancement effects on the nonlinear optical response, thus encompassing the phenomena known as epsilon near zero (ENZ) effect. In particular, our theory describes the impact of the light-matter coupled states on the high frequency generation process, relevant for recent experimental investigations with polaritonic metasurfaces. We unveil the limitations of traditional single-particle approaches and propose novel design principles to optimize nonlinear conversion efficiencies in dense, microcavity-coupled electronic systems. The theoretical framework developed here provides an efficient tool for the development of advanced quantum optical applications in the mid-infrared and terahertz spectral domains. Furthermore, it establishes a foundation for exploring the quantum properties of the ultra-strong light-matter regime through frequency-converted polariton states. 
\end{abstract}

\maketitle

Nonlinear optical phenomena play a very important role in numerous applications, such as advanced bioimaging \cite{Review_Parodi_2020}, sensing, communications, and for the conception of novel laser-pumped coherent sources of light \cite{Review_Garmire_2013, book_HandbookNLOptics}. They also play a paramount role in quantum optics, where nonlinear effects are used, for instance, for the generation of non-classical states of light, such as squeezed states \cite{book_GuideQuantOptics}, as well as for building sources of indistinguishable photons, which are at the heart of quantum communications \cite{Review_QuantCom_2024}.  Specifically, in the mid-infrared (MIR, $\lambda \approx 3\mu m-30\mu m$) and terahertz (THz, $\lambda \approx 30\mu m-300\mu m$) frequency ranges, optical nonlinearities can be strongly enhanced by the quantum engineering of semiconductor nano-structures  \cite{Fejer_PhysRevLett_1989, Capasso_1994, Rosencher_Science_1996, book_Berger_1999, book_Sirtori_2000}.  In the past four decades there have been numerous studies on second \cite{Rosencher_Bois_EL_1989,Sirtori_1991,Julien1992} and third \cite{Sirtori_PRL1992} harmonic generation, difference frequency generation \cite{Dupont_2006}, as well as optical rectification \cite{Rosencher_1989, K_Unterrainer_1996} and nonlinear detectors based on two-photon absorption \cite{Dupont_1994NLdetector, DUPONT_2002NLdetector, Schneider_2005} with such systems. Difference frequency generation of THz frequencies starting from MIR quantum cascade lasers is still a hot research topic, as a means to achieve tuneable, room-temperature THz sources \cite{Belkin2007}.

Most of the aforementioned works exploit propagating geometries in which the semiconductor wafer containing the quantum wells is shaped into a waveguide and light is coupled through a polished wedge \cite{book_Helm}.  In the MIR range, the most recent studies on nonlinear frequency generation exploit quantum heterostructures combined with multimode photonic microcavities and metamaterials \cite{Lee2014nature, kim_giant_2020, sarma_strong_2021, sarma_all-dielectric_2022,Sarma_2022, chung_electrical_2023, yu_broadband_2024, Park2024}. This approach is free from the phase-matching constraints required by propagating geometries \cite{boyd_nonlinear_2008} while, additionally, providing a strong enhancement of the nonlinear signal \cite{Gomez-Diaz_PhysRevB.92.125429}.  

Similar metamaterial architectures allow achieving very high 
light-matter coupling strengths with electronic transitions, and reaching the so-called regime of ultra-strong light-matter interaction \cite{Todorov_PRL2010, delteil_charge-induced_2012}, which sets new frontiers for cavity quantum electrodynamics \cite{ciuti_quantum_2005,RevUSC_1_2019, RevUSC_2_2019}. The ultra-strong light-matter coupling regime (USC) occurs when the Rabi frequency $\Omega_R$ describing the reversible exchange of energy between the material excitation and an optical mode of the micro-cavity becomes of the order of magnitude of the transition frequencies of the electronic system \cite{ciuti_quantum_2005}. This regime predicts novel quantum phenomena, where fundamental quantum vacuum fluctuations can play important role in devices \cite{Ebbesen_2015, Feist_2015, Hagenmuller_2017, Ciuti_2018, Faist_2018, Kontos_2021, Faist_2022}. In the USC regime, the light-matter coupled states, the polaritons, also acquire non trivial quantum optical properties, such as squeezing, and the ground state of the system becomes populated with virtual photons \cite{ciuti_quantum_2005,RevUSC_1_2019, RevUSC_2_2019}. Revealing such properties is not an easy experimental task, as it requires detector performance beyond the current technological level for the MIR and THz frequency range \cite{Rogalski_2012, TodorovDhillonMangeney2024}. In that respect, concepts from nonlinear optics can be very useful as they would allow converting MIR or THz polarition states at higher or lower frequencies where detectors are more sensitive and allow advanced functions such as photon counting \cite{book_GuideQuantOptics}. 

Recently, nonlinear MIR polaritonic metasurfaces have been reported for control of chirality \cite{kim_giant_2020} and phase \cite{chung_electrical_2023} of reflected waves, optical switching \cite{Cotrufo2024}, and leveraging on the increased electromagnetic confinement to enhance nonlinear effects \cite{sarma_all-dielectric_2022}. In these studies one typically employs the ability of the strong coupling regime to control the photonic resonances of the structure by acting on the electronic transition, which is treated in the single-particle approximation. The impact of the strong coupling regime on nonlinear phenomena such as saturation and bistability has been reported in Ref. \cite{jeannin_unified_2021}. However, the regimes of strong and ultra-strong coupling are achieved in systems comprising very high electronic densities \cite{Todorov_PRL2010, delteil_charge-induced_2012}, where collective effects that arise from electron-electron interactions play very important role \cite{todorov_intersubband_2012, pegolotti_quantum_2014, todorov_dipolar_2015}.In particular, electron-electron interactions lead to a strong renormalization of the frequencies and oscillator strengths of the optical resonances, which, as a consequence, become very different from what is predicted by the single-particle picture \cite{delteil_charge-induced_2012}. On a very basic level, it is expected that the nonlinear susceptibilities are enhanced at higher electronic densities, just like the linear absorption process. However, so far the study of nonlinear conversion processes in this regime has been completely overlooked, as it is not obvious how collective effects would alter the nonlinear susceptibility functions. 

\begin{figure}[h!]
    \centering
    \includegraphics[width=1\columnwidth]{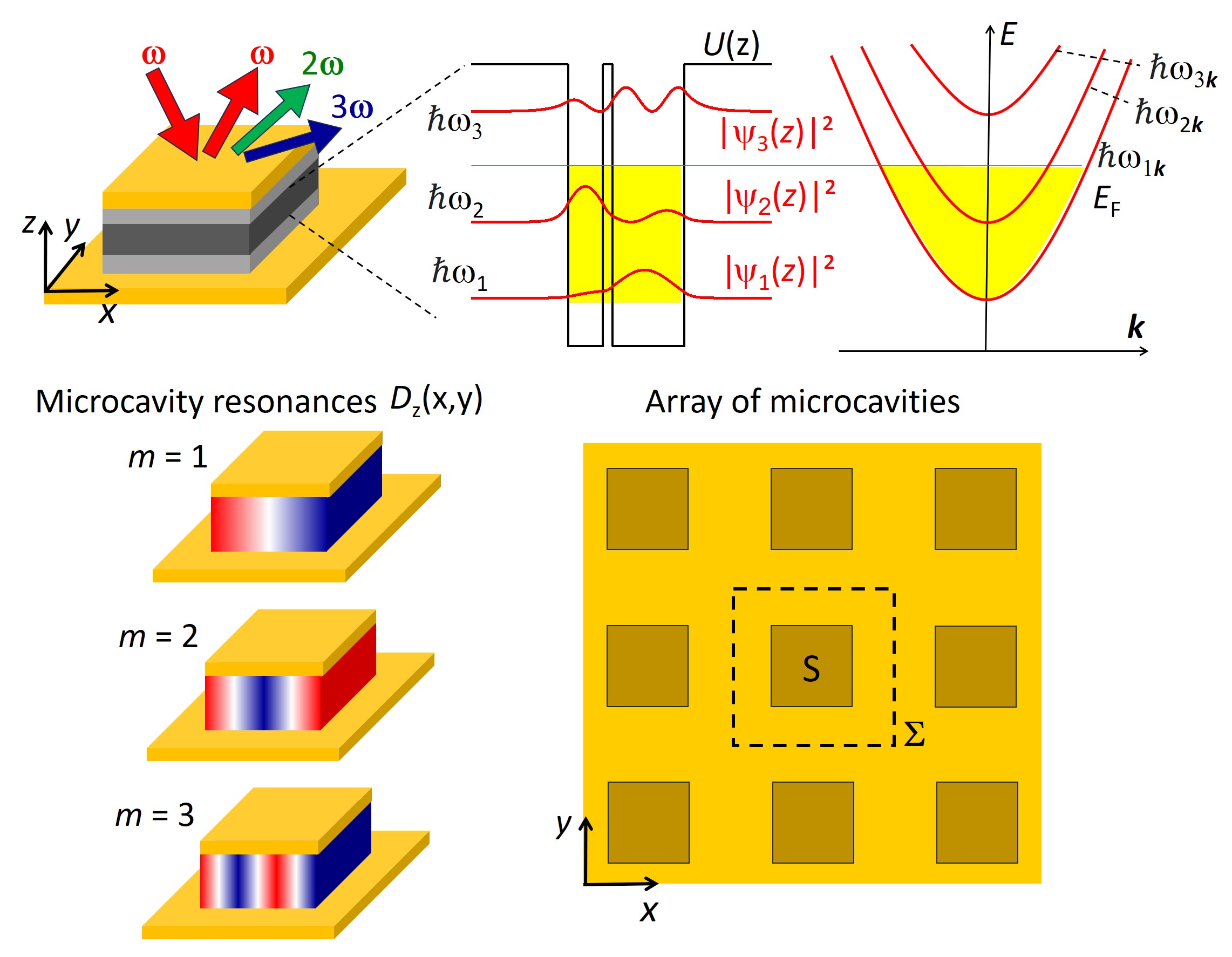}
    \caption{Illustration of the system under investigation. A highly doped  semiconductor heterostructure is embeded into a double-metal electromagentic resonator  \cite{Todorov_PRL2010} (up). The heterostructure potential $U(z)$ confines the electrons in one direction, while they are free to move in the plane of the heterostructure. The electronic states then appear as an ensemble of quasi-parabolic subbands \cite{Rosencher_Vinter_2002} occupied up to a Fermi level $E_F$.  The microcavity sustains multiple resonant modes (bottom) \cite{todorov_optical_2010}. Rather than a single unit, we consider a periodic arrangement of micorocavities, where $\Sigma$ is the array init cell, while the surface of each patch resonator is $S$. This arrangement allows for a full control of the array interaction with free space radiation.  We derive the nonlinear response of the system in the presence of polaritonic and collective electronic effects when it is pumped with a monochromatic field with a frequency $\omega$, and namely the higher frequency generation, $n\omega$, $n=2,3,...$. }
    \label{fig: global}
\end{figure}

In the present work we provide a fully microscopic theory of the nonlinear phenomena in quantum well hetero-structures, where electronic transitions can be tuned from the THz to the MIR spectral domain \cite{book_Helm}. Our approach fully accounts for microcavity effects, as well as for dipole-dipole interactions that are known to completely reshape the optical resonances of the system \cite{delteil_charge-induced_2012}. One of the consequences of our study is that quantum designs that optimize nonlinear effects in the single-particle picture are no longer valid in the presence of collective and polaritonic states. Our model enables one to fully leverage from the enhancement of the light-matter interaction strength through the higher electronic densities. Besides, the theoretical framework that is developed here is grounded in a fully microscopic quantum description that would allow further investigations of the quantum correlations that appear in the regime of ultra-strong light-matter coupling. This framework will allow designing experiments based on nonlinear optics in order to uncover these correlations in frequency converted polariton states. 

Previously reported description of nonlinear optical phenomena in quantum well heterostructures are essentially based on a single-particle approach \cite{ahn_calculation_1987}, following the general approach developed for atomic systems \cite{boyd_nonlinear_2008}. Electronic interactions have been considered in the framework of a time-varying Hartree potential \cite{Khurgin1991, Zaluzny_NL_spectra_PhysRevB.47.3995, Zaluzny_SHG_PhysRevB.51.9757, Zaluzny_THG_10.1063/1.361496, Cominotti_2023}, where nonlinear effects arise essentially from population dynamics. Instead, dipole-dipole interactions, which lead to strong renormalization of the optical response, have not been considered. Our theory is formulated in the Power-Zienau-Wooley, PZW, representation of quantum electrodynamics \cite{cohen-tannoudji_photons_1986}, which was previously employed to describe dipole-dipole interactions in quantum confined electron gas \cite{todorov_intersubband_2012, todorov_dipolar_2014, todorov_dipolar_2015}, as well as its interaction with phonons \cite{Rousseaux_2023}. 
We argue that this is the ideal framework for the description of quantum optical phenomena in condensed matter physics, as it provides the best separation between matter and field degrees of freedom \cite{cohen-tannoudji_photons_1986}. Furthemore, it allows for a global vision on the matter-matter interactions of the systems, which, in the present case allows considering the most relevant ones to describe optical nonlinearites in microcavities. 

In the PZW representation, a central role is played by the polarization vector field $\mathbf{P} (\mathbf{r})$  that is fully expressed from the electronic microscopic variables.  All matter-matter interactions are contained in a single term $\mathcal{H}_{P2}$ of the quantum Hamiltonian of the system, that is expressed as a spatial integral over the $\mathbf{P}^2(\mathbf{r})$. The photon field is represented by an electric induction vector, $\mathbf{D}(\mathbf{r})$ which keeps the same properties as the vacuum electric field, while the coupling with light becomes linear, and expressed from the spatial integral $\mathbf{D}(\mathbf{r})\mathbf{P}(\mathbf{r})$. In particular, the field creation and annihilation operators obey standard commutation relations, identical to the ones for vacuum fields, even in the presence of matter. This approach allows accommodating naturally for dissipation and dispersion in the system, without any need for auxiliary phenomenological variables \cite{Suttorp_Wubs_2004}, and leads to a microscopic theory for the dielectric function of quantum confined electronic systems \cite{todorov_dipolar_2014, todorov_dipolar_2015}. 

In order to extend the PZW formulation of electrodynamics in condensed matter systems for the case of nonlinear interaction, it is essential to abandon the bosonisation method previously used for the description of ultra-strong light-matter coupling \cite{ciuti_quantum_2005, todorov_dipolar_2014, todorov_dipolar_2015}. We treat fully the 
 fermionic degrees of freedom of matter, starting from an approach that  was previously developed to describe the photo-current in quantum well detectors operating in the strong light-matter coupling regime \cite{pisani_electronic_2023}. The bosonization relies on effective bosonic operators describing the collective electronic excitations between pairs of confined states, while assuming constant electronic populations on the states. In this approach the matter system is replaced by an inherently linear system of coupled harmonic oscillators, where optical nonlinear phenomena are absent. Fully treating the fermionic nature of electrons  implies that both the populations and quantum coherences between electronic levels become dynamical variables. The quantum dynamics stemming from the full quantum Hamiltonian that comprises light-matter as well as dipole-dipole interaction terms lead to inherently nonlinear system of coupled equations, where non-linear effects arise from a microscopic description, as for the case of a two-level atom \cite{boyd_nonlinear_2008, book_OptResonance}. Furthermore, through the inclusion of the dipole-dipole interactions we are also able to take into account the local field effects on the susceptibility functions in our dense electronic systems. In particular, we have been able to provide a rigorous formulation of the "epsilon near zero", ENZ \cite{Fomra_reviewENZ_2024},  effects on the linear and nonlinear susceptibilities. 
 
 Our theory constitutes therefore an example for a truly  microscopic quantum framework to describe optical nonlinearities in a condensed matter system. So far, quantum nonlinear  optical approaches \cite{Drummond_Hillery_2014} have been phenomenological in the sense that the theory is quantized after the nonlinear susceptibility function is provided. Because of the aforementioned property of the PZW representation, our theory is able to  generate frequency-dependent susceptibility functions, while photon operators still obey standard commutation rules, and do not need to be corrected by the presence of matter. The PZW framework thus constitutes the best choice, in terms of conceptual simplicity, for developing quantum theories of light-matter interaction in condense matter systems, where strong dissipation and dispersion are inherently present. 

A generic illustration of the system that we consider is provided  in Fig. (\ref{fig: global}). It consists of a quantum confined electron gas that interacts with multiple electromagnetic modes of a double-metal microcavity \cite{todorov_optical_2010}. In this system, electrons are quantum confined in one direction by a heterostructure potential $U(z)$ which leads to the formation of energy subbands \cite{book_Helm, Rosencher_Vinter_2002}. Here we consider only the case of inter-subband transitions. As illustrated in Fig. (\ref{fig: global}), rather than a single cavity, we consider a periodic array of micro-cavities with an array unit cell $\Sigma$. This configuration provides full control of the radiative coupling of the system, as discussed in \cite{todorov_optical_2010, Feuillet-Palma2013, Rodriguez_22}. This type of system is both relevant for the study of ultra-strong light-matter coupling \cite{todorov_intersubband_2012} and can be related to most of the experimental work on nonlinear optical interactions reported in the mid-infrared range \cite{Lee2014nature, Jeannin_2023}.  Similar systems have been also used for quantum detectors and modulators with superior performance \cite{palaferri2018room, Bigioli_2020, Hakl2021, Lin_23, Malerba_2024} or the study of strong light-matter coupling in quantum detectors \cite{pisani_electronic_2023,Vigneron_2019, Lagree_PRA_2022}.  
 
Our paper is organized as follows. In part \ref{FullModelpart} we start by establishing the quantum Hamiltonian of the system  in section \ref{secSystemHamiltonian}.  Electrons are still free to move in the plane perpendicular to the confinement axis. This movement is described by nearly parabolic energy bands (Fig. (\ref{fig: global})), where  the good quantum number is provided by the single-particle momentum $\hbar \mathbf{k}$ \cite{book_Helm, Rosencher_Vinter_2002}. However, the proper description of nonlinear optical phenomena  requires to adopt a new basis where the quantum operators are labeled by a position on a discreet lattice $\mathbf{r}_i$, as described in section \ref{secPositionLattice}. Such operators no longer describe electronic states with a well defined in-plane kinetic energy, but they provide a basis where spatial inhomogeneities of the optical response can be correctly described. Indeed,  the fact that nonlinear dipoles oscillate with different spatial distributions depending on their frequencies is at the origin of phase matching conditions in nonlinear optics \cite{boyd_nonlinear_2008}. Later on, this approach  will allow for rigorous definition of nonlinear geometrical overlap coefficients between various cavity modes, and leads to a proper description of inter-mode coupling induced by the matter nonlinearities.  After writing the Hamiltonian equation of motions in section \ref{EvolutionEqs}, we perform a semiclassical decoupling of the photonic and electronic degrees of freedom \cite{book_OptResonance, Haug_2009, Kira_Koch_2011}, and we introduce dissipation. For the case of photonic modes, we show how radiative dissipation rates must be properly incorporated into the input-output relations such as energy conservation is satisfied for the case of multi-mode system. As a result, we obtain consistent expressions for the absorption and reflectivity of the system for an arbitrary values of the resonance linewidths. Our approach thus provides simple and analytical alternative to more complex theories for non-Hermitian optical quasi-modes interacting with matter \cite{gigli_2020}, and can be readily applied in a broader context than considered here.

For the electronic degrees of freedom, we discuss a dissipation model where decoherence rates are introduced on the quadrature operators, rather than the elements of the density matrix operator, which has, to our opinion, more physical relevance than traditional models. In our work, the treatment of dissipation is free from the rotating wave approximation, both for the photonic and electronic degrees of freedom, and thus more relevant for the description of the system dynamics in the ultra-strong light-matter coupling regime. Our theory of the nonlinear optical susceptibilities is then detailed in section \ref{General theory chi}. In section \ref{secMicrocavity_effects} our dynamical equations are completed with the evolution equations for the microcavity modes, and we treat the problem of  frequency harmonic generation. In part \ref{partExamples} we illustrate our general theory with two examples: a two-subband system, section \ref{SecTwosubband_system}, and a three-subband system \ref{sec3subExample}. For the first example, we revisit saturation effects and bi-stability that have been recently discussed \cite{jeannin_unified_2021}. We also discuss the third order susceptibility $\chi^{(3)} (3\omega; \omega, \omega, \omega)$ and show how the local field enhancement, also refereed to as ENZ effect \cite{Fomra_reviewENZ_2024}, appears naturally in our theory. We also discuss the third harmonic generation, THG, in the ultra-strong coupling regime and illustrate how polariton effects affect the THG power spectrum. For the second example in section \ref{sec3subExample} we discuss the modifications of the second order susceptibility function $\chi^{(2)} (2\omega; \omega, \omega)$ due to collective electronic effects, and we illustrate the second harmonic generation, SHG, with such systems. In particular, we show that the doubly resonant designs are not the optimal choice for high electronic densities and we discuss how more efficient designs can be obtained. Multi-photon absorption in the presence of micro-cavities is also discussed. In the final part \ref{Conclusion} we discussed further developments that are enabled by our microscopic theory of nonlinear interactions. 

\section{Full model} \label{FullModelpart}

\subsection{System  Hamiltonian} \label{secSystemHamiltonian}

As described in Fig. \ref{fig: global}, we consider a quantum system where a bi-dimensional electron gas is hosted in an epitaxially grown semiconductor hetero-structure, with $z$ being the growth axis. We suppose that the conduction band is parabolic and electrons are free to move in the plane of the epitaxial layers, whereas quantum confinement arises owe to the artificially designed one-dimensional potential $U(z)$ of the hetero-structure. We suppose that the Schrodinger equation that corresponds to the single-particle problem is solved  \cite{SemiconductorNanostructures_Ihn}, and the single-particle confinement energies $\hbar \omega_\lambda$ as well as the corresponding envelope wavefunctions $\psi_\lambda (z)$ are known. As indicated in Fig. \ref{fig: global} the system is inserted into a microcavity, which sustains optical resonances with an electric field essentially polarized along the $z$ direction, such as the selection rule for intersubband electronic transitions is satisfied \cite{book_Helm, Rosencher_Vinter_2002}. The microcavity sustains multiple resonances as illustrated in Fig. \ref{fig: global}  that can be depicted as standing waves confined by the openings of the double-metal regions \cite{todorov_optical_2010}. We will be interested in the nonlinear response of the system, which arises when the system is pumped with strong coherent infrared radiation of a frequency $\omega$ \cite{yu_third-harmonic_2019}. We will consider the case where high electronic density is introduced by doping, and the Fermi level 
lies above one or several subbands. 

Our starting point is the full electrodynamics Hamiltonian of the system expressed in the Power-Zienau-Wooley (PZW) picture \cite{cohen-tannoudji_photons_1986}, supplemented with a driving term that describes the incident radiation \cite{bookLouisell}: 

\begin{eqnarray} 
\hat{H}  =    \sum_{\lambda, \mathbf{k}} \hbar \omega_{\lambda \mathbf{k}} c_{\lambda \mathbf{k}}^\dagger c_{\lambda \mathbf{k}} + \frac{1}{2\epsilon \epsilon_0}\int \hat{P}^2(z, \mathbf{r}) dzd^2\mathbf{r}\nonumber \\ 
+ \sum_\mathbf{m}\hbar \omega_{cm} (\hat{a}_m^\dagger\hat{a}_m + 1/2) -\frac{1}{\epsilon \epsilon_0}\int  \hat{P}\hat{D} dzd^2\mathbf{r} \nonumber \\
+\sum_m \frac{2V_{cav}}{\epsilon \epsilon_0 \hbar \omega_m }\sqrt{\Gamma_r^m} \hat{D}_m\hat{I}_{in} \label{EqfullH}
\end{eqnarray}

The first line of Eq. (\ref{EqfullH}) corresponds to the Hamiltonian of the electron gas. Here $\mathbf{k}$ is the planar wavevector that corresponds to the free electron movement in the plane of the epitaxial layers (Fig. \ref{fig: global}). The single-particle electronic energies are provided by  $\hbar \omega_{\lambda \mathbf{k}} = \hbar \omega_{\lambda} + \frac{\hbar^2 \mathbf{k}^2}{2m^*}$ where $\hbar \omega_{\lambda}$ is the $\lambda^{th}$ confined state of the heterostructure potential $U(z)$ and $m^*$ is the effective electron mass. The operators 
$\hat{c}_{\lambda \mathbf{k}}$, $\hat{c}^\dagger_{\lambda \mathbf{k}}$ are fermionic annihilation/creation operators for the single-particle state $|\lambda \mathbf{k} \rangle$. Here  $\epsilon$ stands for the relative permittivity of the material, and $\epsilon_0$ is the electric constant. 

A central role the PZW description of electromagnetic interactions is played by the polarization density operator $\hat{P} (z, \mathbf{r})$ \cite{todorov_dipolar_2014, todorov_dipolar_2015}. This quantity describes the matter-matter interactions that lead to  collective electronic excitations through the quadratic part (second term in Eq. (\ref{EqfullH})) and  our aim is to evaluate its effect for the nonlinear susceptibility functions. We discuss in detail the expression of $\hat{P} (z, \mathbf{r})$ in the next paragraph. 

The second line of Eq. (\ref{EqfullH}) describes the Hamiltonian of the microcavity photons and the light-matter coupling term. Here $a_m^\dagger$ and $a_m$ are the bosonic creation and annihilation operators for the $m^{th}$ cavity mode of a frequency $\omega_m$. The relevant quantity in the PZW theory is the electrical displacement operator that we denote  $\hat{D}(z, \mathbf{r})$:

\begin{eqnarray}
 \hat{D}(z, \mathbf{r}) = \sum_m \hat{D}_m \Theta_{L_{cav}}(z) u_m(\mathbf{r}) \label{DD}
 \\
\hat{D}_m = \sqrt{\frac{\hbar \omega_{m} \epsilon \epsilon_0}{2 V_{cav}}}i(\hat{a}_m - \hat{a}^\dagger_m)
\end{eqnarray}

We consider specifically the case of double-metal cavities which are based the lateral confinement of the $TM_0$ mode (Fig. \ref{fig: global}), and the electric filed/displacement is homogeneous along the cavity thickness $L_{cav}$ \cite{todorov_optical_2010}. This is expressed from the box function $\Theta_{L_{cav}}(z)$  which is unity inside the double-metal region and zero everywhere else. 
The quantity $V_{cav} =  SL_{cav}$ is the micro-cavity effective volume, which coincides with the geometrical volume of the cavity in the case of laterally confined $TM_0$ modes \cite{todorov_optical_2010}. Here $S$ is the surface of the patch, as indicated in Fig. \ref{fig: global}. The functions $u_m(\mathbf{r})$ describe the lateral confinement of the microcavity resonances, as illustrated in Fig. \ref{fig: global}, and they obey the normalization condition:

\begin{equation}
    \iint u_m(\mathbf{r}) u_{m'}(\mathbf{r}) d^2\mathbf{r} = S \delta_{mm'}
\end{equation}

The last line of Eq.(\ref{EqfullH}) describes the pump by an incident field $\hat{I}_{in}$ that can be an arbitrary function of time \cite{bookLouisell}, and $\Gamma_r^m$ is a coefficient homogeneous to a frequency that quantifies the radiation coupling of the $m^{th}$ cavity mode to the external radiation, as discussed later. 

\subsection{Polarization  field}\label{secPolField}

Here, we recall the derivation of the polarization field $\hat{P} (z, \mathbf{r})$ for our system, by providing more details with respect to previous works \cite{todorov_intersubband_2012}. In particular, we explicit both dynamic and static contribution of the polarization field. This is important as it allows distinguishing our approach with previous works that treat the optical nonlinearites of intersubband transitions in the presence of electron-electron interactions \cite{Khurgin1991, Zaluzny_NL_spectra_PhysRevB.47.3995, Zaluzny_SHG_PhysRevB.51.9757, Zaluzny_THG_10.1063/1.361496, Cominotti_2023}. In particular, we show that previous work treated only effects related to the static contribution, while the dynamic dipole-dipole interactions are fully taken into account in the present work for the first time, to our knowledge.

Since we consider intersubband transitions only $\hat{P} (z, \mathbf{r})$ is the $z$-component of a more general polarization field of the electron gas \cite{todorov_dipolar_2015}, where the in-plane component describes intra-subband transitions. Since the microcavity photon field is exclusively polarized along  the $z$-axis the latter do not play role in this study. In the absence of magnetic moments the intersubband polarization density obeys the constitutive relations \cite{cohen-tannoudji_photons_1986}:

\begin{eqnarray}
  \frac{d\hat{P}}{dt}   = \frac{1}{i\hbar} [\hat{P} ,\hat{H}_{e0}] = \hat{j} \label{gendefP} 
  \\
  \frac{\partial \hat{P}}{\partial z} = \hat{\rho} \label{gendefPrho} 
\end{eqnarray}

Here $\hat{H}_{e0} = \sum_{\lambda, \mathbf{k}} \hbar \omega_{\lambda \mathbf{k}} c_{\lambda \mathbf{k}}^\dagger c_{\lambda \mathbf{k}}$ is the single-particle electronic Hamiltonian (first term on the right-hand side of Eq. (\ref{EqfullH})) with  $\hat{j}$ and $\hat{\rho}$ respectively the current and density operators that for intersubband excitations:

\begin{eqnarray}
 \hat{j} = \frac{ie\hbar}{2m^*S}  \sum_{\lambda > \mu}  \xi_{\lambda \mu}(z) \Bigg{(} \sum_{\mathbf{q}, \mathbf{k}}  \hat{c}_{\lambda \mathbf{k+q}}^\dagger \hat{c}_{\mu\mathbf{k}} e^{-i\mathbf{q r}} - h.c. \Bigg{)} \label{defCurrent}
 \\
 \hat{\rho} = \frac{e}{S}\sum_{\lambda \geq \mu}\eta_{\lambda \mu} (z) \Bigg{(} \sum_{\mathbf{q}, \mathbf{k}}  \hat{c}_{\lambda \mathbf{k+q}}^\dagger \hat{c}_{\mu\mathbf{k}} e^{-i\mathbf{q r}} + h.c. \Bigg{)}
\end{eqnarray}

Here $\xi_{\lambda \mu}(z)$ is the microcurrent between the states $\lambda$ and $\mu$ that can be expressed from the confined wavefunctions as  $\xi_{\lambda \mu}(z) = \psi_\mu \partial_z \psi_\lambda - \psi_\lambda \partial_z \psi_\mu$ \cite{todorov_intersubband_2012}, and $\eta_{\lambda \mu} (z) =\psi_\mu (z)\psi_\lambda(z)$. It is easy to verify that the two operators $ \hat{\rho}$ and $\hat{j}$ satisfy continuity equation $d \hat{\rho}/d t + \partial_z  \hat{j} =0$. Then, using the definition (\ref{gendefP}) we arrive at the following expression for the dynamic part of the polarization field:

\begin{eqnarray}\label{EqP}
\hat{P}_{dynamic}(z, \mathbf{r}) = \frac{e \hbar}{2 m^* S} \sum_{\lambda > \mu} \frac{\xi_{\lambda \mu}(z)}{\omega_{\lambda\mu}} \nonumber \\
\times \Bigg{(} \sum_{\mathbf{q}, \mathbf{k}}  \hat{c}_{\lambda \mathbf{k+q}}^\dagger \hat{c}_{\mu\mathbf{k}} e^{-i\mathbf{q r}} + h.c. \Bigg{)}
\end{eqnarray}

Only terms $\lambda > \mu$ contribute to this term, as the intersubband microcurrent vanishes for $\lambda = \mu$, $\xi_{\lambda \lambda}(z) = 0$.
The polarization field (\ref{EqP}) has a non-vanishing time derivative under $\hat{H}_{e0}$ and describes electrons oscillating between several subbands, and with that respect is considered as "dynamic". However, the constitutive relations (\ref{gendefP}), (\ref{gendefPrho}) allow for a static term with $\lambda = \mu$ which does not lead to an intersubband current as it commutes with the single-particle Hamiltonian $\hat{H}_{e0}$. This term is derived from Eq.(\ref{gendefPrho}) by setting $\lambda = \mu$ and $\mathbf{q} =0$. Indeed, the terms with $\lambda = \mu$ and $\mathbf{q} \neq 0$ are linked to the in-plane component of the total polarization field that describe intra-subband contributions (density collective excitations) \cite{todorov_dipolar_2015} that do not couple to the photonic resonators considered here which is polarized along the $z$ axis. The static contribution $\hat{P}_{static}(z)$ along $z$ thus do not depend on the in-plane wavevector and is provided by the constitutive relation:

\begin{equation} \label{Pzstatic}
    \frac{d\hat{P}_{static}(z)}{dz} = \frac{e}{S} \Big[ \sum_\lambda \psi^2_\lambda(z) \hat{N}_\lambda - N_d (z)  \hat{N} \Big]
\end{equation}

Here we introduced the number operators  $\hat{N}_{\lambda} = \sum_\mathbf{k}\hat{c}_{\lambda \mathbf{k}}^\dagger \hat{c}_{\lambda \mathbf{k}}$ that counts the number of electrons on the subband $\lambda$ and the total number operator $\hat{N} = \sum_\lambda \hat{N}_\lambda$ which plays the role of identity operator in the electronic Hilbert space. In the above expression we added a function $N_d(z)$ which describes the distribution of the positively charged donors \cite{SemiconductorNanostructures_Ihn} that is normalized such as $\int N_d(z) dz =1$. This contribution is essential to ensure the overall charge neutrality of the system, and also plays a role as a reference charge in the PZW representation of electrodynamics \cite{cohen-tannoudji_photons_1986}.  Eq. (\ref{Pzstatic}) expresses the fact that, within a factor $\varepsilon_0$, the polarization field  $\hat{P}_{static}(z)$ is the static electric field owe to the charge distribution across the quantum hetero-structure. The corresponding electrostatic potential obeys the Poison equation with a Green function $G(z-z') = |z-z'|/2$ \cite{SemiconductorNanostructures_Ihn} which allow  providing an integral expression for $\hat{P}_{static}(z)$:

\begin{eqnarray}
 P_{static}(z)   = \frac{e}{2S} \sum_\lambda \hat{N}_\lambda \int \psi_\lambda^2 (z') \partial_z |z-z'| dz' \nonumber  \\
 - \frac{e}{2S}\hat{N} \int N_d{(z')}\partial_z |z-z'| dz'
\end{eqnarray}

This static term does not contribute to the light-matter interaction Hamiltonian as the average over the plane of the basis functions $u_m(\mathbf{r})$ is zero. More generally, the phonon field described in Eq. (\ref{DD}) is a transverse field while the static polarization is a longitudinal field according to the definitions from Ref. \cite{cohen-tannoudji_photons_1986}, and therefore their spatial overlap is identically null.

We can now expand the square polarization term in the PZW Hamiltonian (Eq. \ref{EqfullH}): 

\begin{eqnarray}
  \frac{1}{2\epsilon \epsilon_0}\int \hat{P}^2(z, \mathbf{r}) dzd^2\mathbf{r}\nonumber = \nonumber \\  
  \frac{1}{2\epsilon \epsilon_0}\int \hat{P}_{dynamic}^2(z, \mathbf{r})dzd^2\mathbf{r} +\nonumber \\
  \frac{S}{2\epsilon \epsilon_0}\int \hat{P}_{static}^2(z)dz+\nonumber \\
   \frac{1}{\epsilon \epsilon_0}\int \hat{P}_{static}(z)\hat{P}_{dynamic}(z, \mathbf{r})dzd^2\mathbf{r} \label{GeneralP2}
\end{eqnarray}

In the above equation, the second line expresses the contributions of the dipole-dipole interactions that have been considered so far to describe the linear response of collective electronic excitations in highly doped quantum wells \cite{todorov_intersubband_2012, pegolotti_quantum_2014, todorov_dipolar_2015}. The third line is a polarization self-energy which is quadratic in the electron number operators. This term provides a correction to the stationary heterostructure wavefunctions and energy states that can be treated in the framework of the Hartree approximation \cite{SemiconductorNanostructures_Ihn, cohen-tannoudji_quantum_2020},  by introducing an additional Hartree potential $V_H(z)$ expressed form the electronic density and solving self-consistently for the Schrodinger and Poisson equations. This term thus yields corrections to the single-particle energies computed with the hetero-structure potential $U(z)$ alone. The self-consistent energies and wavefunction can then be used in the expression of the dynamic term, Eq. (\ref{EqP}).  Finally, The last line in Eq. (\ref{GeneralP2}) corresponds to a density-dipole coupling.

In the bosonization approach, the number operators $\hat{N}_\lambda$ are replaced with their average values $\langle \hat{N}_\lambda \rangle$ in the ground state of the electronic system and considered constant. However, in a more general theory targeted here the number operators become dynamical variables.  Applying a harmonic perturbation in the system  will then induce nonlinear effects stemming from the static term $\hat{P}_{static}$ through oscillating contributions in the populations. So far, the role of electron-electron interactions for nonlinear optical effects has been analyzed through this term \cite{Khurgin1991, Zaluzny_NL_spectra_PhysRevB.47.3995, Zaluzny_SHG_PhysRevB.51.9757, Zaluzny_THG_10.1063/1.361496, Cominotti_2023}. 
However, it turns out that the nonlinear effects stemming from the static polarization follow different selection rules than the dipole-dipole terms when considering photonic systems with well defined spatial modes as the ones described in Fig. \ref{fig: global}. Mathematically, this is consequence from the fact that the static term depends on $z$-only, and is homogeneous in the heterostructure plane, while the photonic eignenfunctions have zero in-plane average  $\int \int u_m(z) d^2 \mathbf{r} = 0$. As a result, the effect of the static terms is averaged away. Note that such terms do produce important contributions in planar geometries, and relevant physical phenomena such as optical rectification and second order susceptibility in asymmetric quantum wells \cite{Rosencher_1989, K_Unterrainer_1996} actually derive from this terms. These effects can be optimized in carefully designed microcavities and studied in the context of ultra-strong coupling regime; this topic will be discussed elsewhere. 

Thus, in the following, we will consider the  nonlinear optical effects in the presence of dipole-dipole interactions, stemming from the second line in \ref{GeneralP2} while the static polarization has been taken into account by a self-consistent computation of the single-particle eigenenergies and wavefunctions. Without ambiguity the subscript "dynamic" will be dropped from now on for the polarization field from Eq. (\ref{EqP}).

\subsection{Position lattice} \label{secPositionLattice}

Under strong monochromatic pump $\hat{I}_{in} (t) \propto cos (\omega t)$ the electron gas responds locally to the electric field, which leads to nonlinear local contributions of the electronic polarization:

\begin{eqnarray}
 \langle \hat{P}(\mathbf{r}, t) \rangle =   \Re [\chi^{(1)} (\omega) e^{i\omega t}] \langle  \hat{D}(\mathbf{r}) \rangle \nonumber 
 \\
+ \Re [\chi^{(2)} (2\omega; \omega, \omega)  e^{2 i\omega t}] \langle  \hat{D}(\mathbf{r}) \rangle^2 \nonumber 
 \\
+ \Re[\chi^{(3)} (3\omega; \omega, \omega, \omega)  e^{3 i\omega t} ]\langle  \hat{D}(\mathbf{r}) \rangle ^3+...
\end{eqnarray}

Here $\langle ... \rangle$ denotes the average over a suitably chosen quantum state. As seen from Eq. (\ref{DD}) the cavity field  $\langle  \hat{D}(\mathbf{r}) \rangle$ is spanned on a basis of  well-defined spatial modes, $u_m(\mathrm{r})$, while the planar dependence of the polarization field,  Eq. (\ref{EqP}) is spanned on the basis of plane waves $e^{-i\mathbf{qr}}$. In the linear regime, and within the long-wavelength approximation, one can re-express polarization field on the basis of the micro-cavity spatial modes $u_m(\mathrm{r})$ by a suitable linear combination of excitation operators \cite{todorov_intersubband_2012}, such as the polarization field and the cavity field follow the same spatial dependence. This is fundamentally related to the fact that electrons are free to move in the plane of the hetero-structure, and the corresponding plane-wave wavefunctions provide a suitable plane wave basis for the polarization field.  On the other hand, nonlinear phenomena induce mixing between the spatial modes of the microcavity field, as seen from the above equation, and the various spatial components that enter the polarization field no longer correspond to well defined frequencies, as for the linear regime. Rather, one needs to develop a description based on the local response of the polarization based on the above equation. For that, we need to span the polarization field on a basis of operators that are local in space but can have an arbitrary time- dependence. Clearly the set of delocalized electronic states $|\lambda \mathbf{k} \rangle$ and the corresponding fermionic basis $\hat{c}_{\lambda \mathbf{k}}$, $\hat{c}^\dagger_{\lambda \mathbf{k}}$ are not suitable to provide such description.

To overcome this problem, we introduce another basis of states where electronic states are no longer labeled by their momentum $\hbar \mathbf{k}$, but by their position $\mathbf{r}_i$ on a bi-dimensional lattice which will be considered as discreet. In principle, it is possible to develop a theory where the position $\mathbf{r}$ is a continuous variable \cite{cohen-tannoudji_quantum_2020}. However the discreet lattice model leads to simplified commutation relations for the relevant fermionic operators and calculations are easier with the help of a few simple rules. In this picture, the microscopic theory for nonlinear interactions of the electron gas bears close resemblance with the theory for diluted atomic media \cite{boyd_nonlinear_2008} while allowing accommodating the effects of electron-electron interactions. 

\begin{figure}[h!]
    \centering
    \includegraphics[scale=0.3]{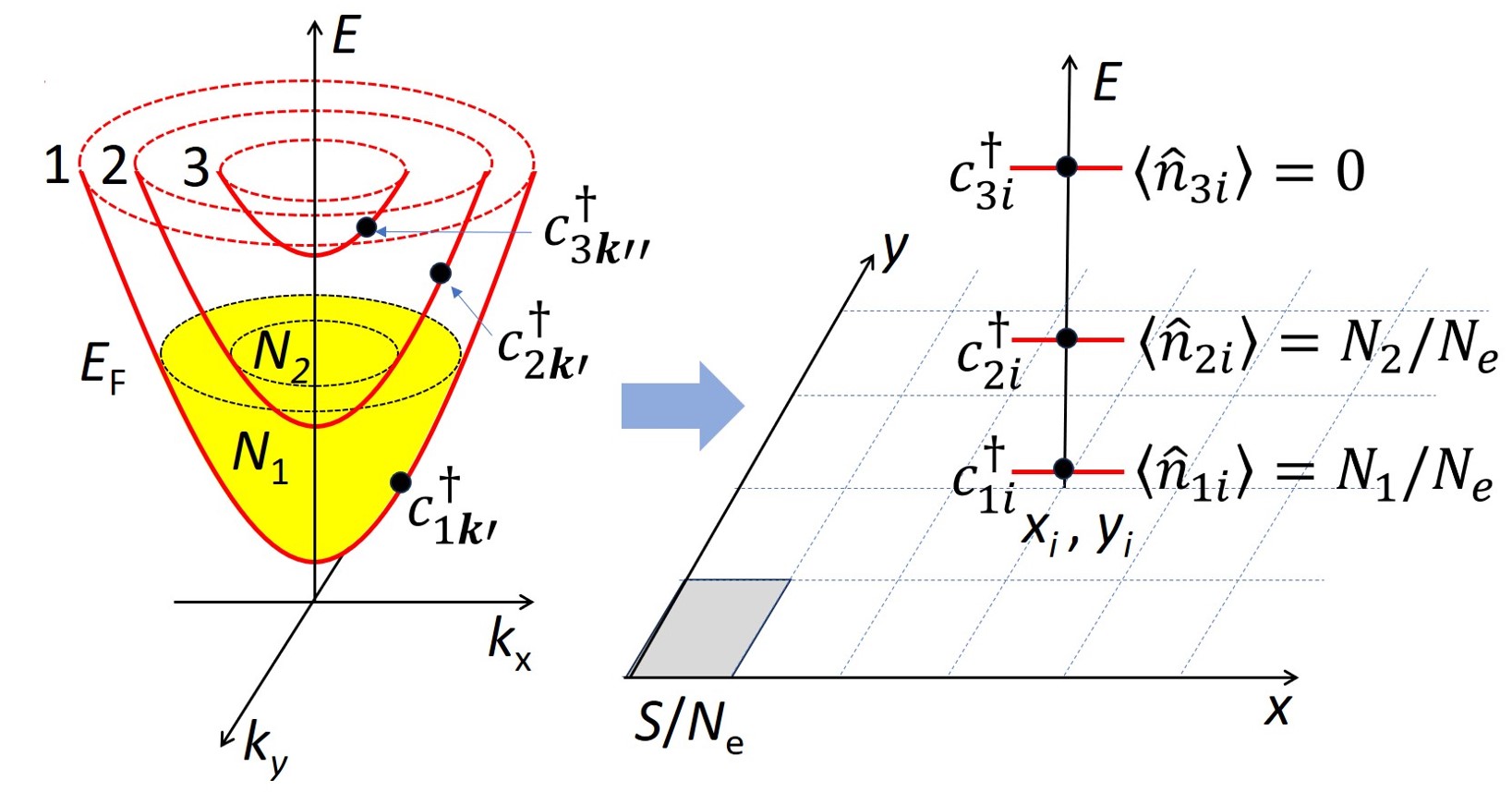}  
    \caption{Mapping of the $\mathbf{k}$ space of electrons into states defined on a discreet planar lattice with a total area $S$. Each case of the lattice has an area $S/N_e$. The operators $c^\dagger_{\lambda i}$ create an electron on the $i^{th}$ case of the lattice in the subband state $\lambda$.}
    \label{fig: lattice}
\end{figure}

Let $N_e$ be the total number of electrons on the sample of surface $S$. The surface $S$ is divided into $N_e$ identical cases, each with a surface a $S/N_e$ (Fig. \ref{fig: lattice}). We introduce the creation $c^\dagger_{\lambda i}$ and annihilation $c_{\lambda i}$ operators that create/destroy an electron in the case $i$, in the subband $\lambda$:

\begin{equation} \label{defci}
    \hat{c}^\dagger_{\lambda i} = \frac{1}{\sqrt{N_e}}\sum_\mathbf{k} \hat{c}^\dagger_{\lambda \mathbf{k}} e^{i \mathbf{k}\mathbf{r}_i}
\end{equation}

Here the vector $\mathbf{r}_i$ can be chosen at any point of the case $i$. We will chose the vectors $\mathbf{r}_i$ to point at the lower left corner of the case $i$. Then, as shown in Appendix A the polarization density Eq.(\ref{EqP}) can be rewritten as:

\begin{eqnarray}
\hat{P}(z, \mathbf{r}) = \frac{e \hbar}{2 m^*} \sum_{\lambda > \mu} \frac{\xi_{\lambda \mu}(z)}{\omega_{\lambda\mu}} \sum_i \hat{S}_{\lambda \mu i} \delta(\mathbf{r} - \mathbf{r}_i)  \label{DP1}
\\
\hat{S}_{\lambda \mu i} = c^\dagger_{\lambda i}c_{\mu i}+c^\dagger_{\mu i}c_{\lambda i}  \label{DP2}
\end{eqnarray}
    
One can easily recognize in the Eqs. (\ref{DP1}) and (\ref{DP2}) the polarization density of an ensemble of two level systems  written in the long wavelength approximation. The latter is justified from the fact that for a very large number of electrons we can suppose that the typical size of each case is much smaller than the wavelength of the infrared radiation interacting with the system, $\sqrt{(S/N_e)}<<\lambda$.  

Clearly, the operators defined by Eq.(\ref{defci}) mix different wavevectors $\mathbf{k}$, and thus the kinetic energy of the state $|i \lambda \rangle$ is no longer well-defined. However, we are interested in the time evolution of the polarization which can be obtained from the commutator of the operator $\hat{S}_{\lambda \mu i}$ with the single-particle Hamiltonian $\hat{H}_{e0} = \sum_{\lambda, \mathbf{k}} \hbar \omega_{\lambda \mathbf{k}} c_{\lambda \mathbf{k}}^\dagger c_{\lambda \mathbf{k}}$. In the  long wavelength approximation the following commutators are obtained:

\begin{eqnarray}
 [ \hat{H}_{e0}, \hat{S}_{\lambda \mu i}] = i \hbar \omega_{\mu \lambda} \hat{J}_{\lambda \mu i} \label{EqComS} 
 \\
 \hat{J}_{\lambda \mu i} = i(c^\dagger_{\lambda i}c_{\mu i}- c^\dagger_{\mu i}c_{\lambda i}) 
 \\
 \left[ \hat{H}_{e0}, \hat{J}_{\lambda \mu i} \right]  =  -i \hbar \omega_{\mu\lambda} \hat{S}_{\lambda \mu i}   \label{EqComJ}
 \end{eqnarray}

In the following, we will refer to the quantities $\hat{S}_{\lambda \mu i}$ as "polarizations", and to the quantities $\hat{J}_{\lambda \mu i}$ as "currents". 
Indeed, according to our definitions, the polarizations are expressed from the symmetric combinations of excitation operators, that are proportional to the microscopic dipole operator, while the anti-symmetric combinations provide the current operator (\ref{defCurrent}). Eq. (\ref{EqComS}) and (\ref{EqComJ}) indicate that in the long-wavelength approximation, the evolution of the polarizations and the currents under the single-particle electronic Hamiltonian are oscillations at the transition frequencies $\omega_{\mu\lambda} = \omega_\mu - \omega_\lambda$ which are independent from the in-plane wavevector $\mathbf{k}$, thus rendering possible the use of the position lattice. The electronic subsystem at each case thus behaves as a multi-level zero-dimensional system, with energy levels that correspond to the confinement energies at the bottom of the subbands $\hbar \omega_\lambda$.

These relations must be completed with the commutator relations between the quantities $\hat{S}_{\lambda \mu i}$ and $\hat{J}_{\lambda \mu i}$: 

\begin{eqnarray}
\frac{1}{i}[\hat{S}_{\lambda \mu i}, \hat{J}_{\lambda' \mu' i} ]   &=& \hat{S}_{\lambda \mu' i} \delta_{\mu \lambda'} - \hat{S}_{\lambda' \mu i} \delta_{\lambda \mu'} \nonumber
\\
&+& \hat{S}_{\mu \mu' i} \delta_{\lambda \lambda'} ( 1 - \delta_{\mu \mu'} ) \nonumber
\\
&-&  \hat{S}_{\lambda \lambda' i} \delta_{\mu \mu'} (1 - \delta_{\lambda \lambda'}) \nonumber 
\\
&+& \hat{n}_{\mu i}\delta_{\mu \mu'} - \hat{n}_{\lambda i}\delta_{\lambda \lambda'} \label{EqComSJ}
\\
i \left[\hat{S}_{\lambda \mu i}, \hat{S}_{\lambda' \mu' i} \right]   &=& \hat{J}_{\lambda \mu' i} \delta_{\mu \lambda'} + \hat{J}_{\mu \lambda' i} \delta_{\lambda \mu'} \nonumber \\ 
&+& \hat{J}_{\lambda \lambda' i} \delta_{\mu \mu'} + \hat{J}_{\mu \mu' i} \delta_{\lambda \lambda'} \label{EqComSS}
\\
-i \left[ \hat{J}_{\lambda \mu i}, \hat{J}_{\lambda' \mu' i} \right]   &=& \hat{J}_{\lambda \mu' i} \delta_{\mu \lambda'} + \hat{J}_{\mu \lambda' i} \delta_{\lambda \mu'} \nonumber \\ 
&-& \hat{J}_{\lambda \lambda' i} \delta_{\mu \mu'} - \hat{J}_{\mu \mu' i} \delta_{\lambda \lambda'} \label{EqComJJ}
\end{eqnarray}

In the last line of Eq. (\ref{EqComSJ}) we introduced the number operators $\hat{n}_{\lambda i} = c^\dagger_{\lambda i}c_{\lambda i}$. Since  $\hat{n}_{\lambda i}$ commutes with $\hat{H}_{e0}$ the number operator $\hat{N}_{\lambda}$ can be written as:

\begin{equation}
\hat{N}_{\lambda} = \sum_{\mathbf{k}} c_{\lambda \mathbf{k}}^\dagger c_{\lambda \mathbf{k}} = \sum_{i} c^\dagger_{ \lambda i}c_{\lambda i} \label{NumberOPs}
\end{equation}

With the help of the expression Eq. (\ref{DP1}) we can now recast the light-matter coupling Hamiltonian $\hat{H}_{LM}$ (second term in the second line of Eq.(\ref{EqfullH})) in the form:

\begin{eqnarray}
  \hat{H}_{LM}  = -\frac{1}{\epsilon \epsilon_0} \sum_{m, \alpha, i} d_\alpha  \hat{S}_{\alpha i} \hat{D}_m u_m (\mathbf{r}_i) \nonumber
   \\
  = - i\sum_{m, \alpha, i} d_\alpha E_{0m} u_m (\mathbf{r}_i) \hat{S}_{\alpha i} (a_m^\dagger - a_m) \label{eqLMintLattice}
\end{eqnarray}

Here we introduced the dipole of the intersubband transition $\alpha = \lambda \mu$:

\begin{equation}
d_\alpha = \frac{\hbar e}{2m^* \omega_\alpha} \int_0^{L_{cav}} \xi_\alpha (z) dz  
\end{equation}

In the last expression, we have assumed that the wavefunctions vanish at the walls of the cavity. One verifies easily that $d_\alpha$ is homogeneous to a dipole (charge times distance). We also introduced the amplitude of the vacuum electric field:

\begin{equation}
  E_{0m} =  \sqrt{\frac{\hbar \omega_{m}}{2 \epsilon \epsilon_0 V_{cav}}}
\end{equation}

The coupling constant $d_\alpha E_{0m}/\hbar$ that appears in Eq.  (\ref{eqLMintLattice}) is thus the well known vacuum Rabi frequency \cite{ciuti_quantum_2005, todorov_intersubband_2012, dutra_cavity_2005}. 

Using the expression Eq. (\ref{DP1}) we can further re-express the first line of Eq.(\ref{EqfullH}) in a way that was made familiar from Ref. \cite{todorov_intersubband_2012}:

\begin{equation}\label{ElectronicH}
\hat{H}_{e0} + \frac{\hbar N_e}{4} \sum_{\alpha, \alpha'} 
\Xi_{\alpha \alpha'} \sum_{i} \hat{S}_{\alpha, i} \hat{S}_{\alpha ', i}   
\end{equation}

Here for commodity we have labeled the pair $\lambda \mu$ with a single greek index $\alpha$. The coupling coefficients $\Xi_{\alpha \alpha'}$ are expressed as:

\begin{equation}
  \Xi_{\alpha \alpha'} = \frac{\omega_{P1, \alpha} \omega_{P1, \alpha'}}{\sqrt{\omega_\alpha \omega_{\alpha'}}} C_{\alpha \alpha'},  
\end{equation}

Here we introduced the plasma frequencies for a single electron:

\begin{equation} \label{SingleWP}
 \omega_{P1, \alpha}^2 = \frac{e^2}{m^*\epsilon \epsilon_0 S L_{\alpha}}   
\end{equation} 

and $L_\alpha$ is the extension of the electronic polarization that is defined from the microcurrents $\xi_\alpha(z)$ (See Ref. \cite{todorov_intersubband_2012}), and the overlap coefficients $C_{\alpha \alpha'}$ are defined as:

\begin{equation}
 C_{\alpha \alpha'} = \frac{\int \xi_\alpha(z)  \xi_{\alpha'}(z) dz}{\sqrt{\int \xi_\alpha^2(z) dz \int \xi_{\alpha'}^2(z) dz}}
\end{equation}

Eq. (\ref{NumberOPs}) shows that transformation defined by Eq. (\ref{defci}) does not change the total number of electrons on each subband $\lambda$.  Using Eq. (\ref{defci}) we can show that the fermionic operators from different cases $i \neq j$ commute. The total electronic Hamiltonian thus does not couple the electrons from different cases $i \neq j$, at least in the long wavelength approximation which is well satisfied in the optical experiments. Thus the multilevel systems from various cites are independent, and we can assume that the level occupancy for each  lattice cite is identical such as the occupancy of the $| i\lambda 
 \rangle$ state is $N_\lambda/N_e$. This assumption is made for commodity, as in the following we will be interested  in the evolution of the total number of populations only.  

 We will also make systematic use of the following relation for an arbitrary function $f(\mathbf{r})$ defined in the plane:

 \begin{equation} \label{faveraging}
     \frac{1}{N_e} \sum_i f(\mathbf{r}_i) = \frac{1}{S} \iint f(\mathbf{r}) d^2 \mathbf{r}
 \end{equation}

 This equality expresses the fact that the mean value of $f(\mathbf{r})$ should be identical whether it is computed on the discrete lattice or from  a continuous integration over the sample surface $S$.

\subsection{Evolution equations} \label{EvolutionEqs}

We establish here the evolution equations of the system. We start by writing the quantum-mechanical equations of motion that are induced by the Hamiltonian from Eq. (\ref{EqfullH}). Next, we consider the response under  monochromatic driving field. The incident wave is thus in a coherent state \cite{bookLouisell}, and eventually the microcavity modes will also evolve in coherent states inducing a semiclassical dynamics has the form:

\begin{equation} \label{EqGeneralO}
    \frac{d \langle \hat{O}_i \rangle }{dt} = \frac{1}{i\hbar} \langle [ \hat{O}_i, \hat{H} ] \rangle +  \Lambda (\langle \hat{O}_i \rangle, \langle \hat{O}_{j\neq i} \rangle )
\end{equation}

As we employ the Heisenberg picture the average is taken over the initial state of the system that is  a tensor product between a coherent photon state and the ground state of the electron gas. The  expression $\Lambda (\langle \hat{O}_i \rangle, \langle \hat{O}_{j\neq i} \rangle)$ describes dissipation and is considered to contain a damping term $-\gamma\langle \hat{O}_i \rangle$, as well as eventually a linear combination with other variables $\langle \hat{O}_{j\neq i} \rangle$ that are coupled to $\langle \hat{O}_i \rangle$. 

\subsubsection{Hamiltonian evolution} \label{secHamiltonian_evolution}

The Hamiltonian evolution of the quantum variables describing the system read:

\begin{widetext}
\begin{eqnarray}
\label{a}
\frac{d \hat{a}_{m}}{d t}&=&-i \omega_{m} a_{m} - \frac{1}{\hbar  } \sum_{\alpha, i}  (E_{0m} d_\alpha) u_m(\mathbf{r}_i ) \hat{S}_{\alpha i} + \sqrt{\frac{2V_{cav}}{\hbar \epsilon \epsilon_0 \omega_{cm}}}\sqrt{\Gamma_r^m} \hat{I}_{in}
\\
\label{n1_1}
\frac{d\hat{n}_{\lambda i}}{dt} &=& \frac{1}{\hbar \epsilon \epsilon_0}\sum_{m} \hat{D}_{m} u_{m}\left(\mathbf{r}_{i}\right) \sum_{\mu \neq \lambda} d_{\lambda \mu} \hat{J}_{\mu \lambda i} + \frac{1}{4} N_e \left(\sum_\alpha \sum_{\mu \neq \lambda} \Xi_{\alpha, \lambda  \mu} \left\{\hat{S}_{\alpha i}, \hat{J}_{\mu \lambda i}\right\} \right)
\\
\label{dSdt}
\frac{d\hat{S}_{\lambda \mu i}}{dt} &=& -\omega_{\mu \lambda} \hat{J}_{\lambda \mu i} \nonumber \\
&-&\frac{1}{\hbar \epsilon \epsilon_0} \sum_{\gamma \neq \lambda \mu} \sum_{m} \hat{D}_{m} u_{m}\left(\mathbf{r}_i\right) \left[ d_{\gamma \lambda} \hat{J}_{\gamma \mu i} - d_{\mu \gamma} \hat{J}_{\lambda \gamma i} \right] \nonumber
\\
&+& \frac{1}{4} N_e \sum_\alpha \sum_{\gamma \neq \lambda, \mu}  \left\{\hat{S}_{\alpha i},  \Xi_{\alpha, \gamma  \lambda} \hat{J}_{\gamma \mu i} - \Xi_{\alpha, \gamma  \mu} \hat{J}_{\lambda \gamma i}\right\}
\\
\label{dJdt}
\frac{d \hat{J}_{\lambda \mu i}}{dt} &=& \omega_{\mu \lambda} \hat{S}_{\lambda \mu i} \nonumber \\
&-& \frac{1}{\hbar \epsilon \epsilon_0} \sum_{m} \hat{D}_{m} u_{m}\left(\mathbf{r}_i\right) \left[ 2 d_{\lambda \mu} \left( \hat{n}_{\lambda i} - \hat{n}_{\mu i} \right)  + \sum_{\gamma \neq \lambda, \mu} d_{\mu \gamma} \hat{S}_{\lambda \gamma i}  - d_{\gamma \lambda} \hat{S}_{\gamma \mu i}\right] \nonumber 
\\
&+& \frac{1}{4} N_e \left(\sum_{\alpha} 2 \Xi_{\alpha, \lambda  \mu} \left\{ \hat{S}_{\alpha i}, \hat{n}_{\lambda i} - \hat{n}_{\mu i} \right\} + \sum_{\gamma \neq \lambda, \mu}  \left\{\hat{S}_{\alpha i},  \Xi_{\alpha, \gamma  \mu} \hat{S}_{\lambda \gamma i} - \Xi_{\alpha, \gamma  \lambda} \hat{S}_{\gamma \mu i}\right\} \right) 
\end{eqnarray}
\end{widetext}

The brackets denote the anti-commutator $\{ \hat{O}_1, \hat{O}_2\} = \hat{O}_1 \hat{O}_2 + \hat{O}_2 \hat{O}_1$. These equations are very general and can be used for further studies of the quantum dynamics of the system. For instance, the spontaneous emission can be obtained from the evolving the system from a vacuum photon state \cite{dutra_cavity_2005}. Another interesting topic would be to study the ground state correlations between photons and electrons that appear in the ultra-strong light-matter coupling regime \cite{ciuti_quantum_2005}. Each of this topics opens perspectives on future works and requires special attention owe to the relative complexity introduced by the dipole-dipole coupling terms. Here we pursue with  the semi-classical dynamics of the system.  

\subsubsection{State averaging} \label{parStateAve}

We suppose that the cavity mode fields are in coherent states, $a_m \ket{\alpha_m}=\alpha_m \ket{\alpha_m}$, and that there are no quantum correlations between photonic and electronic states, such as the  following   factorization is valid:

\begin{eqnarray}
\expval{(a_m^\dagger - a_m) \hat{c}^\dagger_{\lambda i}\hat{c}_{\mu i} } \approx  \expval{a_m^\dagger - a_m} \expval{\hat{c}^\dagger_{\lambda i}\hat{c}_{\mu i}  } \nonumber
\\
= (\alpha_m^* - \alpha_m) \expval{\hat{c}^\dagger_{\lambda i}\hat{c}_{\mu i} }
\label{factorization}
\end{eqnarray}

In the following, instead of the photon creation and annihilation operators we will work with the real-valued field quadratures defined as:

\begin{eqnarray}
D_m = \langle \alpha_m |\hat{D}_m | \alpha_m \rangle 
\\
A_m = -i\langle \alpha_m | [ a_m^\dagger a_m, \hat{D}_m] | \alpha_m \rangle 
\end{eqnarray}

The evolution equations lead to terms with four fermion operators. In order to obtain their state average we apply the Wick's theorem \cite{kira_semiconductor_2012}. It is convenient to use the transform from Eq.(\ref{defci}) in order take explicitly advantage of the long-wavelength approximation. Furthermore, we will neglect coherences between states with different momenta. We will assume that momentum exchange processes lead generally to dissipation, and thus their effect will be incorporated in the dissipation rates for the electronic degrees of freedom. We thus obtain:

\begin{eqnarray}
    \langle \hat{c}_{\lambda i}^\dagger \hat{c}_{\mu i} \hat{c}_{\alpha i}^\dagger \hat{c}_{\beta i} \rangle = \nonumber
    \\
    \frac{1}{N_e^2} \langle \sum_{\mathbf{k}, \mathbf{q}}  \hat{c}_{\lambda \mathbf{k}}^\dagger \hat{c}_{\mu \mathbf{k} + \mathbf{q}} e^{i \mathbf{q} \mathbf{r}_i} \sum_{\mathbf{k'} \mathbf{q'}}  \hat{c}_{\alpha \mathbf{k'}}^\dagger \hat{c}_{\beta \mathbf{k'} + \mathbf{q'}} e^{i \mathbf{q'} \mathbf{r}_i} \rangle  \nonumber  
    \\
    \approx 
    \frac{1}{N_e^2} \sum_{\mathbf{k}, \mathbf{k'}} \langle \hat{c}_{\lambda \mathbf{k}}^\dagger \hat{c}_{\mu \mathbf{k}} \hat{c}_{\alpha \mathbf{k'}}^\dagger \hat{c}_{\beta \mathbf{k'}} \rangle \sum_{\mathbf{q} \mathbf{q'}} e^{i \mathbf{q} \mathbf{r}_i} e^{i \mathbf{q'} \mathbf{r}_i} \nonumber 
    \\
    = \frac{1}{N_e^2} \sum_{\mathbf{k}, \mathbf{q}} \langle \hat{c}_{\lambda \mathbf{k}}^\dagger \hat{c}_{\mu \mathbf{k}} \rangle e^{i \mathbf{q} \mathbf{r}_i} \sum_{\mathbf{k'} \mathbf{q'}} \langle\hat{c}_{\alpha \mathbf{k'}}^\dagger \hat{c}_{\beta \mathbf{k'}} \rangle e^{i \mathbf{q'} \mathbf{r}_i} \nonumber 
    \\
     = \langle \hat{c}_{\lambda i}^\dagger \hat{c}_{\mu i} \rangle \langle\hat{c}_{\alpha i}^\dagger \hat{c}_{\beta i} \rangle
\end{eqnarray}

The four fermions correlation functions are thus reduced to products of two fermion correlations, which are respectively the ground state averaged populations, polarizations and currents:

\begin{equation}
N_{\lambda i} =\expval{\hat{n}_{\lambda i}},
 S_{\lambda \mu i} =\expval{\hat{S}_{\lambda \mu i}}, J_{\lambda \mu i} =\expval{\hat{J}_{\lambda \mu i}}
\end{equation}

We now examine the dissipation dynamics of the system that is contained in the $\Lambda$ term of Eq. (\ref{EqGeneralO}). We discuss separately the photonic and electronic degrees of freedom. 

\subsubsection{Losses and dynamics for a multi-mode cavity} \label{SecCavity}

\begin{figure}[h!]
    \centering
    \includegraphics[scale=0.25]{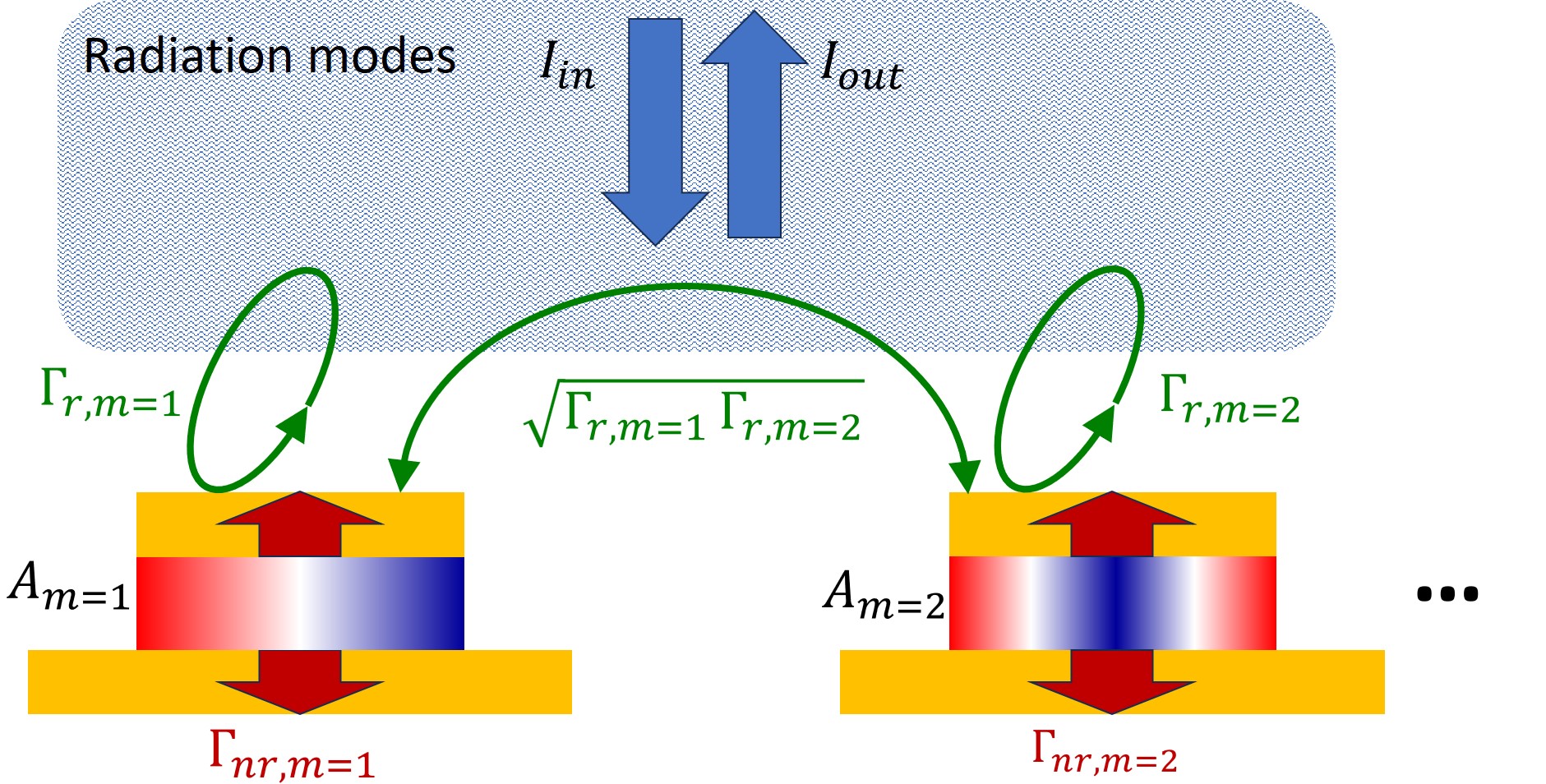}  
    \caption{Cavity loss mechanisms. Each mode is characterized by a non-radiation loss rate $\Gamma_{nr, m}$ and a radiation loss rate $\Gamma_{r, m}$. The later characterize its coupling to the continuum of radiation modes, and in particular to the incoming field $I_{in}$ and the out-coming field $I_{out}$. The coupling to the radiation continuum also induces coupling between modes with a characteristic strengths $\sqrt{\Gamma_{r, m} \Gamma_{r, m'}}$, as illustrated here for the case of the modes $m=1$ and $m=2$.}
    \label{fig: CavLoss}
\end{figure}

As illustrated in Fig. \ref{fig: CavLoss}  each  microcavity resonance is characterized by two types of loss mechanisms: radiation loss and non-radiation loss. The non-radiation loss, $\Gamma_{nr,m}$ describes the irreversible dissipation in the metal and dielectric parts of the cavity: i.e. ohmic loss owe to currents induced in the metal layers, heat dissipated in a lossy dielectric owe to electronic or phononic transitions other than those considered in the polarization field operator Eq.(\ref{EqP}). The coupling to the radiation continuum is quantified by loss rates $\Gamma_{r,m}$. The microcavity interacts with an incoming field $I_{in}$, and, as a result, there is out-coming (reflected) field $I_{out}$. Here we consider the case where only the reflected channel is present, and there is no transmission channel. Furthermore, we consider the period of the cavity array depicted in Fig. \ref{fig: global} to be sufficiently small such as diffracted orders do not propagate in the frequency range of interest. 

The $\Lambda$ operator for the cavity modes from Eq. (\ref{EqGeneralO}) must thus account for the input-output relations for the microcavity modes, such as the ones retrieved from the coupled-mode theory, CMT \cite{Zanotto2014, Jeannin_NLett2020, Rodriguez_22}. However, in CMT the cavity field is described by a single amplitude instead of two quadratures $D_m$ and $A_m$  and the rotating wave approximation, RWA, plays important role \cite{book_haus1984waves}.  In the present case, where we consider ultra-strongly coupled systems, we adopt an approach which is free from RWA. Finally, in the majority of the literature CMT is developed for a single cavity resonance, eventually coupled to multiple inputs, which is coherent with the RWA that describes the physics of the problem in a narrow frequency band. Here we provide an input-output theory which is valid for arbitrary number of photonic modes that can be excited at the same time at various frequencies, which is the case of nonlinear optical devices, especially in the context of high frequency harmonic generation. Our approach still satisfies the constraint of energy conservation for dissipation rates that can be arbitrary large; in particular the linewidths of consecutive photonic modes can partially overlap in frequency. This approach can thus be an efficient tool to treat non-Hermitian optical quasi-modes \cite{gigli_2020}, 

We temporally discard the coupling with matter and we consider only the photon equations: 

\begin{eqnarray} 
\frac{dD_m}{dt} &=& -\omega_{c m} A_m  \label{eqdDdt}
\\
\frac{dA_m}{dt} &=&  \omega_{c m} D_m - (\Gamma_{nr , m}+\Gamma_{r, m}) A_m \nonumber\\
 &-& \sum_{m' \neq m} \sqrt{\Gamma_{r, m} \Gamma_{r, m'}} A_{m'} +   2\sqrt{\Gamma_{r, m}} I_{in} 
 \label{eqdAdt} \\
 I_{out} &=& -I_{in} + \sum_m \sqrt{\Gamma_{r, m}} A_m
 \label{eqIout}
\end{eqnarray}

 The electromagnetic energy stored in the cavity per mode is $W_m =   \hbar \omega_{cm} \langle a_m^\dagger a_m\rangle$, which can be expressed as $W_m = (V_{cav}/2\epsilon \epsilon_0)(D_m^2+A_m^2)$. 
 We can then show that the above equations lead to the following conservation law:

\begin{equation}\label{eqConsLaw}
\left.\sum_m \frac{dW_m}{dt} \right|_{\Lambda} = \frac{V_{cav}}{\epsilon \epsilon_0} \left( I_{in}^2 - I_{out}^2 - \sum_m \Gamma_{nr , m} A_m^2 \right)
\end{equation}

This equation shows that with our choice of the normalization for the variables $I_{in}$ and $I_{out}$ the incident and outcoming power should be written as:

\begin{equation}
 P_{in} =\frac{V_{cav}}{\epsilon \epsilon_0} I_{in}^2, \phantom{Q}    P_{out} =\frac{V_{cav}}{\epsilon \epsilon_0} I_{out}^2
\end{equation}

Eq. (\ref{eqConsLaw}) has clear physical interpretation, which is concordant with the conservation laws that stem from the Maxwell's equations \cite{jackson_classical_1999}. The variations of the cavity energy is due to the difference between the incoming and out-coming Poynting fluxes ($I_{in}^2 - I_{out}^2$), and decreases owe to the non-radiation loss in the cavity volume ($-\Gamma_{nr , m} A_m^2$). The price to pay in order to obtain this simple picture are two hypothesis. First, we assume that only one of the quadratures dissipate energy, while the other is non-dissipative. This hypothesis is actually realized in most systems, where the two quadratures $(D_m, A_m)$ can be associated with the electric and magnetic (or the vector potential) fields.  Consider, for example, an inductor-capacitor (LC) circuit resonator: the electric potential energy stored in the capacitor represents a conservative potential energy. Energy is dissipated in a form of a resistive loss $-Ri^2$ by the current $i$ flowing through the inductor,  and the current is proportional to the magnetic field in the inductor coil. The loss rate is thus quantified by the magnitude of the magnetic field. For patch microcavities, and more generally for metallic conductors losses arise from the eddy currents on the metal wall, which are also proportional to the tangential magnetic field  \cite{todorov_optical_2010, Jackson_book}. Thus, while the quadrature that corresponds to the electric field does not dissipate energy, the quadrature that corresponds to the vector potential does. This is reflected in the linear system of equations (\ref{eqdDdt}, \ref{eqdAdt}) where  damping term have been included for the time evolution of the quadrature $A_m$ Eq.(\ref{eqdAdt}) which represent the vector potential, but not for the quadrature (\ref{eqdDdt}) which represents the electric field. 

The second hypothesis, as can be seen from the second line of Eq. (\ref{eqdAdt}) is that we now must assume that there is mutual coupling between the cavity modes induced by coupling with the external radiation (term $\sqrt{\Gamma_{r, m} \Gamma_{r, m'}} A_{m'}$). This term is needed to cancel properly the contribution of the radiation loss in the conservation law, Eq. (\ref{eqConsLaw}), so that only non-radiative loss contributes to the energy dissipation inside the cavity volume. We can argue that these cross terms are not of a numerical significance for modes with sufficiently narrow linedwidths. But these terms become significant for modes with strong radiation loss, such as the ones found with antennas \cite{jeannin_unified_2021} or all-dielectric cavities \cite{sarma_all-dielectric_2022}. We found that the presence of these terms is essential for of ultra-strongly coupled systems, where, for instance, single electronic transition is coupled to several photonic modes, and the interaction strength is so important that it can cover large frequency band.
For instance, discarding these terms leads to numerical inconsistencies, such as frequency regions where the reflectivity is larger than unity. 

In order to illustrate the physical relevance of these terms consider the case where the index $m$ does not refer to the various frequency modes of a single frequency resonator, but to different spatial positions of $M$ identical resonators in an array, while each resonator supports a mode with an identical frequency $\omega_{c0}$. For the case of a sub-wavelength array periodicity all resonators are identically excited by an incident wave, $A_m = A$, and all radiations loss factors are identical $\Gamma_{r,m} = \Gamma_r$. Then Eq. (\ref{eqdAdt}) becomes:

\begin{equation}
\frac{dA}{dt} = \omega_{c 0} D - (\Gamma_{nr}+ M\Gamma_{r}) A + 2\sqrt{\Gamma_{r}}  I_{in} 
\end{equation}

The total radiation loss for each resonator is multiplied by the number of resonators, $M\Gamma_r$. As a consequence, the line-width of the photonic mode as well as its ability to radiate energy increases, which is the photonoic counterpart of the superradiance phenomenon that has been observed, for instance, with metamaterial arrays \cite{Rodriguez_22,Singh_APL2010, Wenclawiak_2017, Keller_2018}.

We now consider the equations in the context of multimode cavities for an incident monochromatic field at a frequency $\omega$. Since the system in the absence of matter excitation is linear, all fields will evolve in a harmonic manner, and we define complex amplitudes for all fields as $I_{in} = \tilde{I}_{in}e^{i\omega t} + \tilde{I}^*_{in}e^{i\omega t}$, etc. It can then be shown that the equation set  Eq. (\ref{eqdDdt}-\ref{eqIout}) can be solved analytically to yield the result (see Appendix \ref{AppendixB}):

\begin{eqnarray}
 \tilde{A}_m = 2\tilde{I}_{in} \frac{i\omega \sqrt{\Gamma_{r , m}}}{\Delta_{cnr,m}(\omega)} \frac{1}{1+i\omega Y(\omega)}  \label{Am_nomatter}
 \\
 \Delta_{cnr,m}(\omega) = \omega_{cm}^2 - \omega^2 + i\omega \Gamma_{nr , m}
 \\
 Y (\omega) = \sum_m \frac{\Gamma_{r , m}}{\Delta_{cnr,m}(\omega)} \label{defY}
\end{eqnarray}

We can also define a time-averaged mode energy $\bar {W}_m =(V_{cav}/2\epsilon \epsilon_0)(|\tilde{D}_m|^2+|\tilde{A}_m|^2)$ which is constant after averaging the terms that oscillate at $2\omega$. The conservation law Eq. (\ref{eqConsLaw}) is thus expressed as: 

\begin{equation}
 |\tilde{I}_{in}|^2 = |\tilde{I}_{out}|^2 + \sum_m    \Gamma_{nr , m} |\tilde{A}_m|^2 
\end{equation}

The time-averaged incident and out-coming power are now expressed as follows from the complex amplitudes:

\begin{equation} \label{defPower}
 P_{in} =\frac{2V_{cav}}{\epsilon \epsilon_0} |\tilde{I}_{in}|^2, \phantom{Q}    P_{out} =\frac{2V_{cav}}{\epsilon \epsilon_0} |\tilde{I}_{out}|^2
\end{equation}

We can now define a frequency dependent reflectivity coefficient $R(\omega) = |\tilde{I}_{out}|^2/|\tilde{I}_{in}|^2$ that is expressed analytically as:

\begin{equation}
R(\omega) =  1 - \frac{4\omega^2}{|1+i\omega Y(\omega)|^2} \sum_m \frac{\Gamma_{nr , m}\Gamma_{r , m}}{|\Delta_{cnr,m}(\omega)|^2}   \label{Reflvoidcav}
\end{equation}

It can be further shown (Appendix \ref{AppendixB}) that this coefficient is always comprised between 0 and 1, while the dissipation rates can take arbitrary values.  It is also easily verified that for a single mode and within the RWA this expression allows recovering previous results obtained with CMT, see for instance Ref.\cite{Zanotto2014}.

\subsubsection{Electronic dissipation}\label{parElDissipation}

In the case of the electronic system, dissipation can be treated within the density matrix approach \cite{boyd_nonlinear_2008, ahn_calculation_1987, Iotti_2005}. We will consider two types of dissipation: population relaxation, characterized by transition times $\tau_{\lambda \mu}$ between the level $\mu$ to the level $\lambda$, and decoherence rates $\gamma_{\lambda \mu}$ for the transition $\lambda \rightarrow \mu$.  Typical relaxation mechanisms for intersubband systems are described, for instance in Ref. \cite{ferreira_evaluation_1989}. Instead of the  matrix elements for the density matrix $\langle c^\dagger_{\lambda i} c_{\mu i} \rangle$ we use the electronic quadratures, $S_{\lambda \mu i}$ and $J_{\lambda \mu i}$. In this approach, we can avoid RWA and take into account the anti-resonant terms in the light-matter interaction, in the spirit of the USC regime \cite{ciuti_quantum_2005}. Furthermore, just like for the photonic degrees of freedom, we consider that only one of the electronic quadratures, the current  $J_{\lambda \mu i}$, is dissipative:

\begin{eqnarray}
\left.  \frac{d{N}_{\lambda i}}{dt} \right|_{\Lambda} &=& \sum_{\mu > \lambda} \frac{N_{\mu i}-N_{\mu i}^0}{\tau_{\lambda \mu}} -  \sum_{\mu < \lambda} \frac{N_{\lambda i}-N_{\lambda i}^0}{\tau_{\mu \lambda}} \label{Pop_decay}
\\
\left.  \frac{d{S}_{\lambda \mu i}}{dt} \right|_{\Lambda} &=& 0 \label{S_decay}
\\
\left.  \frac{d{J}_{\lambda \mu i}}{dt} \right|_{\Lambda} &=& - \gamma_{\lambda \mu} J_{\lambda \mu} \label{J_decay}
\end{eqnarray}

In Eq. (\ref{Pop_decay}) we introduced the equilibrium populations $N_{\mu i}^0$ which are obtained by assuming that there is thermal equilibrium in the system prior to the light-matter interaction process.  The interaction with light induces an excess population in certain levels. Eq. (\ref{Pop_decay}) expresses the fact that the population in level $\lambda$ increases through the decay from the excess population from all levels $\mu$ at a higher energy. The excess population of level $\lambda$ decays down towards levels with lower energy. Eventually, these equation can be supplemented with terms that describe electronic injection from external contacts in order to describe the photo-detection process \cite{pisani_electronic_2023}. Spontaneous and stimulated emission can be considered as well, but here we will assume that the photon energy density inside the cavity mode is sufficiently low so that the stimulated emission and population inversion can be discarded. Consequently, we consider that the excess populations $N_{\mu i}-N_{\mu i}^0$ are all positive numbers. Eventually, the spontaneous emission can contribute to the relaxation time and decoherence rates \cite{book_QCL_faist}. Let us note that a proper treatment of the spontaneous emission from the system would require to evolve the system from the vacuum state and performing a calculation beyond the factorization scheme  (\ref{factorization}) \cite{book_OptResonance}. Instead, in the semiclassical approach considered here all output photon fields are proportional to the powers of the incident field $I_{in}$.

In Eqs. (\ref{S_decay}) and (\ref{J_decay}) we introduced of the coherence damping constant, $\gamma_{\lambda \mu}$, only in the evolution equation of the microcurrent quadrature Eq. (\ref{J_decay}), but not for the evolution of the polarization quadrature, Eq. (\ref{S_decay}). This choice has a physical significance and can be justified as follows. The electronics resonances coupled with light are collective excitations of plasmonic character driven by dipole-dipole interactions between electrons. The polarization term $S_{\lambda \mu i}$ is linked to the electrostatic energy, which is conservative, while the current term $J_{\lambda \mu i}$ is linked to the kinetic energy of the electrons (kinetic inductance) which is responsible for energy dissipation \cite{KhurginMRS2012, khurgin_how_2015}. Just like discussed in the previous paragraph for the photonic degrees of freedom, these equations allow treating the problem without invoking RWA.

In summary, in our formulation the dissipation is modeled by damping terms introduced for the quadratures $A_m$ and $J_{\lambda \mu i}$. The light-matter coupling term, Eq. (\ref{eqLMintLattice}) is proportional to the product of the conservative quadratures $D_m S_{\lambda \mu i}$. 

\subsubsection{Full set of equations} \label{par: full set}

Applying the state averaging and the semi-classical factorization described in  paragraphs \ref{parStateAve} on the Hamiltonian evolution equations from paragraph \ref{secHamiltonian_evolution} one obtains the equation of motion that couple the mean values of electronic and photonic quadratures. These equations are supplemented with the linear damping terms and a driving term as described in paragraphs \ref{SecCavity} and \ref{parElDissipation}.  For the following, it is convenient to introduce the quantity:

\begin{equation}
    D_t(\mathbf{r}_i) = \frac{1}{\hbar \epsilon \epsilon_0} \sum_{m} {D}_{m} u_{m}\left(\mathbf{r}_i\right) \label{deftotalD}
\end{equation}

This quantity is proportional to the total displacement field in the resonator. Furthermore, the following quantity appears systematically in the equations to follow:

\begin{equation}
    F_{\lambda \mu}(\mathbf{r}_i) = d_{\lambda \mu}  D_t(\mathbf{r}_i) - \frac{N_e}{2} \sum_{\alpha} \Xi_{\alpha, \lambda \mu} {S}_{\alpha i}
\end{equation}

This quantity can be seen as the light-matter interaction term for the transition between levels  $\lambda$ and $\mu$ corrected with the local field created by all dipoles of the system (including the dipole $\lambda\mu$ itself). By using these notation, we obtain the 
 following semi-classical  Maxwell-Bloch set of equations:

\begin{widetext}
\begin{eqnarray}
 \frac{dA_m}{dt} &=& \omega_{c m} D_m  - \Gamma_m A_m - \sum_{m' \neq m} \sqrt{\Gamma_{r, m} \Gamma_{r, m'}} A_{m'} 
 +  \frac{\omega_{cm}}{V_{cav} } \sum_{\alpha, i} d_\alpha S_{\alpha i}   u_m(\mathbf{r}_i) +  2\sqrt{\Gamma_{r, m}} I_{in} \label{dAdt_av}
 \\ 
\frac{dD_m}{dt} &=& - \omega_{c m} A_m \label{dEdt_av} 
\\
\frac{d{N}_{\lambda i}}{dt} &=& \sum_{\mu > \lambda} \frac{N_{\mu i}-N_{\mu i}^0}{\tau_{\lambda \mu}} -  \sum_{\mu < \lambda} \frac{N_{\lambda i}-N_{\lambda i}^0}{\tau_{\mu \lambda}} 
- \sum_{\mu \neq \lambda} F_{\lambda \mu}(\mathbf{r}_i) {J}_{\mu \lambda i} \label{Popilation_dyn}
\\
\frac{d{S}_{\lambda \mu i}}{dt} &=& -\omega_{\mu \lambda} {J}_{\lambda \mu i}  
-\sum_{\gamma \neq \lambda \mu}  \left[ F_{\gamma \lambda}(\mathbf{r}_i)  {J}_{\gamma \mu i} - F_{\mu \gamma} (\mathbf{r}_i)  {J}_{\lambda \gamma i} \right]
\label{dSdt_av}
\\
\frac{d {J}_{\lambda \mu i}}{dt} &=& \omega_{\mu \lambda} {S}_{\lambda \mu i} - \gamma_{\lambda \mu} J_{\lambda \mu i} -2F_{\lambda \mu} (\mathbf{r}_i)({N}_{\lambda i} - {N}_{\mu i})
+ \sum_{\gamma \neq \lambda \mu}  \left[ F_{\gamma \lambda}(\mathbf{r}_i)  {S}_{\gamma \mu i} - F_{\mu \gamma} (\mathbf{r}_i)  {S}_{\lambda \gamma i} \right]
\label{dJdt_av}
\\
\sum_{\lambda} N_{\lambda i} &=&  N_0 =  \mathrm{const}
\end{eqnarray}
\end{widetext}

Here we introduced the total loss of the resonator, $\Gamma_m= \Gamma_{nr , m}+\Gamma_{r, m}$. The last equation of the set expresses the conservation of total number of electrons in the system, fixed by the constant $N_0$; it is implicitly present in the model, as our Hamiltonian is particle conserving. 
The first two equations, (\ref{dAdt_av}) and (\ref{dEdt_av}), describe the evolution of the quadratures of the cavity modes $(A_m, D_m)$. The last term of the right hand side is the external pump, while the second to the last is the light-matter coupling term. The subsequent equations describe respectively the dynamics of the populations, and the two fermionic quadratures that we have identified as polarizations "$S_{\alpha i}$" and microcurrents "$J_{\alpha i}$". We can notice that the equations describing the photonic degrees of freedom are linear, while those for  the electronic degrees of freedom are nonlinear. This is because of our choice to work in the PZW representation, where the square vector potential term is absent, and the nonlinearities of the optical response arise only from the matter degrees of freedom \cite{cohen-tannoudji_photons_1986}. Our equations have been developed on the set of confined cavity modes, thus they contain naturally the microcavity effects, as well as the possibility to reach the ultra-strong light-matter coupling regime. 
Furthermore, our equation set accounts for many-body effects through the coupling terms $\Xi_{\alpha, \lambda  \mu}$. We will show in particular that the effect of the depolarization shift is contained in these equations, and actually plays an important role for the behavior of nonlinear susceptibilities in the limit of high electronic densities. Furthermore, these equations couple evolution of the populations to the coherences, which is inherent for fermionic systems. The nonlinearities of the optical response are thus derived here from a microscopic picture, which naturally  accounts for many-body interactions and local field effects. 

In the following we consider a monochromatic drive, $I_{in} \propto  e^{i\omega t}+e^{-i\omega t}$ and solve the system by an iterative process. Clearly we can envision more complex situations with multiple frequency inputs, but it is already interesting to examine the new features brought by many-body and cavity effects in the monochromatic case. Since the system is nonlinear, the response to the monochromatic drive at a frequency $\omega$ will contain all multiples of the driving frequency, $\pm n\omega$. We can then expand the time-evolution of the any matter observable $O(t)$ into discrete Fourier series:

\begin{eqnarray}
  O(t) = \sum_n O^{(n)} (t) 
  \nonumber \\ = \sum_n [\tilde{O}^{(n)} e^{n i \omega t} + \tilde{O}^{(n)*} e^{-n i \omega t}] \nonumber \\
  = \sum_n 2\Re [\tilde{O}^{(n)} e^{n i \omega t}] \label{evO}
\end{eqnarray}

This equation is also a general definition of the complex amplitudes $\tilde{O}_{n}$ we systematicaly use in the following. The product terms $\tilde{O}_{1}^{(n)}\tilde{O}_{2}^{(\pm m)}$ in the equation set thus couple the low order terms $n\omega$ and $m\omega$ to the higher and lower  orders $(n \pm m)\omega $  and the system can be solved interratively \cite{Landau1976Mechanics}.  This method allows a systematic derivation of the linear and nonlinear response of the system, as shown in the next section.
 
\subsection{General theory for linear and nonlinear susceptibilities} \label{General theory chi}

We will start the analysis of our equations by considering that the cavity field  $D_t(\mathbf{r}_i)$ is imposed upon the system. As a result we will obtain expressions for the effective linear and nonlinear susceptibilities of the electronic system embedded in the microcavity. The full dynamics of the photon field as a function of the external pump is derived in section \ref{secMicrocavity_effects}. 

The response of the electronic system will be expressed through  the polarization field introduced by Eq.(\ref{EqP}). Since the electric field for the  modes that we consider is independent from $z$, we can average the polarization field  over the microcavity thickness, thus obtaining an  expression that depends only on the planar coordinate:

\begin{equation}
\bar{P}(\mathbf{r}, t)  = \frac{1}{L_{cav}} \sum_{\alpha i} d_\alpha S_{\alpha i} (t) \delta (\mathbf{r}-\mathbf{r}_i) \label{generalP}
\end{equation}

The nonlinear response is thus provided by the successive orders of the polarizations  $\tilde{S}^{(n)}_{\alpha i}$. Considering the response up to third order, we define the nonlinear susceptibilities as:

\begin{eqnarray}
 \bar{P}(\mathbf{r}, t) = \chi^{(1)} (\omega) (\hbar \varepsilon \varepsilon_0)\tilde{D}^{(1)}_t(\mathbf{r})e^{i\omega t}  
 \nonumber \\
 + \chi^{(2)} (2\omega; \omega, \omega) (\hbar \varepsilon \varepsilon_0)^2\tilde{D}^{(1)}_t(\mathbf{r})^2e^{2i\omega t}
 \nonumber \\
 + \chi^{(3)} (3\omega; \omega, \omega, \omega) (\hbar \varepsilon \varepsilon_0)^3\tilde{D}^{(1)}_t(\mathbf{r})^3e^{3i\omega t}
 \nonumber \\
 +  \chi^{(3)} (\omega; \omega, \omega, -\omega) (\hbar \varepsilon \varepsilon_0)^3|\tilde{D}^{(1)}_t|^2 \tilde{D}^{(1)}_t(\mathbf{r})e^{i\omega t} + ...+
 \nonumber \\
 + (n>3) ... + \mathrm{c. c.} \label{Def_chi_n}
\end{eqnarray}

Terms like the one expressed in the forth line correct the first order response and are related to saturation effects in the system stemming from the population dynamics. The example of the two-subband quantum well discussed in section \ref{SecTwosubband_system} is a case where these terms can be treated exactly at any order of the incident field. In the following, we consider specifically the susceptibilities $\chi^{(2)} (2\omega; \omega, \omega)$ and $\chi^{(3)} (3\omega; \omega, \omega, \omega)$ which are important for the second and third harmonic generation. 

The factors $(\hbar \varepsilon \varepsilon_0)^n$ compensate an identical factor in the definition Eq. (\ref{deftotalD}). With that choice the first order susceptibility $\chi^{(1)} (\omega)$ is dimensionless quantity that is discussed in the next paragraph.

\subsubsection{Linear susceptibility and bosonisation}\label{secLinear}

We start with a discussion of the first order response (linear optical susceptibility) and its link to the bozonitaion approach derived previously to treat the many-body interactions in the system \cite{todorov_intersubband_2012}. For the linear susceptibility we retain only the first order and zero order terms in our equations.  The equation for the population dynamics, Eq. (\ref{Popilation_dyn}) indicates that leading correction owe to the optical field is either of second order, or of $0^{th}$ order. Indeed, all the terms if the second line of  equation Eq. (\ref{Popilation_dyn}) involve products of first order terms. The $0^{th}$ order correction to the population correspond to saturation effects \cite{boyd_nonlinear_2008} and can be neglected for sufficiently weak pump (see also paragraph \ref{secBistability}).  To infer the linear susceptibility we neglect the population dynamics assuming that populations are equal to their equilibrium values, $N_{\mu i}^0$. Retaining only the linear terms in Eqs. (\ref{dJdt_av}) and (\ref{dSdt_av}) we have the following equation set:

 \begin{eqnarray}
  \frac{d{S}^{(1)}_{\lambda \mu i}}{dt} = -\omega_{\mu \lambda} {J}^{(1)}_{\lambda \mu i}  \label{dSdt1storder}
\\
\frac{d {J}^{(1)}_{\lambda \mu i}}{dt} = \omega_{\mu \lambda} {S}^{(1)}_{\lambda \mu i} - \gamma_{\lambda \mu} J_{\lambda \mu i}^{(1)} \nonumber \\ -2F_{\lambda \mu}^{(1)} (\mathbf{r}_i)({N}^0_{\lambda i} - {N}^0_{\mu i}) \label{dJdt1storder}
 \end{eqnarray}

 Eliminating the variables ${J}^{(1)}_{\lambda \mu i}$ we obtain the following second order differential equation for the polarizations:

\begin{eqnarray}
\frac{d^2 S^{(1)}_{\lambda \mu i}}{dt^2} + \gamma_{\lambda \mu }\frac{d S^{(1)}_{\lambda \mu i}}{dt} + \tilde{\omega}^2_{\lambda \mu} S^{(1)}_{\lambda \mu i} \nonumber \\ + \sum_{\alpha \neq \lambda  \mu } \omega_{\lambda \mu} \Xi_{\alpha, \lambda  \mu} S^{(1)}_{\alpha i} \left( N^0_{\lambda} - N^0_{\mu} \right) \nonumber
\\
=   \frac{2}{N_e}   \omega_{\lambda \mu} d_{\lambda \mu} \left( N^0_{\lambda} - N^0_{\mu} \right) D_{t}\left(\mathbf{r}_{i}\right) \label{S1equation}
\\
\tilde{\omega}^2_{\lambda \mu} = \omega^2_{\mu \lambda} + \omega^2_{P1, \mu \lambda} (N^0_\lambda - N^0_\mu) \label{depolshift}
\end{eqnarray}

The total equilibrium population on the $\lambda^{th}$ level is noted $N^0_\lambda = N_e N^0_{\lambda i}$. The polarization $S^{(1)}_{\lambda \mu i}$ thus obeys a driven harmonic oscillator equation with a damping constant $\gamma_{\lambda, \mu}$ and a characteristic frequency $\tilde{\omega}_{\lambda \mu}$ provided by Eq. (\ref{depolshift}) where we recognize the expression of the depolarization shifted intersubband transitions \cite{book_Helm, todorov_intersubband_2012}. The polarization $S^{(1)}_{\lambda \mu i}$ is coupled linearly with the polarizations form the other transitions, $\alpha \neq \lambda  \mu$, as described by the second line of Eq. (\ref{S1equation}): this is the plasmon-plasmon coupling that was introduced in Ref. \cite{todorov_intersubband_2012}. Finally, it is driven by the optical field, as expressed from the last line of Eq. (\ref{S1equation}). The expression of the first line of Eq. (\ref{S1equation}) which is exactly that of a damped harmonic oscillator with a frequency $\tilde{\omega}_{\lambda \mu}$ and damping constant $\gamma_{\lambda \mu }$ is a consequence of the damping equations (\ref{S_decay}) and (\ref{J_decay}) which justifies a posteriori our choice for the dissipative dynamics of the electronic quadratures. 

Considering the harmonic evolution of the driving field, $\tilde{D}_t \propto e^{i\omega t}$, the equation set Eq. (\ref{S1equation}) becomes a linear system for the complex amplitudes $\tilde{S}^{(1)}_{\alpha i}$, $\alpha = \lambda \mu, \lambda < \mu$:

\begin{eqnarray}
\sum_{\beta}  M_{\alpha \beta} (\omega)   \Bigg[ \frac{\tilde{S}^{(1)}_{\beta i}}{\sqrt{\Delta n_{\beta}}}\Bigg] =\frac{2S}{N_e} \omega_{\alpha}d_\alpha \sqrt{\Delta n_{\alpha}} \tilde{D}^{(1)}_t(\mathbf{r}_i)
\\
M_{\alpha \beta} (\omega) = \delta_{\alpha \beta} \Delta_{\alpha} (\omega) + \sqrt{\frac{\omega_\alpha}{\omega_\beta}}\omega_{P\alpha}\omega_{P\beta} C_{\alpha \beta}(1-\delta_{\alpha \beta}) \label{defMab}
 \\
 \Delta_{\alpha} (\omega) =  \tilde{\omega}^2_\alpha -\omega^2 + i\omega \gamma_\alpha
 \label{defDeltaa}
\end{eqnarray}

 We introduced the areal population density differences $\Delta n_{\alpha=\lambda \mu} = (N^0_\lambda -N^0_\mu)N_e/S$. Here $\omega_{P\alpha}=\sqrt{e^2 \Delta n_\alpha/m^* \varepsilon \varepsilon_0 L_\alpha}$ is the plasma frequencies of the intersubband transition $\alpha$  defined as in Ref. \cite{todorov_intersubband_2012}. The matrix $M_{\alpha \beta} (\omega)$ introduced in the above equations plays important role in our theory. Note that this matrix becomes diagonal if the electron-electron interaction is neglected. 

By inverting the linear system we obtain the first order polarizations and micro-currents as a function of the driving field:

\begin{eqnarray}
 \tilde{S}^{(1)}_{\beta i} =\frac{S}{N_e}  \chi^{(1)}_\beta (\omega)   \tilde{D}^{(1)}_t(\mathbf{r}_i) \label{S1toD}
 \\
  \tilde{J}^{(1)}_{\beta i} =-\frac{S}{N_e} \frac{i\omega}{\omega_\beta} \chi^{(1)}_\beta (\omega)   \tilde{D}^{(1)}_t(\mathbf{r}_i)
 \\
 \chi^{(1)}_\beta (\omega) =   \sum_{\alpha} M^{-1}_{\beta \alpha} (\omega) 2 \omega_{\alpha} d_\alpha \sqrt{\Delta n_{\alpha} \Delta n_{\beta}} \label{chi1toD}
\end{eqnarray}

By inserting this result in Eq. (\ref{generalP}) and using the identity  Eq. (\ref{faveraging}) we obtain the expression of the first order susceptibility:

\begin{eqnarray}
 \chi^{(1)} (\omega) = \frac{1}{\hbar \varepsilon \varepsilon_0 L_{cav}} \sum_{\beta} d_{\beta} \chi^{(1)}_\beta (\omega) =
 \nonumber \\
 =\frac{1}{\hbar \varepsilon \varepsilon_0 L_{cav} } \sum_{\alpha \beta} M^{-1}_{ \beta \alpha} (\omega) 2\omega_{\alpha} d_\alpha d_\beta \sqrt{\Delta n_\alpha \Delta n_\beta}
\label{def_chi_1}
\end{eqnarray}

The frequencies of the collective electronic states are determined from the poles of the first order susceptibility, $1/\chi^{(1)} (\omega) = 0$. This is identical to the condition:

\begin{equation}
    \mathrm{det} [ M_{\alpha \beta} (\omega)] = 0 \label{condMab}
\end{equation}

The oscillator strengths of the collective excitations are then provided by the weight of the poles of the function $\chi^{(1)} (\omega)$. Our description thus encompasses the various phenomena such the oscillator strength transfer \cite{Delteil_APL_2013}, as well as ENZ description of the collective modes \cite{Fomra_reviewENZ_2024}. Indeed, from Eq. (\ref{def_chi_1}) we can define the collective modes form the condition that they yield non-zero polarization even for vanishing incident field; this condition is fulfilled in the limit $1/\chi^{(1)} (\omega) \rightarrow 0$.

As an example, consider the case of a single intersubband transition where the linear susceptibility becomes:

\begin{equation}
    \chi^{(1)} (\omega) = f_w f_\alpha \frac{\omega^2_{P \alpha}}{\Delta_{\alpha} (\omega)} \label{chi1_2sub}
\end{equation}

Here $f_w = L_\alpha/L_{cav}$ is the filling factor of the quantum well system in the cavity, and $f_\alpha = 2 m^* \omega_\alpha d_\alpha^2/e^2\hbar$ is the oscillator strength of the transition. This expression leads to the effective dielectric function of cavity coupled intersubband transition which was previously discussed in Ref. \cite{todorov_intersubband_2012}, see also section \ref{secMicrocavity_effects}.

Eq. (\ref{dJdt1storder}) indicates that the electronic system is driven by the local field $F_{\lambda \mu}^{(1)} (\mathbf{r}_i)$.
Using Eqs.(\ref{S1toD}, \ref{chi1toD}) to express the first order field  $\tilde{F}^{(1)}_{\alpha}(\mathbf{r}_i)$ we have:

\begin{eqnarray}
 \tilde{F}^{(1)}_{\alpha}(\mathbf{r}_i) = \tilde{d}_\alpha (\omega) \tilde{D}^{(1)}_t(\mathbf{r}_i)
 \\
 \tilde{d}_\alpha (\omega) = d_\alpha - \frac{S}{2} \sum_{\gamma} \Xi_{\gamma \alpha} \chi_{\gamma}^{(1)} (\omega) \nonumber \\
 =d_\alpha - \sum_{\gamma \beta} d_\beta \frac{\omega_\beta}{\sqrt{\omega_\alpha \omega_\gamma}} \omega_{P\alpha}\omega_{P\gamma} C_{\gamma \alpha} M^{-1}_{\beta \gamma}(\omega) \sqrt{\frac{\Delta n_\beta}{\Delta n_\alpha}} \label{dtilde}
\end{eqnarray}

In the above equations, we have defined an effective frequency dependent dipole $\tilde{d}_\alpha (\omega)$ of the $\alpha$ transition that takes into account the local field correction from other dipoles excited in the system, including the dipole $d_\alpha$ itself.  For example, if only one transition is present in the system we have:

\begin{eqnarray}
   \tilde{d}_\alpha (\omega) = d_\alpha \frac{\Delta_{\alpha}^0 (\omega)}{\Delta_{\alpha} (\omega)}  \label{d_tilde}
   \\
  \Delta_{12}^0 (\omega) = \omega_\alpha^2 - \omega^2 + i\gamma_\alpha \omega
\end{eqnarray}

Here $\Delta_{12}^0 (\omega)$ is a characteristic function of the intersubband oscillator without the many-body effects. At low frequency the dipole is reduced by a factor $\omega_\alpha^2/\tilde{\omega}^2_\alpha$, while at high frequency the dipole coincides with the bare dipole $d_\alpha$. At resonance $\omega= \tilde{\omega}_\alpha$ the dipole is enhanced by a factor $\sim \omega_{P\alpha}^2/(\gamma_\alpha \tilde{\omega}_\alpha)$; this is yet another expression for the ENZ effect \cite{Fomra_reviewENZ_2024}.

Finally, we can show that our results for the linear susceptibility are consistent with the bosonization approach previous employed for the description of collective intersubband excitations \cite{ciuti_quantum_2005, todorov_intersubband_2012}. To that end, we introduce the operators:

\begin{equation}
    \hat{b}_{\lambda \mu i} = \hat{c}^\dagger_{\lambda i} \hat{c}_{\mu i}/\sqrt{(N^0_\lambda -N^0_\mu)}
\end{equation}

such as we can replace the fermionic commutation rules (\ref{EqComSJ}), (\ref{EqComSS})  and (\ref{EqComSJ}) with a single bosonic commutation rule:

\begin{equation}
    \left[ \hat{b}_{\lambda \mu i}, \hat{b}^\dagger_{\lambda' \mu' j} \right] = \delta_{\lambda \lambda'} \delta_{\mu \mu'} \delta_{i,j}
\end{equation}

The electronic Hamiltonian (\ref{ElectronicH}) is then expressed as:

\begin{eqnarray}
    \hat{H}_e = \sum_{\alpha, i} \hbar \omega_{\alpha} \hat{b}^\dagger_{\alpha i} \hat{b}_{\alpha i} \nonumber \\
    +\sum_{\alpha, \beta, i}  \frac{\hbar \bar{\Xi}_{\alpha \beta}}{4} \left( \hat{b}^\dagger_{\alpha i} + \hat{b}_{\alpha  i} \right) \left( \hat{b}^\dagger_{\beta i} + \hat{b}_{\beta i} \right) \label{boson3}
    \\
    \bar{\Xi}_{\alpha \beta} = \frac{\omega_{P\alpha}\omega_{P\beta}}{\sqrt{\omega_\alpha \omega_\beta}} C_{\alpha \beta}
\end{eqnarray}

This expression is identical to the electronic Hamiltonian considered in Ref. \cite{todorov_intersubband_2012}, but now written on the position lattice. We now define the polarization and current quadratures in the bosonisation approximation as 

\begin{eqnarray}
    \hat{S}_{\alpha i}^{(b)} &=& \hat{b}_{\alpha i}^\dagger + \hat{b}_{\alpha i} ,\\
    \hat{J}_{\alpha i}^{(b)} &=& i \left( \hat{b}_{\alpha i} - \hat{b}_{\alpha i}^\dagger \right),
\end{eqnarray}

Their equations of motion are derived in a straightforward way from the Heisenberg evolution equation and using the bosonic commutation relations:

\begin{eqnarray}
    \frac{d\hat{S}^{(b)}_{\alpha i}}{dt} &=& -\omega_{\alpha} \hat{J}_{\alpha i}^{(b)} 
   \label{1boson} \\
    \frac{d\hat{J}^{(b)}_{\alpha i}}{dt} &=& \omega_{\alpha} \hat{S}^{(b)}_{\alpha i} +  \sum_{\beta} \bar{\Xi}_{\alpha, \beta} \hat{S}^{(b)}_{\beta i} \label{2boson}
\end{eqnarray}

We derive Eq.(\ref{1boson}) over time and we replace in Eq.(\ref{2boson}) to obtain a second order equation for $\hat{S}^{(b)}_{\alpha i}$ only. Then the Hopfield-Bogoliubov diagonalization procedure used to solve the bosonized Hamiltonian \cite{hopfield_theory_1958} can be shown to be equivalent to the condition $d^2\hat{S}^{(b)}_{\alpha i}/dt^2 = -\omega^2 \hat{S}^{(b)}_{\alpha i}$ for any eigenfrequency $\omega$ of the quadratic Hamiltonian (\ref{boson3}). As a result the diagonalization condition can be cast in a form of a homogeneous linear system:

\begin{equation}
    \sum_{\beta} M_{\alpha \beta} (\omega) \hat{S}^{(b)}_{\beta i} = 0
\end{equation}

Here the matrix $M_{\alpha \beta} (\omega)$ is identical to the one from Eq. (\ref{defMab}), with $\gamma_\alpha = 0$ in Eq. (\ref{defDeltaa}). Thus the condition (\ref{condMab}) is another expression of the Hopfield-Bogoliubov equation that provides the frequencies of the coupled states described in Ref. \cite{todorov_intersubband_2012}.

\subsubsection{Second order susceptibility} \label{Gen2nd_order_chi}

Let us now consider the second order of the polarizaions and microcurrents. At this order the population must still be considered at $0^{th}$ order, as they yield third order dynamics in Eqs. (\ref{dSdt_av}, \ref{dJdt_av}). The second order evolution equations are:

\begin{eqnarray}
 \frac{d{S}^{(2)}_{\lambda \mu i}}{dt}  +\omega_{\mu \lambda} {J}^{(2)}_{\lambda \mu i} 
 \nonumber \\
= -\sum_{\gamma \neq \lambda \mu}  \left[ F_{\gamma \lambda}^{(1)}(\mathbf{r}_i)  {J}^{(1)}_{\gamma \mu i} - F_{\mu \gamma}^{(1)} (\mathbf{r}_i)  {J}^{(1)}_{\lambda \gamma i} \right] 
\\
\frac{d {J}^{(2)}_{\lambda \mu i}}{dt} - \omega_{\mu \lambda} {S}^{(2)}_{\lambda \mu i} + \gamma_{\lambda \mu} J^{(2)}_{\lambda \mu i} + 2F^{(2)}_{\lambda \mu} (\mathbf{r}_i)({N}^0_{\lambda i} - {N}^0_{\mu i})
\nonumber \\
= \sum_{\gamma \neq \lambda \mu}  \left[ F^{(1)}_{\gamma \lambda}(\mathbf{r}_i)  {S}^{(1)}_{\gamma \mu i} - F^{(1)}_{\mu \gamma} (\mathbf{r}_i)  {S}^{(1)}_{\lambda \gamma i} \right] 
\\
 F^{(2)}_{\lambda \mu}(\mathbf{r}_i) =  - \frac{N_e}{2} \sum_{\alpha} \Xi_{\alpha, \lambda \mu} {S}^{(2)}_{\alpha i} \label{F2nd}
\end{eqnarray}

From Eq.(\ref{F2nd}) we note that the local field now contains second order correction which arises from the second order evolution of the coherences. We will further consider  the harmonic evolution of the variables according to Eq. (\ref{evO}). Expressing the linear system for the variables $\tilde{S}^{(2)}_{\alpha i}$ and $\tilde{J}^{(2)}_{\alpha i}$ we obtain:

\begin{eqnarray}
 \tilde{S}^{(2)}_{\beta i}  = \frac{S}{N_e} \chi^{(2)}_\beta (\omega) \tilde{D}^{(1)}_t(\mathbf{r}_i)^2
 \\
 \chi^{(2)}_\beta (\omega) = \sum_{\alpha} M_{\beta \alpha }^{-1} (2\omega) \sqrt{\frac{\Delta n_\beta}{\Delta n_\alpha}} G_\alpha^{(2)} (\omega)
 \\
  G_{\lambda \mu}^{(2)} (\omega) = \nonumber
  \\
  \sum_{\gamma \neq \lambda \mu} \frac{ \tilde{d}_{\mu \gamma} (\omega) \chi^{(1)}_{\lambda \gamma}(\omega)}{\omega_{ \gamma \lambda}}  
  (\omega_{\gamma\lambda } \omega_{\mu \lambda } + 2\omega^2 - i\omega \gamma_{\lambda \mu} ) \nonumber
  \\
   -\sum_{\gamma \neq \lambda \mu}  \frac{\tilde{d}_{\gamma \lambda}(\omega) \chi^{(1)}_{\gamma \mu}(\omega)}{\omega_{\mu \gamma }} (\omega_{\mu \gamma} \omega_{\mu \lambda} + 2\omega^2 - i\omega \gamma_{\lambda \mu}) 
  \label{defG2_lambda_mu} 
\end{eqnarray}

The expression of the second order susceptibility thus becomes:

\begin{eqnarray}
\chi^{(2)} (2\omega; \omega, \omega) = \frac{1}{(\hbar \varepsilon \varepsilon_0)^2 L_{cav}} \sum_{\beta }d_\beta \chi^{(2)}_\beta (\omega) \nonumber \\
 =\frac{1}{(\hbar \varepsilon \varepsilon_0)^2 L_{cav}} \sum_{\alpha \beta} M_{\beta \alpha }^{-1} (2\omega) d_\beta \sqrt{\frac{\Delta n_\beta}{\Delta n_\alpha}} G_\alpha^{(2)} (\omega)
\end{eqnarray}

The resonances of the second order susceptibility are now provided by the half values for the poles of the matrix $M_{\alpha \beta} (2\omega)$ introduced previously. The effects of the electron-electron interactions can thus be traced into the structure of the matrix $M_{\alpha \beta}(\omega)$, the corrected frequency dependent dipoles $\tilde d_\alpha (\omega)$ as well as the functions $\chi^{(1)}_{\alpha}(\omega)$ which also incorporate the effect of the collective electronic states. We will provide an example for this expression in section \ref{sec3subExample}. 

If we discard the many-body effects the matrix $M_{\alpha \beta} (\omega) = \delta_{\alpha \beta}\Delta_{\alpha} (\omega)$ and the dipoles $\tilde d_\alpha (\omega)$ are replaced with $d_\alpha$. The function $\chi^{(2)} (2\omega; \omega, \omega)$ is then expressed as a sum of terms proportional to  $d_\alpha d_\beta d_\gamma/\Delta_{\alpha} (2\omega)\Delta_{\beta} (\omega)$. We can thus recognize the expression from the theory of nonlinear interactions developed for atomic vapors \cite{boyd_nonlinear_2008}; a more direct comparison requires to perform rotating wave approximation on the denominators $\Delta_{\beta} (\omega)$ and $\Delta_{\alpha} (2\omega)$. We will come back to this point for our illustration of the $\chi^{(2)} (2\omega; \omega, \omega)$ function for a three-subband system. 

\subsubsection{Third order susceptibility}

The third order susceptibility involves the dynamics of the populations $N_{\lambda i}$, as the last term in Eq. (\ref{Popilation_dyn}) involves the products $F_{\lambda \mu}(\mathbf{r}_i) {J}_{\mu \lambda i}$. There are two types of physical phenomena related to this product. The first is a correction of the $0^{th}$ order populations through the terms $\tilde{F}^{(1)}_{\lambda \mu}(\mathbf{r}_i) \tilde{J}^{(1)*}_{\mu \lambda i}+ c.c.$. These terms appear as rectification terms between the driving field with local field corrections, $F_{\lambda \mu}(\mathbf{r}_i)$, and the microcurrents ${J}_{\mu \lambda i}$, and they contribute to the nonlinear susceptibility indicated in the fourth line of Eq. (\ref{Def_chi_n}). As these type of terms count the number of excited electrons on the electronic levels upon illumination they are important for the photocurrent generation in detectors \cite{pisani_electronic_2023}. 

Dual to these terms, the products  $\tilde{F}^{(1)}_{\lambda \mu}(\mathbf{r}_i) \tilde{J}^{(1)}_{\mu \lambda i} e^{2i\omega t}+ c.c.$ yield second order dynamics in the populations through an additional contribution that oscillates at twice the driving frequency. As a result, in Eq.(\ref{dJdt_av}) there appear terms that oscillate at three times the driving frequency, leading to a contribution of the third order susceptibility $\chi^{(3)} (3\omega, \omega, \omega, \omega)$ (third line of Eq. (\ref{Def_chi_n})).

It is difficult to provide a general expression for the population dynamics as it depends on the particular configuration of the electronic levels.  On the other hand, it is clear that all second order terms $\tilde{N}^{(2)}_{\lambda i}$ will be proportional to the square of the driving photon field:

\begin{equation}
 \tilde{N}^{(2)}_{\lambda i} = H_\lambda (\omega)   \tilde{D}^{(1)}_t(\mathbf{r}_i)^2
\end{equation}

Here $H_\lambda (\omega)$ is a function of the frequency only. For instance, if we assume that the population $N_{\lambda i}$ can only relax to lower levels with a relaxation time $\tau_\lambda$, then the function $H_\lambda (\omega)$ is written explicitly:

\begin{equation}
  H_\lambda (\omega) = \frac{1}{2i\omega + 1/\tau_\lambda} \sum_{\mu \neq \lambda} \frac{i\omega}{\omega_{\lambda \mu}} \tilde{d}_{ \mu \lambda}  (\omega) \chi^{(1)}_{\lambda \mu} (\omega)
\end{equation}

In the most general case, where there is a possibility to refill the level $\lambda$ through relaxation from higher energy levels the quantity $H_\lambda (\omega)$ should be obtained from the inversion of a linear system for the populations $N_{\lambda i}$ which is written according to the particular level configuration.

The equation of movement for the third order now become:

\begin{eqnarray}
\frac{d{S}^{(3)}_{\lambda \mu i}}{dt}  +\omega_{\mu \lambda} {J}^{(3)}_{\lambda \mu i} 
 \nonumber \\
= -\sum_{\gamma \neq \lambda \mu}  \left[ F_{\gamma \lambda}^{(1)}(\mathbf{r}_i)  {J}^{(2)}_{\gamma \mu i} - F_{\mu \gamma}^{(1)} (\mathbf{r}_i)  {J}^{(2)}_{\lambda \gamma i} \right] 
 \nonumber \\
-\sum_{\gamma \neq \lambda \mu}  \left[ F_{\gamma \lambda}^{(2)}(\mathbf{r}_i)  {J}^{(1)}_{\gamma \mu i} - F_{\mu \gamma}^{(2)} (\mathbf{r}_i)  {J}^{(1)}_{\lambda \gamma i} \right] 
\\
\frac{d {J}^{(3)}_{\lambda \mu i}}{dt} - \omega_{\mu \lambda} {S}^{(3)}_{\lambda \mu i} + \gamma_{\lambda \mu} J^{(3)}_{\lambda \mu i}
\nonumber 
\\+ 2F^{(3)}_{\lambda \mu} (\mathbf{r}_i)({N}^0_{\lambda i} - {N}^0_{\mu i})
\nonumber \\
 = -2F^{(1)}_{\lambda \mu} (\mathbf{r}_i)[H_\lambda (\omega) - H_\mu (\omega) ]\tilde{D}^{(1)}_t(\mathbf{r}_i)^2
\nonumber \\
 +\sum_{\gamma \neq \lambda \mu}  \left[ F^{(2)}_{\gamma \lambda}(\mathbf{r}_i)  {S}^{(1)}_{\gamma \mu i} - F^{(2)}_{\mu \gamma} (\mathbf{r}_i)  {S}^{(1)}_{\lambda \gamma i} \right] 
\nonumber \\
 +\sum_{\gamma \neq \lambda \mu}  \left[ F^{(1)}_{\gamma \lambda}(\mathbf{r}_i)  {S}^{(2)}_{\gamma \mu i} - F^{(1)}_{\mu \gamma} (\mathbf{r}_i)  {S}^{(2)}_{\lambda \gamma i} \right] 
\\
 F^{(3)}_{\lambda \mu}(\mathbf{r}_i) =  - \frac{N_e}{2} \sum_{\alpha} \Xi_{\alpha, \lambda \mu} {S}^{(3)}_{\alpha i} \label{F3nd}
\end{eqnarray}

The third order susceptibility is eventually expressed through an equation set that is build analogously to the those for the second order:

\begin{eqnarray}
 \tilde{S}^{(3)}_{\beta i}  = \frac{S}{N_e} \chi^{(3)}_\beta (\omega) \tilde{D}^{(1)}_t(\mathbf{r}_i)^3
 \\
 \chi^{(3)}_\beta (\omega) = \sum_{\alpha} M_{\beta \alpha}^{-1} (3\omega) \sqrt{\frac{\Delta n_\beta}{\Delta n_\alpha}} G_\alpha^{(3)} (\omega)
 \\
G_{\lambda \mu}^{(3)} (\omega) = 2\tilde{d}_{\lambda \mu} (\omega)\omega_{\lambda \mu}[H_\lambda (\omega) - H_\mu (\omega)]\frac{N_e}{S} 
\nonumber \\
 - \sum_{\gamma \neq \lambda \mu}  \tilde{d}_{\gamma \lambda}(\omega) \Big[ \omega_{\mu \lambda}\chi^{(2)}_{\gamma \mu} + \frac{3\omega^2 - i\omega\gamma_{\mu \lambda} }{\omega_{\mu \gamma }} \eta^{(2)}_{\gamma \mu} \Big]
\nonumber \\
+ \sum_{\gamma \neq \lambda \mu}  \tilde{d}_{\mu \gamma}(\omega) \Big[ \omega_{\mu \lambda }\chi^{(2)}_{\lambda \gamma} + \frac{3\omega^2 - i\omega\gamma_{\mu \lambda} }{\omega_{\mu \lambda }} \eta^{(2)}_{\lambda \mu}\Big]
\nonumber \\
+\sum_{\gamma \neq \lambda \mu} d^{(2)}_{\lambda \gamma} (\omega)\chi^{(1)}_{\gamma \mu} (\omega)\Big[ \omega_{\mu\lambda } + \frac{3\omega^2 - i\omega\gamma_{\mu \lambda} }{\omega_{\mu \gamma }} \Big]
\nonumber \\
-\sum_{\gamma \neq \lambda \mu} d^{(2)}_{\mu \gamma} (\omega)\chi^{(1)}_{\lambda \gamma} (\omega) \Big[ \omega_{\mu \lambda } + \frac{3\omega^2 - i\omega\gamma_{\mu \lambda} }{\omega_{\gamma \lambda }} \Big]
\end{eqnarray}

Here we introduced a "second order dipole" through the definition of $F^{(2)}_{\lambda \mu}(\mathbf{r}_i)$: 

\begin{eqnarray}
 F^{(2)}_{\alpha}(\mathbf{r}_i) =  -d^{(2)}_{\alpha} (\omega) \tilde{D}^{(1)}_t(\mathbf{r}_i)^2
 \\
 d^{(2)}_{\alpha}  (\omega) = \frac{N_e}{2} \sum_{\beta} \Xi_{\beta \alpha} \chi^{(2)}_{\beta} (\omega)
\end{eqnarray}

This term clearly arises only from electron-electron interactions. The quantity $\eta^{(2)}_{\gamma \mu}(\omega)$ is defined as:

\begin{eqnarray}
 \eta^{(2)}_{\lambda \mu}(\omega) = 2\chi^{(2)}_{\lambda \mu} (\omega) \nonumber \\
 - \sum_{\gamma \neq \lambda \mu} \Big[ \frac{\tilde{d}_{\gamma \lambda} (\omega)\chi^{(1)}_{\gamma \mu}(\omega)}{\omega_{\mu \gamma }}
 -\frac{\tilde{d}_{\mu \gamma}(\omega) \chi^{(1)}_{\lambda \gamma }(\omega)}{\omega_{\gamma \lambda }} \Big]
\end{eqnarray}

This quantity expresses the proportionality between the second order microcurrents and the square of the driving field:

\begin{equation}
\tilde{J}^{(2)}_{\beta i} =-\frac{S}{N_e} \frac{2i\omega}{\omega_\beta} \eta^{(2)}_\beta (\omega)   \tilde{D}^{(1)}_t(\mathbf{r}_i)^2    
\end{equation}

Finally, the third order susceptibility is expressed as:

\begin{eqnarray}
\chi^{(3)} (3\omega; \omega, \omega, \omega) = \frac{1}{(\hbar \varepsilon \varepsilon_0)^3 L_{cav}} \sum_{\beta}d_\beta \chi^{(3)}_\beta (\omega) \nonumber \\
 =\frac{1}{(\hbar \varepsilon \varepsilon_0)^3 L_{cav}}  \sum_{\beta \alpha } M_{\beta \alpha}^{-1} (3\omega) d_\beta \sqrt{\frac{\Delta n_\beta}{\Delta n_\alpha}} G_\alpha^{(3)} (\omega)
\end{eqnarray}

It now involves the inverse of the matrix $M_{\alpha \beta} (3 \omega)$ at the triple of the driving frequency. Clearly, the expressions are now more complex, however one can clearly track the contributions of the population dynamics and the dynamics of the coherences; furthermore it is easy to spot the new terms that arise from the electron-electron interactions. 

The approach can be applied to explicit the terms in the fourth line of Eq. (\ref{Def_chi_n}). The iterative procedure outlined above can be applied to obtain expressions for high order susceptibilities; the complexity of the formulas obtained will depend on the specificity of the electronic level configuration. 

\subsection{Microcavity effects} \label{secMicrocavity_effects}

We now consider the first two lines of the Maxwell-Bloch equations, Eqs (\ref{dAdt_av}) and (\ref{dEdt_av}) that couple the cavity fields with the polarizations and the incident pump. Thus the cavity fields $A_m$ and $D_m$ are no longer imposed but become dynamical variables that follow the harmonic decomposition Eq. (\ref{evO}).  We perform an order by order analysis of the Maxwell-Bloch equations starting with the first order.

\subsubsection{First order effects: polaritons}\label{1storder}

The first order equations are written as:

\begin{eqnarray}
(i\omega +\Gamma_{nr, m})\tilde{A}^{(1)}_m +  \sqrt{\Gamma_{r, m}} \sum_{m'}\sqrt{\Gamma_{r, m'}}\tilde{A}^{(1)}_{m'} 
\nonumber \\ 
= \omega_{cm}\tilde{D}^{(1)}_m + \frac{\omega_{cm}}{V_{cav}} \sum_{\alpha i} d_\alpha \tilde{S}^{(1)}_{\alpha i}u_m(\mathbf{r}_i) \nonumber \\
+ 2\sqrt{\Gamma_{r, m}}\tilde{I}_{in}
\\
-i\omega\tilde{D}^{(1)}_m = \omega_{cm}\tilde{A}^{(1)}_m \label{AmtoDm}
\\
\tilde{S}^{(1)}_{\alpha i} = \frac{S}{N_e \hbar \varepsilon \varepsilon_0}\chi_{\alpha}^{(1)} (\omega)\sum_m \tilde{D}^{(1)}_m u_m(\mathbf{r}_i)
\label{DmtoSalpha}
\\
\tilde{I}_{out}^{(1)} =  - \tilde{I}_{in} + \sum_{m}\sqrt{\Gamma_{r, m}}\tilde{A}^{(1)}_{m} \label{AmtoIout}
\end{eqnarray}

We have a system of linear equations that can be solved analytically following the method outlined in Appendix \ref{AppendixB} for void cavities. Analytical solutions are obtained as follows:

\begin{eqnarray}
 \tilde{A}_m^{(1)} = 2\tilde{I}_{in} \frac{i\omega \sqrt{\Gamma_{r, m}}}{\Delta^p_{cnr,m}(\omega)} \frac{1}{1+i\omega Y_p(\omega)}  \label{Am_matter}
 \\
 \Delta_{cnr,m}^p(\omega) = \omega_{cm}^2 (1-\chi^{(1)} (\omega)) - \omega^2 + i\omega \Gamma_{nr , m} \label{Delta_nrp}
 \\
 Y_p (\omega) = \sum_m \frac{\Gamma_{r , m}}{\Delta_{cnr,m}^p(\omega)}
\end{eqnarray}

The quantities $\tilde{D}^{(1)}_m$ and $\tilde{S}^{(1)}_{\alpha i}$ can then be easily inferred from Eqs. (\ref{AmtoDm}) and (\ref{DmtoSalpha}). The expressions for  $\tilde{A}_m^{(1)}$ in the presence of matter has a very similar structure  compared to the case of a void cavity, except that now the frequency $\omega_{cm}^2$ in Eq. (\ref{Delta_nrp}) is corrected with a factor $(1-\chi^{(1)} (\omega))$. We can recognize in this substitution the contribution to the effective dielectric function for the micro-cavity photon field  owe to the electronic system. Indeed, for the case of vanishing loss the poles of the denominator of Eq.(\ref{Am_matter}) are provided by the following equation:

\begin{eqnarray}
\varepsilon (\omega) \omega^2 = \varepsilon_0 \omega_{cm}^2
\\
\varepsilon (\omega) = \frac{\varepsilon_0}{1-\chi^{(1)} (\omega)}
\end{eqnarray}

We recover here the dispersion equation and the effective dielectric function for polariton waves in a planar waveguide that were discussed in Ref.\cite{todorov_intersubband_2012}. In the case of a single intersubband transition we can further use Eq.  (\ref{chi1_2sub}) and cast the polariton dispersion relation in the well-known form \cite{todorov_intersubband_2012}:

\begin{equation}
    (\omega^2 -  \omega_{cm}^2) (\omega^2 - \omega_\alpha^2) = f_w f_\alpha \omega^2_{P \alpha}\omega_{cm}^2
\end{equation}

Our current theory allows to correct this equation for the presence of losses. Let us consider the denominator of Eq. (\ref{Am_matter}) which is written in the form:

\begin{eqnarray}
\omega_{cm}^2 (1-\chi^{(1)} (\omega)) - \omega^2 + i\omega (\Gamma_{nr , m} + \Gamma_{r , m}) \nonumber \\ + i\omega \sum_{m' \neq m} \frac{\Gamma_{r , m'}}{\Delta_{cnr,m'}^p(\omega)} \label{DenAm}
\end{eqnarray}

In the case where the cavity modes have sufficiently narrow linewidths we can neglect the last term in Eq. (\ref{DenAm}) and the complex poles or the cavity amplitudes are provided by the equation:

\begin{equation}
 \omega_{cm}^2 (1-\chi^{(1)} (\omega)) - \omega^2 + i\omega (\Gamma_{nr , m} + \Gamma_{r , m}) \approx 0   \label{Polzeroes1}
\end{equation}

The solutions of this equation are complex valued frequencies. Their real part correspond to the polariton resonances, while their imaginary part corresponds to the linewidths of the coupled states that are now expressed as function of the characteristics of the uncoupled systems.  Zeroing the more general expression Eq. (\ref{DenAm}) allows to take into account the presence of other photonic modes and their losses for the total polariton linewidths.  

Finally, we can use our results together with Eq. (\ref{AmtoIout}) in order to obtain the reflectivity and absorption coefficient of the polaritonic system. We will come back to this point in section \ref{secEnCons}.

\subsubsection{High order frequency generation} \label{sec HF generation}

Let us now consider the dynamics of the $n^{th}$ order $(n=2,3)$, which is derived from the Maxwell-Bloch equations as follows:

\begin{eqnarray}
(in \omega +\Gamma_{nr, m})\tilde{A}^{(n)}_m +  \sqrt{\Gamma_{r, m}} \sum_{m'}\sqrt{\Gamma_{r, m'}}\tilde{A}^{(n)}_{m'} 
\nonumber \\ 
= \omega_{cm}\tilde{D}^{(n)}_m + \frac{\omega_{cm}}{V_{cav}} \sum_{\alpha i} d_\alpha \tilde{S}^{(n)}_{\alpha i}u_m(\mathbf{r}_i)
\\
-in\omega\tilde{D}^{(n)}_m = \omega_{cm}\tilde{A}^{(n)}_m \label{AmtoDm_n}
\\
\tilde{I}_{out}^{(n)} =  \sum_{m}\sqrt{\Gamma_{r, m}}\tilde{A}^{(n)}_{m} \label{AmtoIout_n}
\end{eqnarray}

Now the cavity field comprises additional terms  oscillating at a frequency  $n \omega$ that were absent in the derivation of the nonlinear susceptibilities from the previous paragraph. These terms will induce a first order response that will contribute to the electronic polarizations and micro-current dynamics. For instance, the terms $d_{\lambda \mu}\tilde{D}^{(2)}_{t} (\mathbf{r}_i)$ and $d_{\lambda \mu}\tilde{D}^{(3)}_{t} (\mathbf{r}_i)$ must have been present in Eqs.(\ref{F2nd}) and (\ref{F3nd}) respectively.
We can amend for this omission  by the following consideration. If we discard all nonlinear terms in the equations linking $\tilde{A}^{(n)}_{m}$, $\tilde{D}^{(n)}_{m}$,$\tilde{S}^{(n)}_{\alpha i}$ and $\tilde{J}^{(n)}_{\alpha i}$ we will end up with equations that match exactly the first order equations (i.e. Eqs.(\ref{dSdt1storder})  and (\ref{dJdt1storder})), but written at the frequency $n\omega$. Since the extra contributions arising from the fields $\tilde{D}^{(n)}_{m}$ are expressed from the first order susceptibility function taken at the frequency $n\omega$, we consider the following ansatz for the full nonlinear polarizations:

\begin{equation} \label{ansatz}
    \tilde{S}^{(n)}_{\alpha i}  = \frac{S}{N_e} [\chi^{(1)}_\alpha (n\omega) \tilde{D}^{(n)}_t(\mathbf{r}_i) +
    \chi^{(n)}_\alpha (\omega) \tilde{D}^{(1)}_t(\mathbf{r}_i)^n]
\end{equation}

The first term of this equation can be seen as a homogeneous solution to the full set of equations written for the $n^{th}$ order, while the second part is a source term. We can this solve for the coefficients $\tilde{A}^{(n)}_m$ exactly in the same way as before, obtaining an analytical expression:

\begin{equation}
 \tilde{A}_m^{(n)} = \tilde{I}_m^n \frac{in \omega }{\Delta^p_{cnr,m}(n\omega)} \frac{1}{1+in\omega Y_p(n\omega)}    
\end{equation}

Here the source terms $\tilde{I}_m^n$ are provided by:

\begin{eqnarray}
\tilde{I}_m^n = \frac{\omega_{cm}}{L_{cav}} \sum_{\alpha j} d_\alpha \chi^{(n)}_\alpha (\omega) \tilde{D}^{(1)}_t(\mathbf{r}_j)^n u_m(\mathbf{r}_j) \nonumber \\
= \omega_{cm} \chi^{(n)} (n\omega; \omega ... \omega) \nonumber \times \\
\frac{1}{N_e} \sum_j u_m(\mathbf{r}_j) \Big[ \sum_{m'} \tilde{D}^{(1)}_{m'} u_{m'}(\mathbf{r}_j)\Big]^n
\end{eqnarray}

The last line can be  expanded using the Newton's multinome formula:

\begin{equation}
   \sum_{|\vec{k}|=n} \binom{n}{\vec{k}} \kappa_{m, [i^{k_i}]}\prod_{i=1}^{m'} [\tilde{D}^{(1)}_{m'}]^{k_i} 
\end{equation}

Here the index vector $\vec{k}$ is defined as $\vec{k} = (k_1, k_2,...k_{m'})$, and we defined the following overlap coefficients by  using the identity from Eq. (\ref{faveraging}): 

\begin{eqnarray}
\kappa_{m, [i^{k_i}]}   = \kappa_{m, [1^{k_1}, 2^{k_2}, ... m'^{k_{m'}}]} \nonumber \\
 = \frac{1}{S}\iint u_m(\mathbf{r})\prod_{i=1}^{m'} u_{i}^{k_i}(\mathbf{r}) d^2 \mathbf{r} \label{defKappa}
\end{eqnarray}

The parameters $\kappa_{m, [i^{k_i}]}$ are dimensionless and depend only on the transverse distribution of the cavity modes. They represent the coupling coefficients between microcavity modes of different frequencies introduced by the nonlinearities of the electronic medium.  Similar overlap coefficients were discussed in Refs. \cite{KRASNOK20188, Gomez-Diaz_PhysRevB.92.125429}. The phase matching condition for nonlinear generation in planar structures is replaced by the requirement for having non zero coefficients $\kappa_{m, [i^{k_i}]}$ for the case of the micro-cavities. 

 The total output field radiated from the microcavity at the photon energy  $n \hbar \omega $ is thus provided by Eq.(\ref{AmtoIout_n}):

\begin{equation}
 \tilde{I}_{out}^{(n)} = \frac{in\omega}{1+in\omega Y_p(n\omega)}   \sum_{m} \frac{\sqrt{\Gamma_{r, m}} }{\Delta^p_{cnr,m}(n\omega)} \tilde{I}_m^n 
\end{equation}

Using the relations (\ref{defPower}) we arrive at the following expression of the power of the $n^{th}$ harmonic generated in the system as a function of the incident pump power:

\begin{eqnarray}
P^{(n)} (n\omega) = 4 \Big( \frac{2\varepsilon_0 \varepsilon}{V_{cav}}\Big)^{n-1} (P_{in})^n
\times \nonumber \\
\frac{(n\omega)^2|\chi^{(n)} (n\omega; \omega ... \omega)|^2}{|1+in\omega Y_p(n\omega)|^2 |1+i\omega Y_p(\omega)|^{2n}} \times \nonumber \\
\Bigg |  \sum_m \Big \{ \Theta_m (n\omega) \sum_{|\vec{k}|=n} \binom{n}{\vec{k}} \kappa_{m, [i^{ k_i}]}\prod_{i=1}^{m'}\Theta_i (\omega)^{k_i} \Big \} \Bigg |^2 \label{P_n_out}
\\
\Theta_m (\omega) = \frac{\omega_{cm} \sqrt{\Gamma_{r,m}}}{\Delta^p_{cnr,m}(\omega)} \label{defTheta_m}
\end{eqnarray}   
  
This expression comprises both many-body effects through the expressions of  $\chi^{(n)} (n\omega; \omega ... \omega)$ as well as microcavity (polaritonic) effects through the quantities $Y_p (\omega)$ and $\Delta^p_{cnr,i}(\omega)$. The latter describe the response of the cavity-coupled electronic system to the first order, where many-body effects appear on the first order susceptibility $\chi^{(1)} (\omega)$. Thus  
our approach requires computing independently the nonlinear response contained in the function $\chi^{(n)} (n\omega; \omega ... \omega)$ using the method in section \ref{General theory chi} and the cavity effects where only the function $\chi^{(1)} (\omega)$ is needed. The function $\chi^{(1)} (\omega)$ must eventually be corrected for saturation effects, that arise from the fourth line of Eq.(\ref{Def_chi_n}).

An interesting aspect of the above formula is the second line, which is a coherent sum of the contribution of all cavity modes. This part can become important in the case where there is finite spectral overlap between the two modes, as in Ref. \cite{Sarma_2022}. Indeed, meta-material systems can be designed to have   spectrally  large modes owe to radiative broadening \cite{Jeannin_NLett2020}. 
The above expression allows examining possible interference effects that might occur in the overlap region between the several resonances, or in large cavities with small spectral range. 
 
As an example let us consider the situation where the incident pump is resonant with a single mode $a$, with $\omega_{ca} \approx \omega$ while the $n^{th}$ harmonic is resonant with a mode $b$, $\omega_{cb} \approx n\omega$. We use the results from the previous section in order to link the amplitude $\tilde{D}^{(1)}_a$ to the incident field $\tilde{I}_{in}$, as well as the the definitions Eq. (\ref{defPower}) for the incoming and out-coming power. As a result we obtain the following expression for the output power generated for the $n^{th}$ harmonic:

\begin{eqnarray}
    P^{(n)} (n\omega) = 4 \Big( \frac{2\varepsilon_0 \varepsilon}{V_{cav}}\Big)^{n-1} (P_{in})^n |\chi^{(n)} (n\omega; \omega ... \omega)|^2  \nonumber \\
    \times  (n \omega)^2 \frac{\omega_{cb}^2 \Gamma_{r,b}}{|\Delta_{cb}^p(n\omega)|^2}|\kappa_{b, na}|^2 \Bigg[ \frac{\omega_{ca}^2 \Gamma_{r,a}}{|\Delta_{ca}^p(\omega)|^2} \Bigg]^n \label{Pn_output}
    \\
    \Delta_{cm}^p(\omega) = \omega_{cm}^2 (1-\chi^{(1)} (\omega)) - \omega^2 + i\omega \Gamma_m \label{Delta_p}
\end{eqnarray}

Here the function $\Delta_{cm}^p(\omega)$ describes the overall response of the polaritonic system. The above expression shows that the nonlinear response bears clear signature of the strong coupling regime, as the denominator $|\Delta_{ca}^p(\omega)|^2$ is minimized at the frequencies of the light-matter coupled states. Examples will be provided further.

Let as consider the case of a weak absorption such as the term $\chi^{(1)} (\omega)$ can be neglected in Eq. (\ref{Delta_p}). We furthermore assume perfect resonant conditions $\omega_{ca} = \omega$ and $\omega_{cb} = n\omega$. Then the above expression of the extracted  power simplifies to:

\begin{eqnarray}
    P^{(n)} (n\omega_{ca}) =  (P_{in})^n |\chi^{(n)} (n\omega_{ca}; \omega_{ca} ... \omega_{ca})|^2  \nonumber \\
    \times 4n\Big( \frac{2\varepsilon_0 \varepsilon}{V_{cav}}\Big)^{n-1}|\kappa_{b, na}|^2 \eta_b Q_b (\eta_a Q_a)^n  \label{Pn_simple}
\end{eqnarray}

Here we introduced the quality factors $Q_{m} = \omega_{cm}/\Gamma_{m}$ and the extraction efficiencies $\eta_m= \Gamma_{m,r}/\Gamma_{m}$. The latter quantify the part of the radiation loss over total loss for a particular cavity mode: the higher the radiation loss the higher the power extracted from or coupled into the cavity. This expression comforts a basic intuition for nonlinear generation in microcavity coupled-systems, for the case of weak coupling where the effect of polariton splitting can be neglected. First, the generated nonlinear power is proportional to the square modulus of the nonlinear susceptibility function taken at the value of the pump. There are $n$ pump photons at an energy $\hbar \omega_a$ that produce a single output photon at the energy $\hbar \omega_b = n \hbar \omega_a$. The extracted power is thus proportional to the $n^{th}$ power of the quality factor $Q_a$ for the mode $a$, which translates the recycling of the $n$ pump photons by the cavity resonance $a$. For a given incident power $P_{in}$ a fraction $\eta_a P_{in}$ is absorbed by the electronic system, and hence the dependence $\eta_a^n$. Similarly, the product $\eta_b Q_b$ quantifies the ability of the mode $b$ to  recycle the photons produced at the energy $\hbar \omega_b$ and to radiate them outside the cavity.  The coefficient $|\kappa_{b, na}|^2$ provides the geometrical overlap between the cavity mode $b$ and the field induced by the $n^{th}$ order nonlinear polarization excited in the mode $a$.   Finally, the nonlinear generated power is inversely proportional to the $n-1$ power of the inverse of the cavity volume $1/V_{cav}^{n-1}$. It is thus apparent from Eq. (\ref{Pn_simple}) that the strong electromagnetic confinement of the micro-cavity  which allows the reduction of $V_{cav}$ is beneficial for the nonlinear generation process. Furthermore, Eq. (\ref{Pn_simple}) naturally allows the expression of the extracted power  as a function of the factor $(Q_b/V_{cav})(Q_a/V_{cav})^n$ that is reminiscent of the Purcell factor for the spontaneous emission from a micro-cavity \cite{Purcell_1946}; our result can thus be regarded as a generalization of the Purcell effect for the case of nonlinear generation process. 

So far, we have not discussed the radiation pattern of the power $P^{(n)} (n\omega)$. The effect of the radiation pattern can be taken into account through an appropriate angular dependence of the radiation loss coefficients $\Gamma_{r,m}$, which can be generally expressed in spherical coordinates as:

\begin{equation}
 \Gamma_{r,m} = \Gamma_{0 r,m} g_m (\theta, \phi) 
\end{equation}

Here $g_m (\theta, \phi)$ is the normalized angular gain of the micro-cavity mode $m$ regarded as an aperture antenna \cite{Orfanidis2024}. For the case of patch antennas discussed here one can provide explicit analytical expressions of the loss coefficients and their angular dependence \cite{Orfanidis2024, Feuillet-Palma2013}. In particular, from Eq. (\ref{Pn_simple}) the radiation pattern of the extracted nonlinear power is expected to follow the radiation pattern of the microcavity resonance $b$.

As a second, less trivial example let us consider the case where the mode $a$ is sufficiently large so that both the pump frequency $\omega$ and the output frequency $n\omega$ fall within the bandwidth of the mode $a$; furthermore there is a second, sharper mode $b$ that is resonant with the nonlinear signal. This corresponds to the case described in Ref. \cite{Sarma_2022}. In accordance with the general expression from  Eq.(\ref{P_n_out})  we can evaluate the power of the nonlinear conversion by making following substitution in Eq. (\ref{Pn_output}):

\begin{eqnarray}
 \frac{\omega_{cb}^2 \Gamma_{r,b}}{|\Delta_{cb}^p(n\omega)|^2}|\kappa_{b, na}|^2 \rightarrow \nonumber \\
 \Bigg{|}\frac{\omega_{cb} \sqrt{\Gamma_{r,b}}}{\Delta_{cb}^p(n\omega)} \kappa_{b, na} + \frac{\omega_{ca} \sqrt{\Gamma_{r,a}}}{\Delta_{ca}^p(n\omega)} \kappa_{a, na} \Bigg{|}^2
\end{eqnarray}

This term now involves coherent sum between the nonlinear signals from both modes, which hints intriguing possibilities in designing interference patterns both in frequency and space for the converted power. 

\subsubsection{Energy conservation}\label{secEnCons}

Let us consider the total electromagnetic energy stored in all modes as defined in section \ref{SecCavity}. By computing the time variation of the total energy from the Maxwell-Bloch equations from section \ref{par: full set} we obtain the following conservation law:

\begin{eqnarray}\label{consLaw}
    \frac{\varepsilon \varepsilon_0}{V_{cav}} \sum_m \frac{d W_m}{dt} + I_{out}^2 = I_{in}^2 - \sum_m \Gamma_{m, nr} A_m^2 \nonumber \\
    + \sum_{m \alpha i} \frac{\omega_{cm}}{V_{cav}} [d_\alpha S_{\alpha i} A_m u_m(\mathbf{r}_i)]
\end{eqnarray}

As long as we have not specified the time dependence of the incident pump, this is a very general exact law that is valid for all types of excitation and an arbitrary time evolution of the system. The last line describes the energy loss owe to the intersubband excitations (both in the linear and nonlinear cases). Considering the case of a single monochromatic pump, using the harmonic decomposition in Eq. (\ref{evO}) as well as the ansatz Eq.(\ref{ansatz}) and performing time averaging in order to keep only the stationary terms we arrive at the following relation: 

\begin{eqnarray}\label{consLawHarmonic}
|\tilde{I}_{in}|^2 = \sum_n|\tilde{I}_{out}^{(n)}|^2 + \sum_{m,n} \Gamma_{nr, m} |\tilde{A}_m^{(n)}|^2 \nonumber \\
- \sum_n (n\omega) \mathrm{Im} [\chi^{(1)} (n\omega)] \sum_m |\tilde{D}_m^{(n)}|^2 \nonumber \\
- \sum_{n>1,m} (n\omega) \mathrm{Im} [\chi^{(n)} (n\omega; \omega, ...\omega) 
\nonumber \\ \times \frac{1}{N_e}
\Big( \sum_{m'} \tilde{D}_{m'}^{(1)}u_{m'}(\mathbf{r}_i)\Big)^n  \tilde{D}_{m}^{(1)*}u_{m}(\mathbf{r}_i)] \label{approxEnBalance}
\end{eqnarray}

The first line we recognize the out-coming electromagnetic power from the system distributed over the all multiples $n\omega$ of the driving frequency $\omega$, as well as the total non-radiative loss in the cavities. The second line describes the linear absorption which sums over all modes and high order fields. Finally, the last two lines describe the absorption due to nonlinear effects. This term is driven by the coupling between optical modes through the nonlinear matter dipoles. While the law Eq. (\ref{consLaw}) is very general, the above Eq. (\ref{consLawHarmonic}) should be considered with care, it overlooks two aspects of the population dynamics: saturation effects as well as the possibility for multiple photon absorption which can be seen as consecutive linear absorption events, as discussed further. Eq. \ref{consLawHarmonic} is thus valid for relatively weak pumps.  

Clearly, the high order terms are typically much smaller than the first order ones, and the above equation can be considered as a perturbative expansion of the energy balance as a function of $n$. Retaining only the first order contributions this equation becomes:

\begin{eqnarray}
|\tilde{I}_{in}|^2 = |\tilde{I}_{out}^{(1)}|^2 + \sum_{m} \Gamma_{nr, m} |\tilde{A}_m^{(1)}|^2 \nonumber \\
+ \omega \mathrm{Im} [\chi^{(1)} (\omega)] \sum_m |\tilde{D}_m^{(1)}|^2 
\end{eqnarray}

This equation allows us to define the intersubband absorption coefficient:

\begin{eqnarray}
\eta (\omega) = -\omega \mathrm{Im} [\chi^{(1)} (\omega)] \sum_m \frac{|\tilde{D}_m^{(1)}|^2}{|\tilde{I}_{in}|^2} \nonumber \\
= -\omega \mathrm{Im} [\chi^{(1)} (\omega)] \sum_m \frac{4\omega_{cm}^2 \Gamma_{r,m}}{|\Delta_{cnr,m}^p (\omega)(1+i\omega Y_p(\omega))|^2} \label{eta_lin}
\end{eqnarray}

Here we used the analytical solution for the first order mode obtained in section \ref{1storder}. Likewise, we can express the reflectivity coefficient $R(\omega) = |\tilde{I}_{out}|^2/|\tilde{I}_{in}|^2$:

\begin{equation}
    R(\omega) = 1- \sum_m\frac{4\omega  \Gamma_{r,m} [\omega \Gamma_{nr,m}-\omega_{cm}^2 \mathrm{Im} (\chi^{(1)} (\omega))] }{|\Delta_{cnr,m}^p (\omega)(1+i\omega Y_p(\omega))|^2}
\end{equation}

These results are the multi-subband and multimode generalizations of previous results \cite{Zanotto2014, Jeannin_NLett2020}.

\section{Examples} \label{partExamples}

\subsection{Two-subband system} \label{SecTwosubband_system}

\subsubsection{Population dynamics}\label{parN20}

\begin{figure}
    \centering
    \includegraphics[width = 6cm]{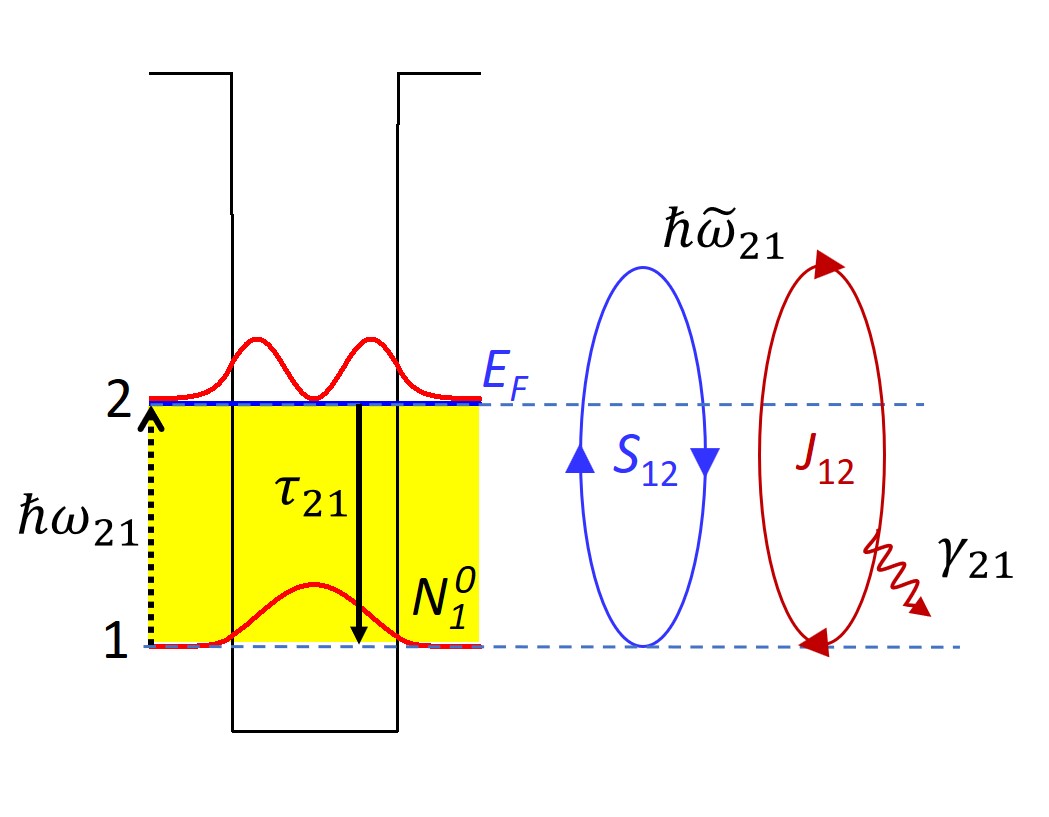}
    \caption{Two-subband quantum well where the Fermi level $E_F$ coincides with the edge of the second subband. The single-particle energy separation is $\hbar \omega_{21}$. The population excited on the second subband has a relaxation time $\tau_{21}$. The ellipses represent the polarization $S_{12}$ and the microcurrent $J_{12}$ oscillating at a frequency $\hbar \tilde{\omega}_{21}$ that is re-normalized by dipole-dipole interactions. $S_{12}$ and $J_{12}$ oscillate in quadratures, and the microccurent relaxes with a rate $\gamma_{21}$.}
    \label{fig:2subF1}
\end{figure}

In this section we apply the model to the case of two electronic subbands, which constitutes the most simple possible case. This system is reminiscent of a two-level system from atomic physics \cite{book_OptResonance, boyd_nonlinear_2008}. It is illustrated in Figure \ref{fig:2subF1} together with the relevant physical parameters. As illustrated in the figure, we consider that there is an important population $N_{1}^0$ on the first electronic subband such as at equilibrium the Fermi level $E_F$ lies at maximum below the second subband edge. For simplicity we consider here the zero-temperature case, whereas our results can be easily modified for a finite temperature by expressing the equilibrium electronic populations with the Fermi-Dirac statistics.  The subband edge separation is $\hbar \omega_{21}$, whereas first order polarization $S_{12i}^{(1)}$ and microcurrent $J_{12i}^{(1)}$ oscillate at the frequency $\tilde{\omega}_{21}= \sqrt{\omega_{21}^2+\omega_{P}^2}$ renormalized by many-body effects as described in the previous sections. As illustrated in Fig. \ref{fig:2subF1}  the energy $\hbar \tilde{\omega}_{21}$ can be larger than the single-particle transition energy $\hbar \omega_{21}$.
Since we consider here a single transition the index "$21$" will be dropped in the following.

Owe to the symmetry of the system the lowest order nonlinear susceptibility is the third order one. Furthermore we can study saturation effects and bi-stability \cite{jeannin_unified_2021}. Both these effects are related to the dynamics of the excited subband $N_2$. The population dynamics is expressed from the following equation:

\begin{equation}
\frac{d{N}_{2i}}{dt} 
=-\frac{{N}_{2i}}{\tau_{21}} 
+  \left(N_e \frac{\omega_{P1}^2}{2\omega_{21}} S_{i} -  d_{12} D_{t} (\mathbf{r}_i) \right) J_i
\label{dN2dt}
\end{equation}

Here $\tau_{21}$ is the relaxation time for the excited level population; we assume here the most simple dynamics where excited electrons can only relax to the first subband. This equation is completed with the equations for the coherences from paragraph \ref{par: full set} that we reproduce here for completeness:

\begin{eqnarray}
\frac{dS_{i}}{dt} &=& -  \omega_{21} J_{i} \label{SatS12 1}
\\
\frac{dJ_{i}}{dt} &=& \left(\frac{\omega_{P1}^2 N_e}{\omega_{21}} S_{i} - 2d_{12} D_t (\mathbf{r}_i)\right) (N_{1i}-N_{2i}) \nonumber \\
&+& \omega_{21} S_{i} -\gamma_{21} J_{i}
\label{SatJ12 1} 
\end{eqnarray}

Here we introduced the microcurrent relaxation rate $\gamma_{21}$ as illustrated in Fig. \ref{fig:2subF1}. 

Assuming time-independent population difference $N_{1i}-N_{2i}$ we can infer the first order polarizations $\tilde{S}^{(1)}_i$ and currents $\tilde{J}^{(1)}_i$ as described in the previous paragraphs. Then from Eq.(\ref{dN2dt}) we obtain the excited subband population dynamics up to the second order:

\begin{eqnarray}
N_{2i} (t)= {N}_{2i}^{(0)} + \tilde{N}_{2i}^{(2)}e^{2i\omega t} + \tilde{N}_{2i}^{(2)*}e^{-2i\omega t}
\\
{N}_{2i}^{(0)} = -\tau_{21} 2\Re [\tilde{d}_{21}(\omega)\tilde{D}^{(1)}_t(\mathbf{r}_i)\tilde{J}^{(1)*}_i] \label{N2_0}
\\
\tilde{N}_{2i}^{(2)} = -\frac{\tau_{21}}{1+ 2i\omega \tau_{21}}\tilde{d}_{21} \tilde{D}^{(1)}_t(\mathbf{r}_i)\tilde{J}^{(1)}_i \label{N2_2}
\end{eqnarray}

The steady state contribution Eq. (\ref{N2_0}) is related to the linear absorption process, as well as saturation effects.  It is discussed in the next paragraph \ref{secBistability}. The second order contribution, Eq. (\ref{N2_2}) leads to third order optical nonlinearites as discussed in paragraph \ref{parN22}. 

The ground state of the system  in Fig. \ref{fig:2subF1} in the absence of light and at zero temperature corresponds to zero population on the second subband  $N_{20} = 0$, and the first subband filled with $N_{10}$ electrons; the corresponding areal density is noted $\Delta n_0 = N_{10}/S$. Under illumination, a finite number of electrons populates the second subband, as expressed from Eq. (\ref{N2_0}). Using the results from section \ref{secLinear} the total number of electrons under illumination is found to be:

\begin{eqnarray}
    N_2 = \sum_i {N}_{2i}^{(0)} = \nonumber \\
     \frac{4\omega^2 d_{12}^2 \Delta n_{12}}{|\Delta_{21} (\omega)|^2} \frac{S}{(\hbar \varepsilon \varepsilon_0)^2} \sum_m |\tilde{D}^{(1)}_m|^2
\end{eqnarray}

The same notations have been used as the ones introduced in paragraph \ref{secLinear}. Using the expression of the first order susceptibility, Eq. (\ref{chi1_2sub}), as well as the expression of the linear absorption coefficient, Eq.(\ref{eta_lin}), together with the definition of the incident power, Eq. (\ref{defPower}) we arrive at the following relation:

\begin{equation}
    N_2 = \eta (\omega) \frac{P_{in} \tau_{21}}{\hbar \omega_{21}} \label{N_2toPin}
\end{equation}

This expression has a clear physical meaning. Under illumination, the extra energy introduced in the electronic system is $N_2 \hbar \omega_{21}$. Since the excited energy lifetime is $\tau_{21}$ the energy lost per unit time is $N_2 \hbar \omega_{21}/\tau_{21}$. The energy per unit time brought into the electronic system from the incident wave is $\eta(\omega)P_{in}$. Equating energy lost and energy brought in per unit time leads to Eq. (\ref{N_2toPin}) which can also be rewritten as $N_2 \hbar \omega_{21}/\tau_{21} = \eta(\omega)P_{in}$.  Eq. (\ref{N_2toPin}) is also important to properly quantify responsivity of detectors, where the photocurrent is extracted from  electrons excited on the second subband. \cite{pisani_electronic_2023}.

It is interesting to note that the elementary energy scale here is $\hbar \omega_{21}$, that is the separation between the single-particle states, and not the re-normalized  energy $\hbar \tilde{\omega}_{21}$ or the energies of  polariton states in a microcavity-coupled system as one would naively expect. Yet, the effect of the many-body interactions and the microcavity effects are still all contained in the parameter $\eta (\omega)$. To illustrate this fact, we consider the case where the two subband system interacts with the mode $m=1$ of the microcavity array. In that case the expressions for the absorption efficiency $\eta (\omega)$ and the reflectivity $R(\omega)$ become:

\begin{eqnarray}
 \eta(\omega) = 4f_wf_{12}\frac{\gamma_{21}\Gamma_{r,1} \omega^2\omega_{c1}^2 \omega_{P}^2}{|\Pi(\omega, x)|^2}   \label{etasinglemode}
 \\
R(\omega) = 1- \frac{4\omega^2  \Gamma_{r,1}  \Gamma_{nr,1}} {|\Pi(\omega, x)|^2} - \eta(\omega) \label{EqRefl1mode}
\end{eqnarray}

Here we introduced the dimensionless variable $x=\Delta n_{12}/\Delta n_{0}$ which describes the population difference normalized on its value in the absence of pump $\Delta n_{0}$. The quantity $f_w = N_{QW} L_{eff}/L_{cav}$ is the geometric filling factor of the electronic polarization \cite{todorov_intersubband_2012}. We have supposed that microcavity is filled with $N_{QW}$ identical quantum wells in order to increase the light-matter interaction strength \cite{Todorov_PRL2010}. The function $\Pi (\omega, x)$ is provided by:

\begin{eqnarray}
  \Pi (\omega, x) = xf_{12}f_w\omega_{cm}^2\omega_{P}^2  \nonumber \\
  -\Delta_{cm} (\omega)(\omega_{21}^2 +\omega_{P}^2x - \omega^2 + i\gamma_{21}\omega )   \label{PidefDn}
  \\
\Delta_{cm} (\omega) = \omega_{cm}^2 - \omega^2 + i\omega \Gamma_m
\end{eqnarray}

\begin{figure}
    \centering
    \includegraphics[width = \columnwidth]{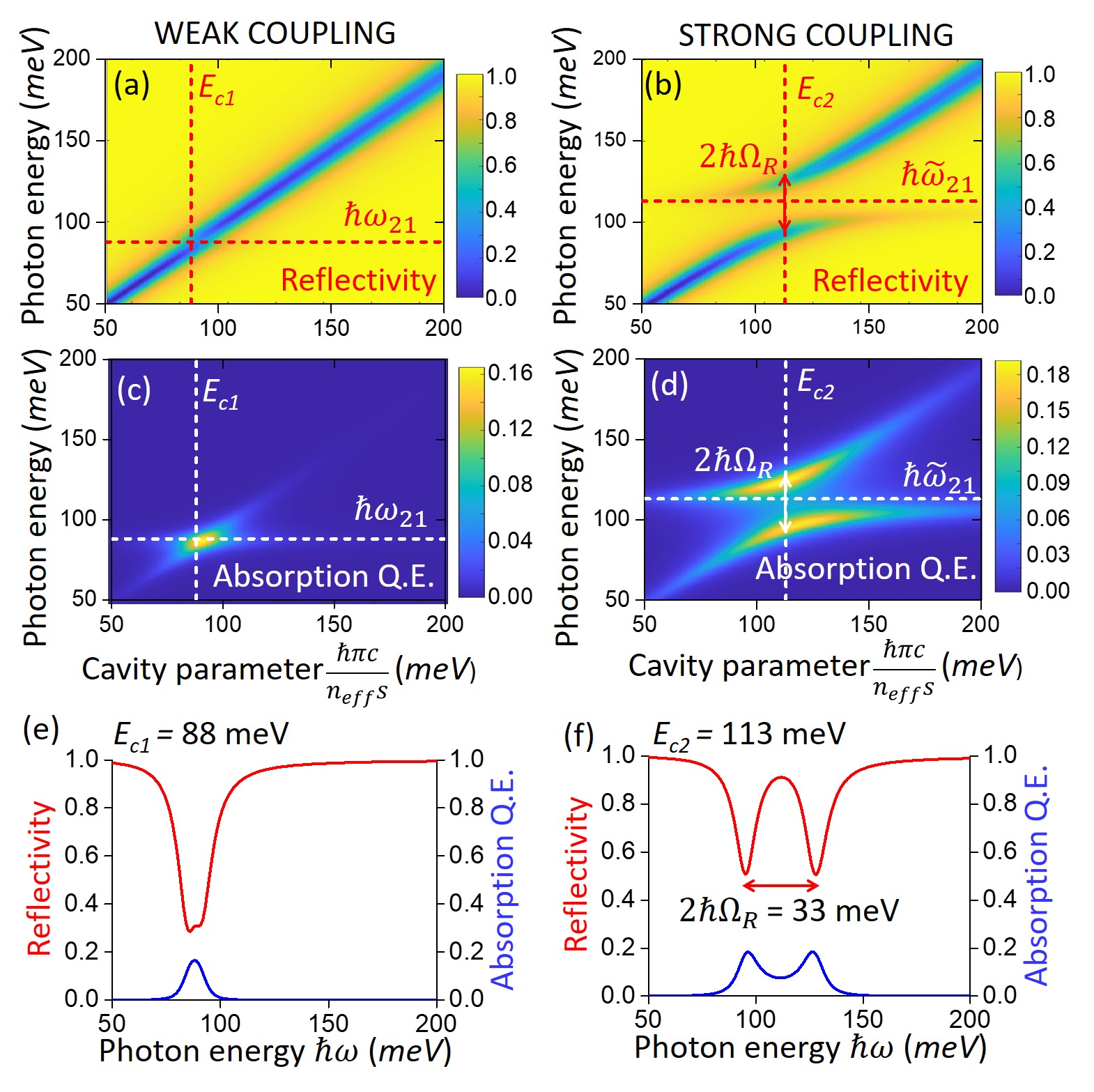}
    \caption{First order electromagnetic response of a two subband quantum well coupled to the $m=1$ mode of the patch microcavities. The left column corresponds to the weak coupling case ($\Delta n_0 =10^{11}$ $cm^{-2}$ and the right column to the strong coupling case ($\Delta n_0 =2.4\times 10^{12}$ $cm^{-2}$)). (a,b) Reflectivity in color-scale as a function of the cavity parameter $\hbar \pi c/n_{eff} s$ in $meV$ and the incident photon energy. (c,d) Absorption efficiency in color-scale as a function of the cavity parameter and the incident photon energy. (e) Reflectivity $R(\omega)$ and absorption quantum efficiency $\eta (\omega)$ spectra for a cavity with the $m=1$ mode resonating at an energy $E_{c1}=88 meV$, nearly resonant with the single photon transition energy $\hbar \omega_{21}$, also indicated by the vertical dashed line in (a,c). (f) Reflectivity $R(\omega)$ and absorption quantum efficiency $\eta (\omega)$ spectra for a cavity with the $m=1$ mode resonating at an energy $E_{c2}=113 meV$, nearly resonant with the collective electric excitation at an energy $\hbar \tilde{\omega}_{21}$, also indicated by the vertical dashed line in (b,d). A $10$ nm wide $\mathrm{GaAs}$ quantum well with $\mathrm{Al_{0.3}Ga_{0.7}As}$ has been considered in this example; other parameters are provided in the main text.}
    \label{fig:2subFcavity}
\end{figure}

The zeroes of this function provide the polariton frequencies just like Eq. (\ref{Polzeroes1}). Here we have written it in such a way that the dependence on the population difference through the variable $x$ is made explicit, by anticipating our further discussion on saturation and bi-stability. In Eq. (\ref{PidefDn}) both the light matter-coupling strength and the depolarization shift are manifestly dependent on $x$.  Let us suppose that the number of excited electrons is weak with respect to the ground state population, $N_2 << N_{10}$ so that saturation effects can be neglected, and $x\approx 1$. In Fig. 
\ref{fig:2subFcavity} we analyze the quantum efficiency $\eta(\omega)$ and the reflectivity $R(\omega)$ for both the weak (left column) and strong coupling (right column of the figure) regimes that can be achieved for respectively low and high electronic density $\Delta n_0$, and in both cases we use our exact expressions without approximations. For these plot we consider a heterostructure composed of $L_w=10$ nm wide $\mathrm{GaAs}$ quantum wells separated by $L_{b} =15$ nm  $\mathrm{Al_{0.3}Ga_{0.7}As}$ barriers, such the well barrier repetition fills at best the thickness of the micro-cavity $L_{cav} \approx N_{QW}(L_w+L_{b})$. The radiation and non-radiation loss are assumed to be both equal to $\Gamma_{r, m=1} = \Gamma_{nr, m=1} = 10$  meV such as the $m=1$ mode is in a critical coupling regime with almost zero reflectivity at resonance in the absence of the hetero-structure; typically cavities with such properties that resonate in the mid-infrared region have a thickness around $L_{cav} \approx 300$ nm \cite{Rodriguez_22} ($N_{QW} \approx 10$). 

In Fig. \ref{fig:2subFcavity}(a) we plot the reflectivity spectra for photon energies between $50$ meV and $200$ meV for a series of cavities where we vary the size of the patch $s=\sqrt{S}$ \cite{todorov_optical_2010}. The frequencies of the microcavity modes $\omega_{c,m}$ are set by the patch width $s$ through the formula $\omega_{c,m}= m \pi c/n_{eff}s$ \cite{todorov_optical_2010}, and here we consider a constant effective index $n_{eff} \approx 3.4$ for simplicity. Fig. \ref{fig:2subFcavity}(a) is color plot of the reflectivity as a function of the incident photon energy $\hbar \omega$ and the parameter $E_s =\hbar \pi c/n_{eff}s$ on the horizontal axis. The $m=1$ resonance is clearly visible along the diagonal of the  plot. In this case the areal electronic density is $\Delta n_0 = 10^{11}$ $cm^{-2}$ and the system is in the weak coupling regime. In Fig. \ref{fig:2subFcavity}(c) we show the color-plot for the absorption quantum efficiency $\eta (\omega)$ which acquires important values only in the case where the bare transition frequency $\hbar \omega_{21} = 87$ meV is close to the energy of the cavity resonance $\hbar \omega_{c,m=1}$. The plasma energy is in that case $\hbar \omega_P = 5$ meV and it is small with respect to the transition energy, and collective effects can be ignored.
The energy $\hbar \omega_{21}$ has been indicated as dotted horizontal lines in Figs. \ref{fig:2subFcavity}(a,c). In Fig. \ref{fig:2subFcavity}(e) we plot the reflectivity $R(\omega)$ and absorption efficiency $\eta(\omega)$ spectra for a cavity resonance with an energy $E_{c1} = 88$ meV, also indicated by dotted vertical lines in Figs. \ref{fig:2subFcavity}(a,c). The function $\eta (\omega)$ has a nearly Lorentzian shape centered at the transition frequency $\hbar \omega_{21}$. In that case, according to Eq. (\ref{N_2toPin}) the excited level population is maximum for a photon frequency $\omega = \omega_{21}$ that matches the energy separation, as expected from a single-particle picture. 

In the case of high electronic density, $\Delta n_0 = 24 \times 10^{11}$ $cm^{-2}$ the reflectivity and absorption efficiency maps shown in Figs. \ref{fig:2subFcavity}(b, d). Now we observe polariton splitting between the microcavity modes and the electronic system, indicating the strong coupling regime. As discussed in Ref. \cite{todorov_intersubband_2012} the splitting is provided by the minimal separation between the polariton resonances obtained for a cavity such as $\omega_c = \tilde{\omega}_{21}$. Here we obtain  $2\hbar \Omega_R = 33$ meV for a cavity energy $E_{c2} \approx  \hbar \tilde{\omega}_{21} = 113$ meV. The plasma energy is in that case $\hbar \omega_P = 70$ meV. We recall that for the two-subband system the polariton splitting is related to the plasma frequency through the relation $2\Omega_R = f_{12}f_w \omega_P^2$ \cite{todorov_intersubband_2012}. The energy $\tilde{\omega}_{21}$ is also indicated by a dotted vertical line in Figs. \ref{fig:2subFcavity}(b, d). In Fig.\ref{fig:2subFcavity}(f) we show the reflectivity $R(\omega)$ and absorption quantum efficiency $\eta (\omega)$ spectra a cavity resonance at the energy $E_{c2}$. The function $\eta(\omega)$ now acquires finite values in the whole frequency range spanned between the two polariton resonances, in a frequency band roughly provided by the Rabi splitting $2\hbar \Omega_R$. Since the absorption efficiency has finite values in that frequency band, in light of Eq. (\ref{N_2toPin}) we conclude that the probability to find electrons excited on the second subband is no longer maximal for the transition frequency $\omega_{21}$ but instead peaks at the polariton resonances and it is non-zero in the whole frequency band in-between. These considerations can be useful for designing polariton devices such as detectors \cite{pisani_electronic_2023} and emitters \cite{Jouy_PhysRevB_2010, Lagree_2024},  in the context of electrical injection of polariton states, since the injection current couples to the populations rather than the dipole of the transition. 

It is thus important to distinguish the dynamics of populations, where the relevant energy separations are those of the single-particle levels, $\hbar \omega_{12}$  is the dynamics of coherences, which oscillate at the re-normalized frequency $\tilde{\omega_{12}}$. This example shows that  dipole-dipole interactions change the oscillation frequency of dipoles between the levels, but do not lead to new electronic levels in the system. This particularity can be captured in the full fermion approach, where populations and coherences are distinct dynamical variables for the electronic system.  

\subsubsection{Saturation and bistability}\label{secBistability}

\begin{figure}
    \centering
    \includegraphics[width = \columnwidth]{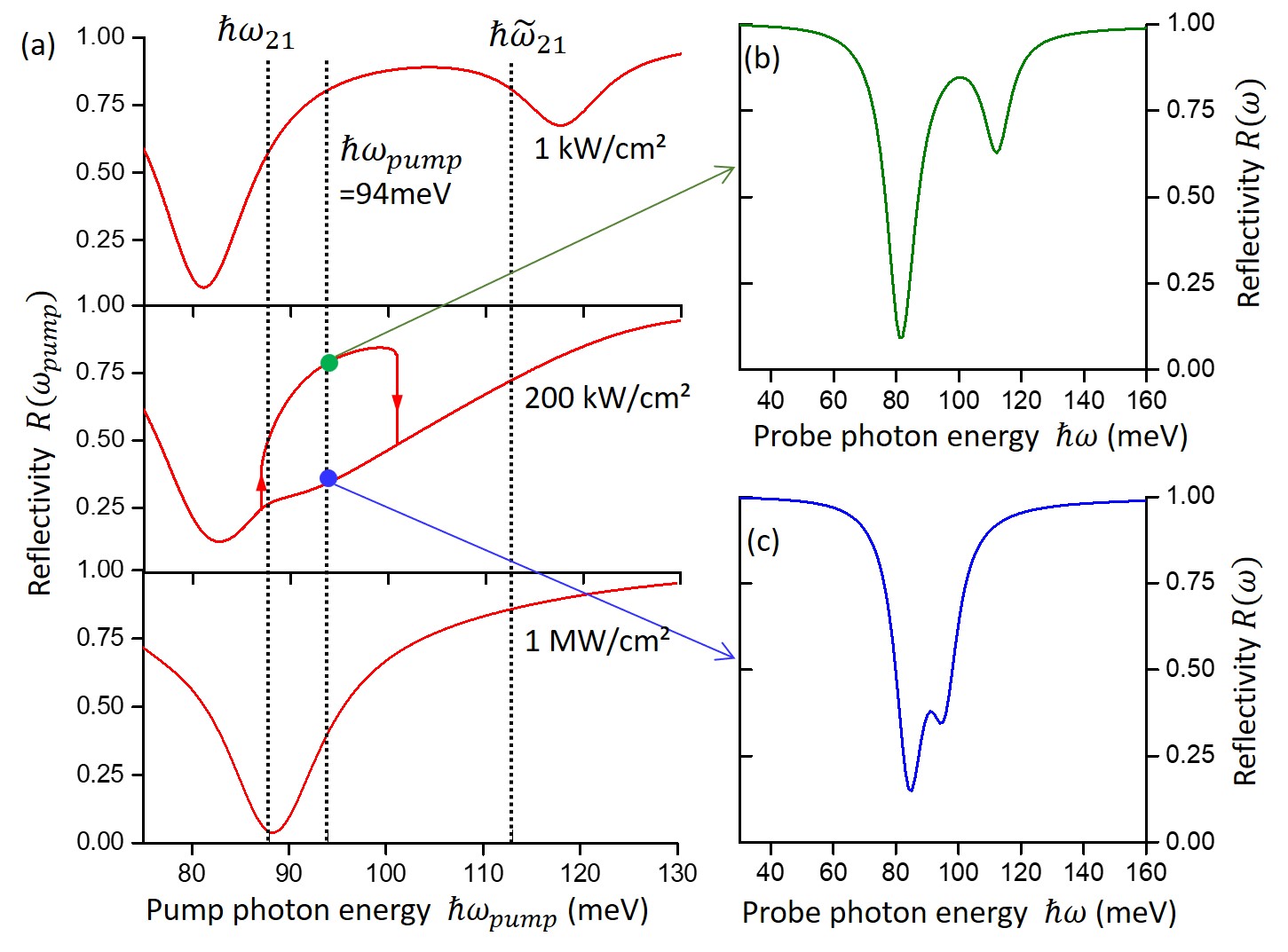}
    \caption{(a) Reflectivity of of the system for the case of a high electronic density, $2.4\times 10^{12}$ $cm^{-2}$ as a function of the photon energy  $\hbar \omega_{pump}$ of the incident pump, for various intensities. (b,c) Reflectivity measured with a weak intensity probe, for an incident pump intensity of $200$ $kW/cm^2$ and a pump photon energy $\hbar \omega = 94$ meV. The system is prepared either in the high population (b) or low population (c) regime, indicated by dots in (a).}
    \label{fig:Bistable2}
\end{figure}

\begin{figure}
    \centering
    \includegraphics[width = \columnwidth]{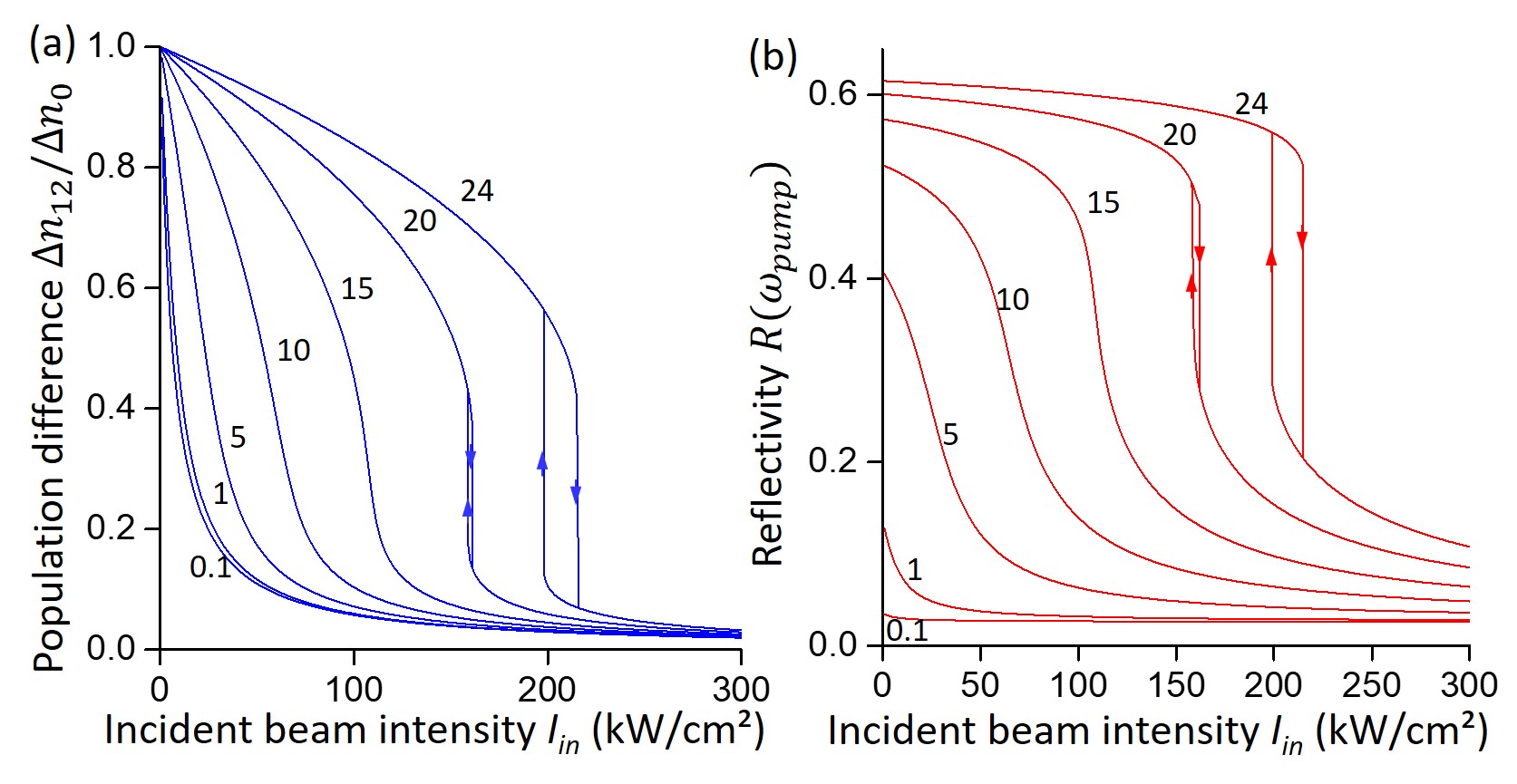}
    \caption{Population difference (a) and reflectivity (b) as a function of the incident beam intensity for various electronic populations $\Delta n_0$. The figures of next to each curve indicates $y$ where $\Delta n_0 = y\times 10^{11}$ $cm^{-2}$. The microcavity is resonant with the single-particle transition frequency $\omega_{c1} = \omega_{21}$ and the pump photon energy is $\hbar \omega = 94$ meV.}
    \label{fig:Bistable1}
\end{figure}

Recently the possibility to obtain  bistability in microcavity-coupled intersubband systems was discussed, and the bistability was linked to the effect of the saturation of the transition \cite{jeannin_unified_2021}. Here, we discuss these effects from the perspective of our model, which takes explicitly into account dipole-dipole interactions alongside with microcavity effects. We still consider a  single cavity mode $m=1$ coupled with the quantum well system, with a pump frequency in the vicinity of the polariton resonances. We treat the areal population difference $\Delta n_{12}$ as an unknown variable in Eq. (\ref{etasinglemode}). Using the constraint $N_1 + N_2 = N_1^0$ of a fixed total population, we can further express the excited level population as $N_2 = S(\Delta n_0 - \Delta n_{12})/2$; using Eq. (\ref{N_2toPin}) then provides an implicit equation for the population difference $\Delta n_{12}$ as a function of the incident power $P_{in}$. The latter can be cast in the form:
 
\begin{equation}
   x =\frac{\Delta n_{12}}{\Delta n_{0}} = \frac{1}{1+ \frac{P_{in}}{P_{sat}(\omega, x)}} \label{EqDeltan12}
\end{equation}

Here we defined a quantity  $P_{sat}(\omega, x)$ homogeneous to a power that can be identified as the  saturation power of the microcavity-coupled system:

\begin{equation}
  P_{sat}(\omega, x)  = \frac{ \hbar \omega_{21}}{\tau_{21}}  
  \frac{|\Pi (\omega, x)|^2}{8f_wf_{12} \gamma_{21}\Gamma_{r,1} \omega^2 \omega_{c1}^2 \omega_{P1}^2} \label{Psat_general}
\end{equation}

In the above expression $\omega_{P1}$ is the single electron plasma frequency introduced in Eq. (\ref{SingleWP}). Together with Eq. (\ref{EqDeltan12}) the above equation provides the population difference $\Delta n_{12}$ for any incident power at any pump frequency, as far as the single mode approximation is satisfied. From Eq. (\ref{PidefDn}) it is clear that $|\Pi (\omega, x)|^2$ is a second order polynomial of $x$. Thus Eq. (\ref{EqDeltan12}) is a third order equation for $x$ that can have two stable and one unstable solutions for a sufficiently high electronic population \cite{jeannin_unified_2021}. 

It will be interesting to establish typical values for the saturation power (\ref{Psat_general}) in light of recent experimental results \cite{Lin_23, Jeannin_2023}. To that end, we will consider the situation where the pump frequency $\omega$ is nearly resonant with the cavity frequency, $\omega_{c1}$, and furthermore the cavity frequency is nearly resonant with the intersubband plasmon energy: $\omega_{c1} \approx \tilde{\omega}_{21}$. Indeed, we can anticipate that, as the pump power is increased, the population difference decreases and the polariton splitting is reduced (see further), thus the final state of the system satisfies the condition stated above. We introduce the following quantities that will allow simplifying the expression of the saturation power and provide a physical insight of its expression. First, we consider the absorption cross section of a single electron:

\begin{equation}
    \sigma_{12} = \frac{f_{12}e^2}{2n\varepsilon_0 m^* c \gamma_{21}} \label{SecEfficace}
\end{equation}

This quantity was introduced in Ref. \cite{Jeannin_NLett2020, Bahrehmand2023-ud} and is deduced from a general theory based on dipolar absorbers \cite{Tretyakov2014}. Here $c$ is the speed of light and $n=\sqrt{\epsilon}$ the background refractive index of the material filling the cavity, without the contribution of the electronic transitions of the quantum wells, that is the refractive index of the "void" cavity. Next, we introduce a characteristic time:

\begin{equation}
    \tau_{tr} = \frac{L_{cav}}{c/n} \label{traverse_time}
\end{equation}

This is the time needed for a photon to traverse the cavity thickness $L_{cav}$. Finally, since most experiments are performed with periodic arrays of microcavities \cite{todorov_optical_2010}, as shown in Fig. \ref{fig: global}, we also employ the array filling factor $f_{array} = S/\Sigma$.  Expanding the numerator $|\Pi (\omega, x)|^2$ in Eq.(\ref{Psat_general})  we write the saturation intensity for a nearly resonant pump to be:

\begin{eqnarray}
    I_{sat} = \frac{P_{sat}}{\Sigma} = f_{array} \frac{\hbar \omega_{21}}{4\tau_{21}} \frac{1}{\eta_r} \times \nonumber \\
    \Big{\{} \frac{\Gamma_1 \tau_{tr}}{4 N_{QW} \sigma_{12}} + \Delta n_{12}  + \frac{N_{QW} \sigma_{12} \Delta n_{12}^2}{\Gamma_1 \tau_{tr}} \Big{\}} \label{Psat_res}
\end{eqnarray} 

Here we make use of the extraction efficiency of the cavity mode $m=1$, $\eta_1 = \Gamma_{1,r}/\Gamma_{1}$ introduced in section \ref{sec HF generation} and $\Gamma_{1}$ is the total linewidth of the void cavity. $\Gamma_{1}$ is related the cavity quality factor, $\Gamma_{1} = \omega_{c1}/Q_1$. The product $\Gamma_1 \tau_{tr}$ is a dimensionless quantity which is a ratio between the traverse time $\tau_{tr}$ of a photon over the lifetime $1/\Gamma_1$ of the microcavity photon. The quantity in the brackets of Eq. (\ref{Psat_res}) is an inverse of an area, which quantifies the effect of the electronic system and the microcavity confinement on the saturation power. The first term of the bracket is a single electron contribution, that can be obtained assuming that there is exactly one electron in each quantum well, and the intersubband transition is treated as a two-level system \cite{book_OptResonance}. The second term corrects for this picture by taking into account the total areal density of electrons. These two terms comfort the expected intuitive behavior for the saturation intensity, as they increase with the areal density, and decrease for longer photon lifetime in the cavity, as well with the absorption cross section for a single electron $\sigma_0$. The last term arises from the contribution $f_{12} f_w \omega_P^2\omega_c^2$  in Eq. (\ref{PidefDn}) that quantifies the light-matter coupling and the polariton splitting in the system \cite{todorov_intersubband_2012}. Combining the second and third terms in the brackets we can define and effective areal density:

\begin{equation}
    \Delta n_{12,eff} = \Delta n_{12} \Big( 1 + \frac{N_{QW} \sigma_{12} \Delta n_{12}}{\Gamma_1 \tau_{tr}} \Big)
\end{equation}

The quantity $\Delta n_{12,eff}$ can be seen as an effective areal density dressed by microcavity and collective effects. Indeed, the light-matter coupling is ultimately related to collective electronic effects \cite{Todorov_PRL2010}, and the interaction strengths is enhanced as the number of electrons in the system increases. However, larger number of electrons makes the saturation regime harder to achieve, as shown by the last term in Eq. (\ref{Psat_res}). 
Taking typical values $f_{12} \approx 1$,  $\gamma_{21} = 10$ meV, $L_{cav}=300$ nm and $n=3.4$ and $\Gamma_1 = 20$ meV we have the following estimations:

\begin{equation}
    \sigma_{12} \approx 10^{-13} cm^2 = 0.1 nm^2, \phantom{S} \Gamma_1 \tau_{tr} \approx 1.6 \times 10^{-3}
\end{equation}

Using a typical initial value of the areal population  $\Delta n_{21} \approx \Delta n_0 \approx \times 10^{11}$ $cm^{-2}$ and a structure with $N_{QW} =10$ quantum wells we find that the last term of the brackets in Eq. (\ref{Psat_res}) dominates the other two and amounts at $\Delta n_{12,eff} \approx 6\times 10^{12}$ $cm^{-2}$. Using a typical array filling factors $f_{array} =0.1$ which provide critical coupling for void cavity, $\eta_1 = 0.5$, a population lifetime $\tau_{21} \approx 1$ $ps$ and a photon energy $\hbar \omega_{21} \approx 100$ $meV$ we estimate a saturation intensity $I_{sat} \approx 5$ $kW/cm^2$. This is the correct order of magnitude that has been reported in experiments \cite{Lin_23, Jeannin_2023}, for typical arrays that  have a surface $10^{-5}$ $cm^{-2}$.

More precise estimations of saturation effects can be obtained by numerically solving Eq. (\ref{Psat_general}).  Generally speaking, there a several characteristic resonances of the system: the cavity resonance $\omega_{c1}$, the bare electronic transition $\omega_{21}$ and depolarization-shifted transition $\tilde{\omega}_{21}$, furthermore, we also have to consider the frequencies of the two polariton states $\omega_\pm$. The absorption of the system is maximized at the  frequencies $\omega_\pm$; at the same time their values depend strongly on the population difference. This large number of parameters leads to rich dynamics. Typically, as the pump power is increased, as the electronic density $\Delta n_{12}$ is reduced which results to red shift of the optical resonances of the system, and eventually the frequency of the optimal absorption is shifted away from the pump frequency, $\omega_{pump}$. 

In Fig. \ref{fig:Bistable2} we consider the case where the system has high electronic density $\Delta n_0 = 24\times 10^{11}$ $cm^{-2}$, and the absorption and reflectivity spectra display two polariton resonances. The cavity frequency  $\omega_{c1}$ is chosen to be slightly above the single-particle transition, $\omega_{c1} \geq \omega_{21}$. The pump frequency is $\omega_{pump}$ is then varied between $75$ meV and $130$ meV for various pump intensities $P_{in}/\Sigma$.  Eq. (\ref{EqDeltan12}) is solved in numerically to extract the  population difference $\Delta n_{12}$ and then the reflectivity is determined from Eq.(\ref{EqRefl1mode}). In Fig. \ref{fig:Bistable2} (a) we show the corresponding plots $R(\omega_{pump})$ for three values of the pump intensity: $1$ $kW/cm^2$, $200$ $kW/cm^2$ and $1$ $MW/cm^2$. The positions of the single-particle transition energy $\hbar \omega_{21}$ and the collective electronic resonance $\hbar \tilde{\omega}_{21}$ have also been indicated. For $I_{in} = 1$$kW/cm^2$ the saturation effects are negligible and the reflectivity curve $R(\omega_{pump})$ is essentially identical to the one that would have been obtained from a weak probe, such as black body source. For an intensity  $200$ $kW/cm^2$ a bistability region develops for a band of pump frequencies between  $\omega_{21}$ and $\tilde{\omega}_{21}$, which correspond to two possible stable solutions for $\Delta n_{21}$ from Eq. (\ref{EqDeltan12}). In Fig. \ref{fig:Bistable2} (b,c) the pump frequency is fixed to a value $\omega_{pump0}$ in the middle of the bistability band, the pump intensity is $200$ $kW/cm^2$, and we show the reflectivity spectra that correspond to the two states when the system is probed by a weak radiation source such as black-body. The two stable values of the electronic population are translated into two separations of the polariton modes, that can be set by forward/backward sweep of the pump frequency $\omega_{pump}$ around the bistable region. Finally, for a high incident intensity  $I_{in} = 1$ $MW/cm^2$ the system is saturated, and the population difference is nearly zero. As the quantum well system is close to transparency in that region, the graph $R(\omega_{pump})$ then corresponds to the void cavity spectrum. This study shows that in the presence of strong collective effects the optimal pump frequency that provides lowest saturation power is to be found in the interval  between the two characteristic frequencies $\omega_{21}$ and $\tilde{\omega}_{21}$. 

For completeness, in Fig. \ref{fig:Bistable1} we also show the population difference $\Delta n_{21}$ and the reflectivity $R(\omega_{pump})$ for a fixed pump photon energy $\hbar \omega_{pump} =  94$ meV, while the incident power is progressively increased, for various values of the initial ground state population. As discussed in \cite{jeannin_unified_2021} the bi-stability develops for high electronic populations, at the onset of the strong coupling regime. 

Our formalism, that is illustrated here for a single transition and single cavity mode allows recovering previous results; however it can be easily implemented for multiple cavity modes and quantum wells systems supporting several intersubband transitions, where effects of multi-stability can be anticipated. Furthermore, in actual heterostructure potentials there high energy levels which can be connected with the lower subbands through multi-photon absorption, as described further. 

Since the two subband system maps to a two-level atomic system that has been extensively studied in the literature \cite{book_OptResonance} it is interesting to compare our results with the former case. For sufficiently low powers equation Eq. (\ref{EqDeltan12}) can be expanded up to the first order in $P_{in}/P_{sat}$, obtaining a first order correction of $\Delta n_{12}$ as a function of the incident power. We then use Eq.(\ref{chi1_2sub}) and the relation $\omega_{P}^2 =\omega_{P1}^2S\Delta n_{12}$ to obtain the leading correction of the first order susceptibility $\chi^{(1)}(\omega)$ as a function of the incident power:

\begin{equation}
 \chi^{(1)}(\omega, P_{in}) \approx  \chi^{(1)}(\omega, 0) \Bigg [ 1 -\frac{P_{in}}{P_{sat} (\omega, 1)} \Bigg ]  
\end{equation}

This first order correction can then be related to the third order susceptibility $\chi^{(3)} (\omega; \omega, \omega, -\omega)$ which appear in Eq. (\ref{defci}). To obtain an explicit expression, we consider the case of weak absorption and low electronic density, thus discarding the depolarization shift,  and a cavity resonant with the bare intersubband transition $\omega_{c1}= \omega_{21}$. We further perform rotating wave approximation, and we introduce the detuning of the pump with the intersubbnad transition $\delta = \omega_{21}-\omega$. After some manipulation, and using the definition of $\chi^{(3)} (\omega; \omega, \omega, -\omega)$ from Ref. \cite{boyd_nonlinear_2008} we arrive at the following formula, expressed for a single quantum well, $N_{QW} =1$ that is comparable to the expressions known for a two-level system \cite{boyd_nonlinear_2008}:

\begin{equation}
\chi^{(3)} (\omega; \omega, \omega, -\omega) \approx \frac{d_{12}^4 S\Delta n_{12}}{3V_{cav}^2 \hbar^3 (\varepsilon \varepsilon_0)^2} \frac{\tau_{21} \gamma_{21} \Gamma_{m,r}}{\delta^4 (\delta+i\gamma_{21}/2)}P_{in} \label{chi3simple}
\end{equation}

In the above expression, besides the  terms which are familiar from Ref. \cite{boyd_nonlinear_2008} there are additional terms related to the microcavity geometry. Indeed, the our model expresses the population difference from  both the electronic and photonic degrees of freedom in the system. The third order susceptibility $\chi^{(3)} (\omega; \omega, \omega, -\omega)$ in our model is not intrinsic,  but rather an effective one which takes into account the properties of the microcavity mode \cite{KRASNOK20188}. In particular, we have a stronger dependence of the detuning, $\delta^{-5}$, with respect to the case of a two level system in free space, $\delta^{-3}$, \cite{boyd_nonlinear_2008}. The additional factor $\delta^{-2}$ can be traced back to the filtering effect of the cavity, as the pump should be made resonant with both the cavity mode and the intersubband transition. 

The third order susceptibility $\chi^{(3)} (\omega; \omega, \omega, -\omega)$ is often considered in the context of the  Kerr effect \cite{boyd_nonlinear_2008} related to the refracting index dependence on the light intensity in the case of propagation in a nonlinear medium \cite{boyd_nonlinear_2008}. In the present case, where the light interaction is mediated by spatially confined electromagnetic resonances this picture does not apply. Rather, a  more pertinent quantity to consider here is the  reflectivity from the system \cite{Cotrufo2024}, as shown in Fig. (\ref{fig:Bistable2}), and namely the phase shift introduced by the polaritonic system on the reflected signal. The latter can be easily inferred from our equations as the argument of the complex function $arg(\tilde{I}^{(1)}_{out}/\tilde{I}_{in})$ and Eq. (\ref{EqDeltan12}) to determine the population difference $\Delta n_{12}$ self-consistently: this method provides the nonlinear phase shift at any order of the incident power. 

\begin{figure}
    \centering
    \includegraphics[width = \columnwidth]{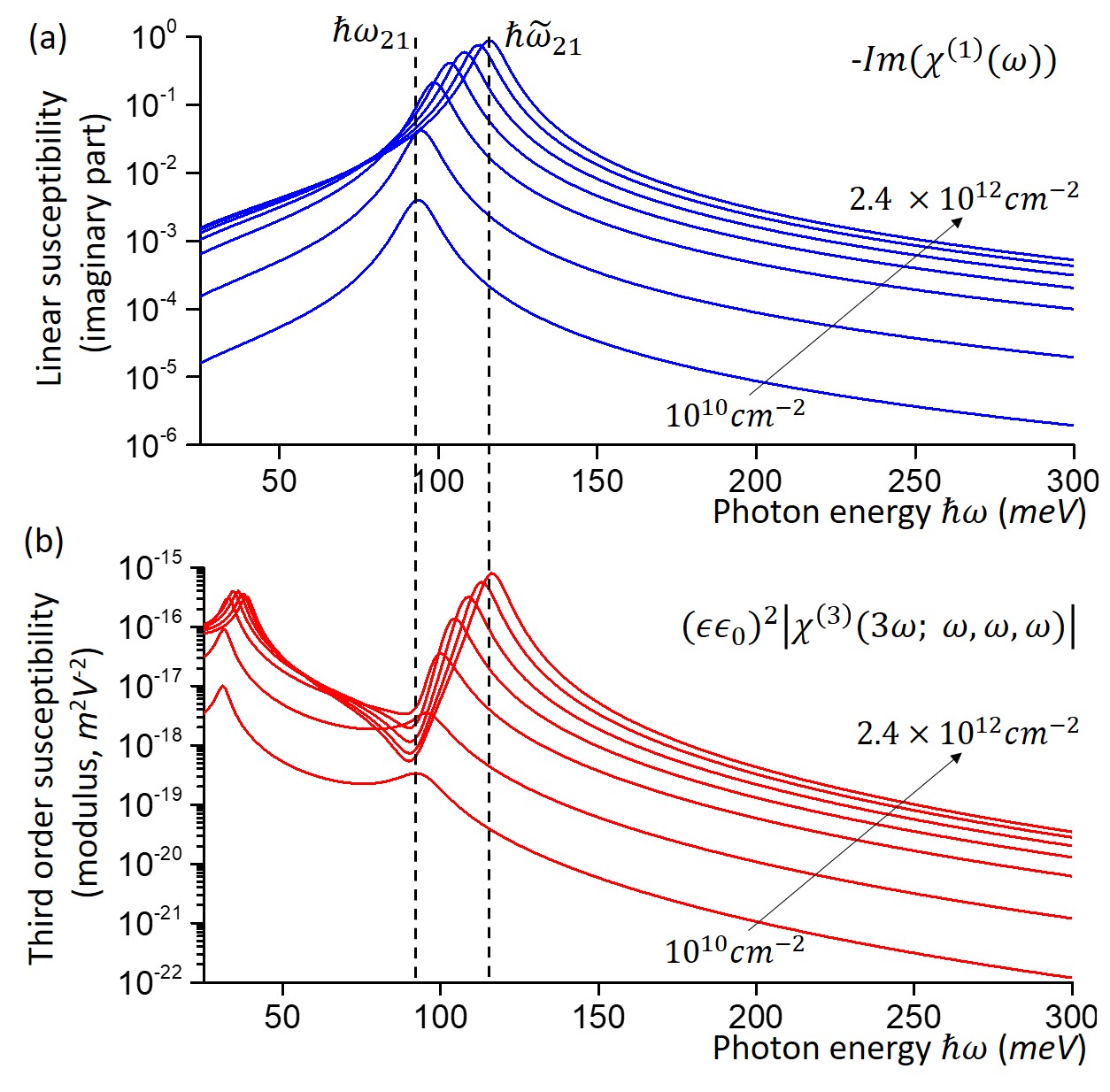}
    \caption{(a) Spectra of the imaginary part $-\Im(\chi^{1} (\omega))$ of the linear susceptibility for a $10$ nm wide $\mathrm{GaAs}$ quantum well with $\mathrm{Al_{0.3}Ga_{0.7}As}$ with two confined subband, and for increasing areal electronic density: $\Delta n_{21} = 10^{10}cm^{-2}$, $10^{11}cm^{-2}$, $5 \times 10^{11}cm^{-2}$, $10^{12}cm^{-2}$, $1.5\times 10^{12}cm^{-2}$, $2.0 \times 10^{12}cm^{-2}$ , $2.4 \times 10^{12}cm^{-2}$. 
    (b) Modulus of the third order nonlinear susceptibility, Eq. (\ref{chi3twosub}),  multiplied by  $(\varepsilon \varepsilon_0)^2$  for increasing areal electronic density as described in panel (a).}
    \label{fig:chi3_2sub}
\end{figure}

\subsubsection{Third order susceptibility and third harmonic generation} \label{parN22}

 We now examine the second order contribution to the excited level population, Eq. (\ref{N2_2}). Proceeding as described in the previous sections, it is straightforward to express $\tilde{S}^{(3)}_i$ from Eqs. (\ref{SatS12 1}) and (\ref{SatJ12 1}). The third order susceptibility is then obtained in a closed form:

\begin{eqnarray}
  \chi^{(3)} (3\omega; \omega, \omega, \omega) =  
  \nonumber \\ -\frac{8}{(\hbar \varepsilon \varepsilon_0)^3L_{cav}}\frac{i\tau_{21}}{1+i2\omega \tau_{21}}\frac{\tilde{d}_{12}^2 (\omega) d_{12}^2 \omega_{12}^2 \Delta n_{21}}{\Delta_{21} (\omega)\Delta_{21} (3\omega)}
  \nonumber \\
  =-\frac{8}{(\hbar \varepsilon \varepsilon_0)^3L_{cav}}\frac{i\tau_{21}}{1+i2\omega \tau_{21}}
  \frac{d_{12}^4  [\Delta^0_{12} (\omega) ]^2\omega_{12}^2 \Delta n_{21}}{[\Delta_{21} (\omega)]^3\Delta_{21} (3\omega)} \label{chi3twosub}
\end{eqnarray}

In the second line of Eq. (\ref{chi3twosub}) we made use of the effective dipole renormalized by collective many-body effects introduced in Eq. (\ref{d_tilde}). An alternative expression is obtained using the overlap factor $f_w$ and the oscillator strength $f_{12}$ of the transition:

\begin{eqnarray}
 \chi^{(3)} (3\omega; \omega, \omega, \omega) =    \nonumber \\ 
 - (f_w f_{12} \omega_{P1}^2)^2  \frac{V_{cav}}{\hbar \varepsilon \varepsilon_0}\frac{i\tau_{21}}{1+i2\omega \tau_{21}}\frac{\Delta N_{21}[\Delta^0_{12} (\omega) ]^2 }{[\Delta_{21} (\omega)]^3\Delta_{21} (3\omega)} \label{chi3twosub_2}
\end{eqnarray}

Here $\Delta N_{21} = S\Delta n_{21}$ is the population difference between subband 1 and subband 2. The advanatge of the above expression is that it can be generalized directly for the case of where the sample contains $N_{QW}$ identical quantum wells, as discussed in the previous section, bearing in mind that now the population difference $\Delta N_{21}$ must be defined per quantum well.

As discussed in section \ref{General theory chi} our definition of the nonlinear susceptibilities differs by a factor $(\varepsilon\varepsilon_0)^{n-1}$ with respect to the definitions in the literature \cite{boyd_nonlinear_2008}. Eqs. (\ref{chi3twosub_2}) must be multiplied by a factor $(\varepsilon\varepsilon_0)^2$ in order to obtain a quantity with the usual units $[m^2V^{-2}]$. Our expressions can be easily compared with known results for the third order susceptibility in quantum wells. For instance, supposing that the upper state lifetime is sufficiently long, $\omega_{21}\tau_{21} >> 1$, discarding many-body effect and assuming RWA we arrive at the following expression:

\begin{eqnarray}
    \chi^{(3)} (3\omega; \omega, \omega, \omega)|_{\omega \approx \omega_{21}} \nonumber \\ \approx
    \frac{N_{QW}\Delta n_{12} d_{12}^4}{(\hbar \varepsilon \varepsilon_0)^3L_{cav}}\frac{1}{\omega_{12}^2 (\delta + i\gamma_{12}/2)}
\end{eqnarray}

The nonlinear susceptibility in that case is maximized at a pumping frequency $\omega = \omega_{21}$ ($\delta = 0$). Similar expressions have been obtained in the Ref. \cite{ahn_calculation_1987}. 

\begin{figure}
    \centering
    \includegraphics[width = \columnwidth]{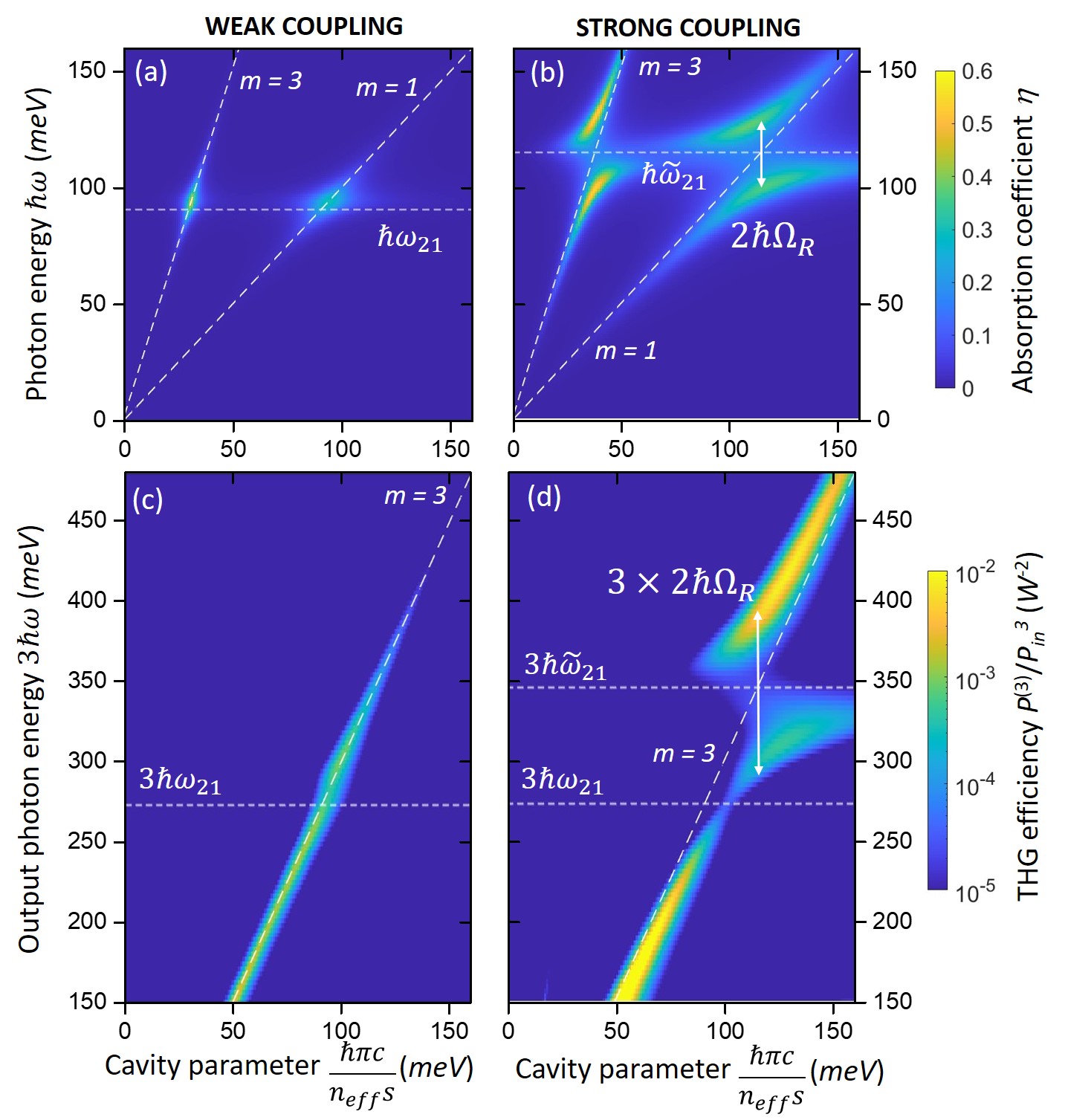}
    \caption{ (a,b) Color plot of the absorption efficiency $\eta (\omega)$) for the case of low population, $10^{10}$ $cm^{-2}$ (a) and high  population,$2.4\times 10^{12}$ $cm^{-2}$ (b). The quantum efficiency is plotted as a function of the photon energy $\hbar \omega$ and the cavity parameter $\hbar \pi c/n_{eff} s$ ($x$- axis shared with panels (c,d).
    (c) Color plot of the THG efficiency for the power exracted from the $m=3$ resonance, Eq. (\ref{THGpower}), as a function of the output photon energy $3 \hbar \omega$, and the cavity parameter $\hbar \pi c/n_{eff} s$,  for an areal population $10^{10}$ $cm^{-2}$. (d) Contour plot of the extracted THG power efficiency for the case of high areal population $2.4\times 10^{12}$ $cm^{-2}$, where microcavity and the intersubband transition are in a strong coupling regime, with a Rabi energy $\hbar \Omega_R$.}
    \label{fig:THG_2sub}
\end{figure}

An interesting aspect of the general expression for the third order susceptibility (\ref{chi3twosub_2}) is that it depends both on the single-particle characteristic function  $\Delta^0_{12} (\omega)$ as well as on the characteristic function of the re-normalized oscillator, $\Delta_{12} (\omega)$. As a result, the spectral behavior of $\chi^{(3)} (3\omega; \omega, \omega, \omega)$ contains both features related to the single-particle transition frequency, $\omega_{12}$, as well as the renormalized one, $\tilde{\omega}_{12}$.
In Fig. \ref{fig:chi3_2sub} we plot the modulus of the full expression Eq. (\ref{chi3twosub}) as a function of the incident photon energy in $meV$, in comparison with the quantity $-\Im \chi^{(1)} (\omega)$ which is representative of the absorption spectrum of the electronic system alone. As mentioned above, we 
 the quantity $\varepsilon_0^2 |\chi^{(3)} (3\omega, \omega, \omega, \omega)|$ which has units $m^2/V^2$.  The values observed are indeed comparable to the ones reported for intersubband systems \cite{Capasso_1994, Sirtori_PRL1992, Park2024}, even if the comparison is not direct as here we consider single intersubband transition instead of triply-resonant structures. In our calculations the parameters of the quantum well are identical to those from the previous paragraph. We vary the population difference $\Delta n_{12}$ between the values $0.1 \times 10^{11}$ $cm^{-2}$ and $24 \times 10^{11}$ $cm^{-2}$ in order to be able to explore the transition between the case of strong collective effects and the case where they are negligible. 

At low densities we observe a peak at $\hbar \omega_{21}$ both for linear and nonlinear susceptibilities. For the nonlinear susceptibility there is a low frequency resonance for $\hbar\omega = \hbar\tilde{\omega}_{12}/3$ which arises from the denominator  $\Delta_{12} (3\omega)$; the latter has a much higher amplitude than the resonance at the intersubband energy $\hbar \omega_{21}$. As the electronic population is increased, the second resonance follows the collective energy $\hbar \tilde{\omega}_{12}$ while acquiring a much higher amplitude, eventually surpassing the resonance $\hbar \tilde{\omega}_{12}/3$ for very high electronic densities. 

This behavior  can be attributed to the local field enhancement  of the effective dipole, Eq. (\ref{d_tilde}), as commented in section \ref{secLinear}. This is precisely an expression of the ENZ effect on the nonlinear susceptibility \cite{Fomra_reviewENZ_2024, Barbet_2023}. To be more specific, we observe from Eq. (\ref{chi3twosub}) that the third order susceptibility is propositional to the ratio $[\Delta^0_{12}(\omega)]^2/[\Delta_{12}(\omega)]^3$. In the case where collective effects are small, we have $\Delta_{12}(\omega) \approx \Delta^0_{12}(\omega)$ and this ratio becomes $\approx 1/\Delta^0_{12}(\omega)$. This factor produces the weak resonance at $\hbar \omega \approx \hbar \omega_{21}$ seen in Fig. \ref{fig:chi3_2sub}(b) for the low doping case. This resonance appears in the background  of $\hbar \omega \approx \omega_{21}/3$ resonance provided by the factor $1/\Delta_{12}(3\omega)$, as the latter has $3^2$-fold larger amplitude. As the electronic density increases, and the intersubband resonance is blue shifted at a  higher energy  $\hbar \tilde{\omega}_{21}$ the factor $\Delta^0_{12}(\omega)^2$ no longer compensates for the pole provided by $1/\Delta_{12}(\omega)^3$ which is triply degenerated and acquires now a much higher spectral weight. At the same time, the local minimum at the energy  $\hbar \omega_{12}$ that is brought by the factor $|\Delta^0_{12}(\omega)|^2$ becomes more and more apparent. Clearly, the spectral dependence of the function $\chi^{(3)} (3\omega, \omega, \omega, \omega)$ contains more information than the linear susceptibility, as it displays features that depend both on the collective and single-particle characteristic energies. 

Next, we consider the third harmonic generation from the system. To that end we now consider both the $m=1$ and $m=3$ modes and we assume $\omega_{c3} = 3\omega_{c1}$ for simplicity, thus ignoring the spectral dependence of the effective refractive index.  For pump and signal under normal incidence of the array the $m=2$ resonance is not excited \cite{todorov_optical_2010} and can be excluded from our discussion. We further consider a linear polarization of the pump along the $x$ direction, such as only the resonances $TM_{m0}$ are excited. The spatial distributions of the  $\mathrm{TM}_{10}$ and $\mathrm{TM}_{30}$ modes are then provided by the expressions (see also Fig. \ref{fig: global}):

\begin{equation}
    u_1 (x) = \sqrt{2} \cos (\pi x/s), \phantom{Q} u_3 (x) = \sqrt{2} \cos (3\pi x/s)
\end{equation}

We must then compute the coefficients $\kappa_{1[1^{k_1}3^{k_3}]}$ from Eq.(\ref{defKappa}). There is a total of 8 coefficients: 

\begin{center} 
\begin{tabular}{|m{5em} | m{1cm}|m{5em} | m{1cm}|} 
\hline
 \multicolumn{4}{|c|}{Intermode coupling coefficients} \\
 \hline
  $\kappa_{1[1^{3}3^{0}]}$ & 3/2 &  $\kappa_{3[1^{3}3^{0}]}$ & 1/2 \\ 
  \hline
  $\kappa_{1[1^{2}3^{1}]}$ & 1/2 &  $\kappa_{3[1^{2}3^{1}]}$ & 1 \\ 
  \hline
 $\kappa_{1[1^{1}3^{2}]}$ & 1 &  $\kappa_{3[1^{1}3^{2}]}$ & 0 \\ 
  \hline
$\kappa_{1[1^{0}3^{3}]}$ & 0 &  $\kappa_{3[1^{0}3^{3}]}$ & 3/2 \\ 
  \hline
\end{tabular}
\end{center}

Using Eq.(\ref{P_n_out}) for $n=3$ we have the following expression for the extracted power:

\begin{eqnarray}
\frac{P^{(3)} (3\omega)}{P_{in}^3} = \Big( \frac{\varepsilon_0 \varepsilon}{V_{cav}}\Big)^2 \frac{36 \omega^2|\chi^{(3)} (3\omega; \omega, \omega,\omega)|^2}{|1+i3\omega Y_p(3\omega)|^2 |1+i\omega Y_p(\omega)|^{6}} 
\nonumber \\ 
\times \Big| [3 \Theta_1 (3 \omega) + \Theta_3 (3 \omega)] \Theta_1^3 (\omega) 
\nonumber \\ 
+[3 \Theta_1 (3 \omega) + 6\Theta_3 (3 \omega)] \Theta_1^2 (\omega) \Theta_3 (\omega) 
\nonumber \\ 
+ 6 \Theta_1 (3 \omega) \Theta_1 (\omega) \Theta_3^2 (\omega) + 3 \Theta_3 (3 \omega) \Theta_3^3 (\omega) \Big|^2 \label{THGpower}
\end{eqnarray}

The functions $\Theta_m (\omega)$ have been defined in Eq. (\ref{defTheta_m}). In particular, there is a coherent sum involving the two modes inside the expression of the square modulus. 

In Fig. \ref{fig:THG_2sub} we provide the color plots for the linear absorption quantum efficiency, $\eta (\omega)$ and the quantity $P^{(3)} (3\omega)/P_{in}^3$ as a function of the photon energy $\hbar \omega$ and the inverse cavity size $\hbar \pi c/n s$. Since now there are several modes involved we use our general expression Eq. (\ref{eta_lin}) for the computation of the quantum efficiency. Identical parameters $\Gamma_{m, nr}=\Gamma_{m, r}=10$ $meV$ have been used for both modes. We consider both the weak coupling case (left panels, Fig. \ref{fig:THG_2sub}(a,c)), electronic density $10^{10}$ $cm^{-2}$, and the strong coupling case, right panels, (right panels, Fig. \ref{fig:THG_2sub}(b,d)), electronic density $2.4 \times10^{12}$ $cm^{-2}$. In the weak coupling regime, Fig. \ref{fig:THG_2sub}(a) the absorption is strong only when the electronic transition is matched with either cavity resonance, $\omega_{21} = \omega_{c1}$ or $\omega_{21} = \omega_{c3}$. The single-particle photon energy $\hbar \omega_{21}$ has been indicated as a reference. For a high electronic density, Fig. \ref{fig:THG_2sub}(b) the system is in strong coupling regime, where two polariton states arise from either the $m=1$ and $m=3$ modes. For both modes the polariton splitting is identical, $2\hbar \Omega_R = 33$ meV. The photon energy of the collective electronic resonance $\hbar \tilde{\omega}_{21}$ is indicated as a reference. 

Fig.\ref{fig:THG_2sub} (c,d) are the vertical axis is the output photon energy, $3\hbar \omega$, for a photon pump energy $\hbar \omega$. The power is collected from the $m=3$ resonances of the micro-cavities.  For the weak coupling case, Fig. \ref{fig:THG_2sub} (c), we observe that THG power can be efficiently extracted from micro-cavities with a broad range of parameters. Indeed, as seen from Fig. \ref{fig:chi3_2sub} (b) the function $\chi^{(3)} (3\omega;\omega,\omega,\omega)$ is rather broad in that case, and dominated by the low frequency resonance $\omega_{21}/3$. 
For the strong coupling case, Fig. \ref{fig:THG_2sub} (d) we observe a pattern that is determined both from the spectral dependence of the function $\chi^{(3)} (3\omega;\omega,\omega,\omega)$, which has now a strong resonance at $\tilde{\omega}_{21}$ and the polaritonic resonances of the system. Indeed, now the pattern with maximum THG power follows the tripled version of the polariton dispersion arising from the $m=1$ mode (Fig. \ref{fig:THG_2sub}(b)). In particular, in  Fig. \ref{fig:THG_2sub} (d) the separation between the two polariton branches is tripled and equal to $3\times 2\hbar \Omega_R$. The node of the nonlinear susceptibility that develops at the single-particle frequency, $\omega_{21}$  (Fig. \ref{fig:chi3_2sub} (b)) now translates as a drop in the THG efficiency for a single-particle photon energy $3\hbar \omega_{21}$. These results hint the possibility to observe and study intersubband polariton states by frequency tripling, in a spectral range where detectors are typically more efficient than those in the MIR region. Thus, the quantum statistics of the upconverted polarion states can bear signatures of the squeezing properties that have been predicted in the ultra-strong coupling regime \cite{ciuti_quantum_2005}. 

The plots in Fig.\ref{fig:THG_2sub} (c,d) display rather high THG generation efficiencies, on the order of $10^{-2} W^{-2}$ for the high doping case. Here we have neglected saturation effects discussed in the previous section, and thus the typical pumping powers can be taken to be on the onset of saturation, on the order of $P_{in} \approx 10$ mW. The outcoupled power is in that case on the order of $100$ nW. For instance, in a recent experimental study \cite{Park2024} peak THG powers of the same order of magnitudes are reported with triply resonant InGaAs/AlInAs quantum wells, with pump powers in the range of $200$ mW. Difference with respect to our model could arise from the fact that our estimations are done for critically coupled cavities $\eta_r =0.5$ and we use relatively narrow linewidths, $10$ meV in our calculations. Furthermore, the relaxation time parameter, here taken to be $\tau_{21} \approx 1$ ps also plays important role. These differences could explain the fact that we estimate higher THG efficiencies with respect to one experimentally observed in Ref. \cite{Park2024}.

\subsection{Three-subband system} \label{sec3subExample}

\begin{figure} 
    \centering 
    \includegraphics[width = 7cm]{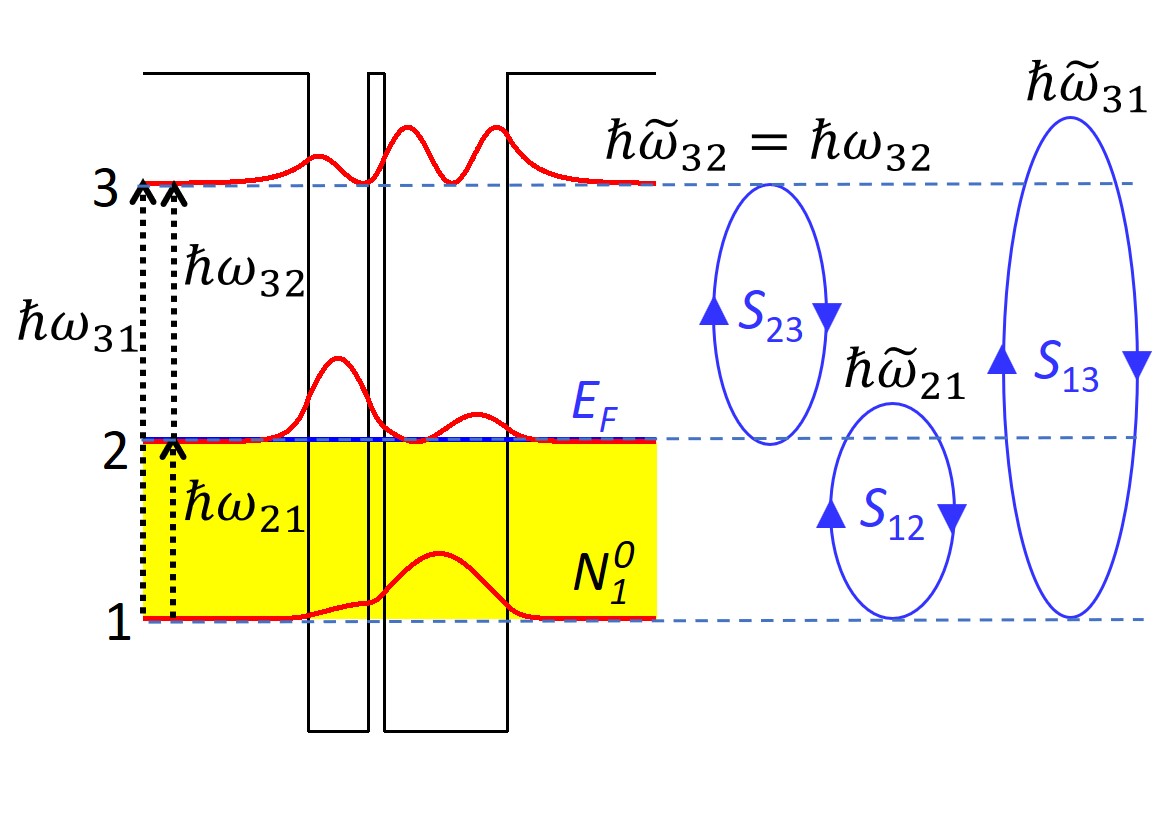}
    \caption{Typical hetero-structure potential yielding three confined subbands that is used for the study of second order nonlinear processes. The Fermi level $E_F$ does note exceed the second subband edge, thus only the first subband is populated at $N_1^0$. The ellipses indicate the polarizations $S_{ij}$ considered for this system, while their larger axes of the ellipses indicate the energies of the collective electronic excitations $\hbar \tilde{\omega}_{ji}$.}
    \label{fig:3subF1}
\end{figure}

As a second example we consider a three subband system depicted in Fig. \ref{fig:3subF1}. Such systems have been considered for the second order generation in semiconductors \cite{Capasso_1994, book_Berger_1999,  book_Sirtori_2000}, whereas so far they have been described in the single-particle approximation in the context of experimental studies with metasurfaces \cite{sarma_all-dielectric_2022}. Here we consider the simplest case where only the first subband is populated at $N_1^0$ at equilibrium, however the population can be sufficiently high in order to observe strong coupling phenomena and manybody effects \cite{delteil_charge-induced_2012}. In the Fig. \ref{fig:3subF1} we have indicated the three single-particle transitions between the quantized subbands $\hbar \omega_{21}$, $\hbar \omega_{31}$ and $\hbar \omega_{32}$ as well as a schematics of the corresponding polarizations $S_{ij}$ that oscillate at renormalized frequencies.  

In this example the transitions $1\rightarrow 2$ and  $1\rightarrow 3$ are renormalized by many-body effects, with the frequencies of the collective electronic modes noted as $\hbar \tilde{\omega}_{21}$ and $\hbar \tilde{\omega}_{31}$, while the transition $2\rightarrow 3$ occurs at the single-particle value $\hbar \omega_{32}$. 

\subsubsection{First and second order susceptibility}

\begin{figure}
    \centering
    \includegraphics[width = \columnwidth]{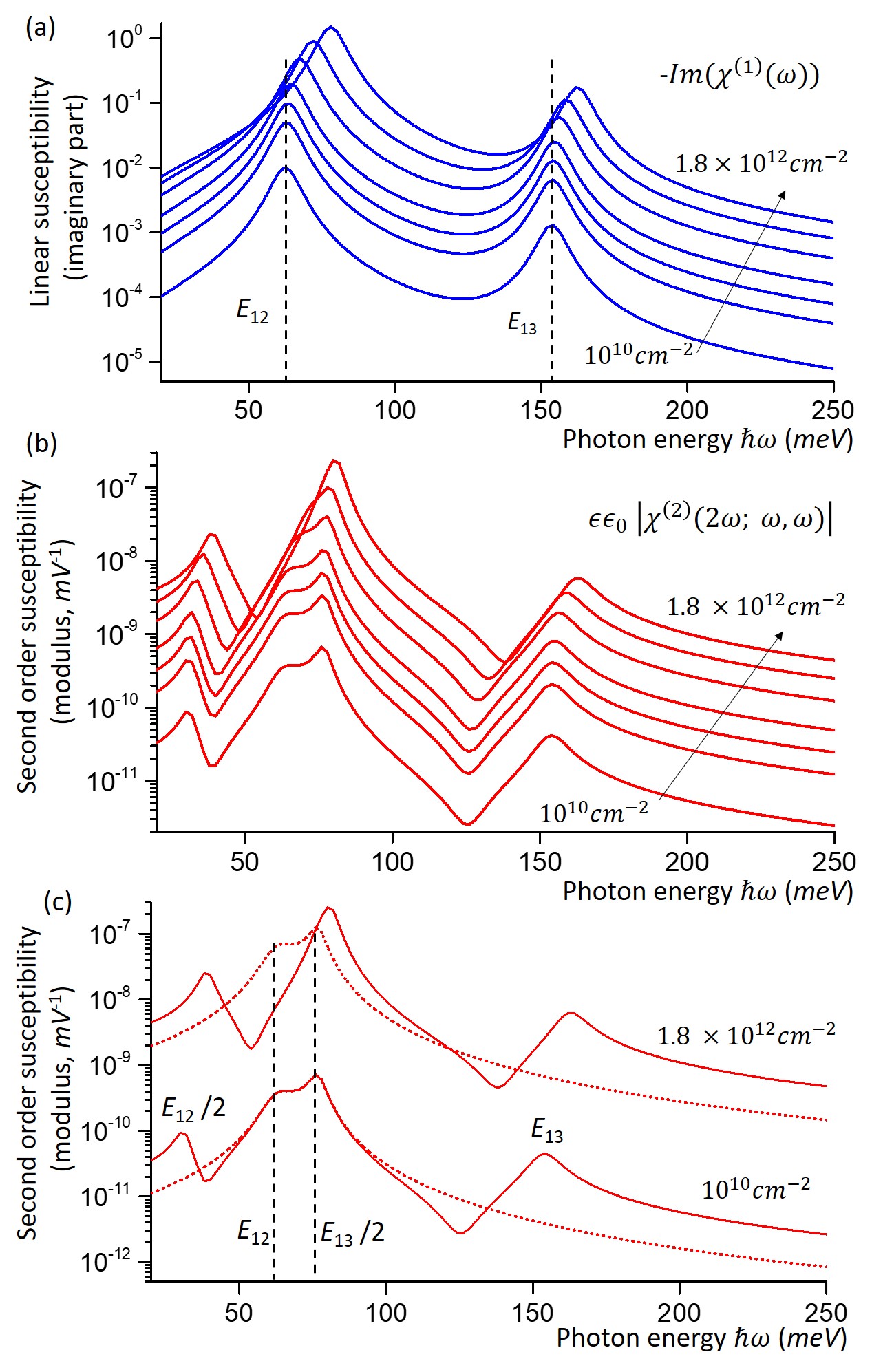}
    \caption{Linear and nonlinear susceptibility for a  $\mathrm{GaAs}/\mathrm{Al_{0.3}Ga_{0.7}As}$ two coupled quantum wells system with thicknesses $3.6/\mathbf{1}/7.4$ in nanometers. (a) Imaginary part of the linear susceptibility (Eq. (\ref{chi1_3sub})) as a function of the photon frequency $\hbar \omega$ for various values of the electronic density: $0.1, 0.5, 1, 2, 5, 10, 18$ $\times 10^{11}$ $cm^{-2}$ (from bottom curve to top). Here $E_{12}$ and $E_{13}$ label the energies of the single-particle transitions $1 \rightarrow 2$ and  $1 \rightarrow 3$. (b) Corresponding values of the modulus of the second order susceptibility (Eq. (\ref{Eq_chi2_3subfull})) multiplied by $\varepsilon \varepsilon_0$. (c) Comparison between the full expression (full curve), Eq. (\ref{Eq_chi2_3subfull}), and the single-particle result (dotted curve), Eq.(\ref{chi2_SP}), as derived in the literature so far (i.e. Ref \cite{Rosencher_Bois_EL_1989}), for the lowest and highest electronic density considered in this example.}
    \label{fig:chi2_3sub}
\end{figure}

We start our analysis by providing the matrix elements of the inverse of the matrix $M_{\alpha\beta}^{-1} (\omega)$ for the case depicted in Fig. \ref{fig:3subF1}. At equilibrium and without external illumination there is an areal population difference $\Delta n_0 = N^0_{1}/S$ for the transitions $1 \rightarrow 2$ and $1 \rightarrow 3$, while the transition $2 \rightarrow 3$ is not populated. The relevant matrix elements for the matrix  $M_{\alpha\beta}^{-1}(\omega)$ are then:

\begin{eqnarray}
M_{\alpha\beta}^{-1}(\omega) = \frac{1}{\mathrm{Det} (\omega)} \times \nonumber \\
\begin{array}{c| c c} 
\alpha \downarrow,  \beta \rightarrow & 12 & 13 \\ 
\hline \\[-8pt]
12 & \Delta_{13} (\omega) & -W^2 \sqrt{\frac{\omega_{21}}{\omega_{31}}}\\[10pt]
13 & -W^2 \sqrt{\frac{\omega_{31}}{\omega_{21}}} & \Delta_{12} (\omega) \\
\end{array} 
\\
W^2 = C_{12,13} \omega_{P,12}\omega_{P,13} \label{DefC1213}
\\
\Delta_{12} (\omega) = \tilde{\omega}^2_{21} -  \omega^2 + i\gamma_{12}\omega
\\
\Delta_{13} (\omega) = \tilde{\omega}^2_{31} -  \omega^2 + i\gamma_{13}\omega
\\
\mathrm{Det} (\omega) = \Delta_{12} (\omega)\Delta_{13} (\omega) - W^4
\end{eqnarray}

The quantity $W^2$ introduced in Eq. (\ref{DefC1213}), homogeneous to a frequency squared, quantifies the coupling between the intersuband plasmons from the transitions  $1 \rightarrow 2$ and $1 \rightarrow 3$ \cite{todorov_intersubband_2012}. Using Eq. (\ref{def_chi_1}) from section \ref{secLinear} we obtain the following expression for the linear susceptibility:

\begin{eqnarray}
    \chi^{(1)} (\omega) =  \frac{2\Delta n_0}{\hbar \varepsilon \varepsilon_0 L_{cav} \mathrm{Det} (\omega)}
    \nonumber \\\times \Big \{ \Delta_{13} (\omega) \omega_{21} d_{12}^2 + \Delta_{12} (\omega) \omega_{31} d_{13}^2 \nonumber \\
    -2\sqrt{\omega_{21}\omega_{31}}d_{12}d_{13}W^2 \Big\}   \label{chi1_3sub}
\end{eqnarray}

This equation can be recast in a different form by using the expressions of the single transition susceptibilities, Eq. (\ref{chi1_2sub}):

\begin{equation}
 \chi^{(1)} (\omega) = \frac{\chi^{(1)}_{1\rightarrow 2} (\omega) + \chi^{(1)}_{1\rightarrow 3} (\omega)  - 2\frac{W^2\sqrt{\chi^{(1)}_{1\rightarrow 2} (\omega) \chi^{(1)}_{1\rightarrow 3} (\omega) }}{\sqrt{\Delta_{12} (\omega)\Delta_{13} (\omega)}}}{1- \frac{W^4}{\Delta_{12} (\omega)\Delta_{13} (\omega)}} 
\end{equation}

With this form, it is apparent that the overall first order susceptibility is not merely the sum of the susceptibilities for individual transitions, as one would expect from a single-particle picture. Dipole-dipole interactions induce blue shifts and transfer of oscillator strength, as previously shown for instance in Ref. \cite{Delteil_APL_2013}. These effects can be easily inferred from the analytical expressions above, as the collective resonances are provided by the poles of the function $\chi^{(1)} (\omega)$ that are provided by the equation $\mathrm{Det} (\omega)= 0$, and their respective oscillator strengths are proportional to the residues of the poles. 

Next, we apply the results from  section \ref{Gen2nd_order_chi} in order to compute the second order susceptibility. We start by providing the expressions of the functions  $G_{\lambda \mu}^{(2)} (\omega)$ from Eq.(\ref{defG2_lambda_mu}). The only non-zero contributions are:

\begin{eqnarray}
 G_{12}^{(2)} (\omega) =  \frac{2d_{23}\Delta n_0}{\mathrm{Det} (\omega)}\Bigg[ d_{13} \Delta_{12} (\omega) - d_{12} W^2 \sqrt{\frac{\omega_{21}}{\omega_{31}}} \Bigg] \nonumber \\
\times  (\omega_{21}\omega_{31} + 2\omega^2 - i\omega \gamma_{31})
\\
G_{13}^{(2)} (\omega) =  \frac{2d_{23}\Delta n_0}{\mathrm{Det} (\omega)}\Bigg[ d_{12} \Delta_{13} (\omega) - d_{13} W^2 \sqrt{\frac{\omega_{31}}{\omega_{21}}} \Bigg] \nonumber \\
\times  (\omega_{21}\omega_{31} + 2\omega^2 - i\omega \gamma_{21})
\end{eqnarray}

The second order susceptibility then becomes:

\begin{eqnarray}
    \chi^{(2)} (2\omega; \omega, \omega) = \frac{1}{(\hbar \varepsilon \varepsilon_0 )^2 L_{cav}}\frac{1}{\mathrm{Det} (2\omega)} \nonumber \\
    \times \Bigg \{ G_{12}^{(2)} (\omega) \Bigg[ d_{12} \Delta_{13} (2\omega) - d_{13} W^2 \sqrt{\frac{\omega_{31}}{\omega_{21}}} \Bigg]
    \nonumber \\
    + G_{13}^{(2)} (\omega) \Bigg[ d_{13} \Delta_{12} (2\omega) - d_{12} W^2 \sqrt{\frac{\omega_{21}}{\omega_{31}}} \Bigg] \Bigg \} \label{Eq_chi2_3subfull}
\end{eqnarray}

The second order susceptibility is thus proportional to the dipole $d_{23}$ of the empty transition, which translates the general requirement of symmetry breaking for second order nonlinear effects \cite{boyd_nonlinear_2008, book_Berger_1999}.  Just like for the case of the third order susceptibility the second order nonlinear effects depend both on the characteristics of the collective resonances and the single-particle transitions. 

It is interesting to examine the various limits of the above expression depending on the strength of dipole-dipole interactions. To simplify the discussion we will assume similar values of the linewidth broadening $\gamma \approx \gamma_{21} \approx \gamma_{31}$ \cite{Rosencher_Bois_EL_1989}. At the lowest order, where dipole-dipole interactions can be ignored, Eq. (\ref{Eq_chi2_3subfull}) simplifies to:

\begin{eqnarray}
\chi^{(2)} (2\omega; \omega, \omega) \approx \nonumber \\
\frac{2d_{12}d_{13}d_{23}\Delta n_0}{(\hbar \varepsilon \varepsilon_0 )^2 L_{cav}}(\omega_{21}\omega_{31} + 2\omega^2 - i\omega \gamma) \nonumber \\ 
\times \Big \{ \frac{1}{\Delta_{12}(\omega) \Delta_{13}(2\omega)} +\frac{1}{\Delta_{12}(2\omega) \Delta_{13}(\omega)} \Big \} \label{chi2_SP_full}
\end{eqnarray}

For the case $\omega_{31} \approx 2\omega_{21}$ we can perform RWA, thus arriving at the well-known expression for the second order susceptibility of doubly-resonant intersubband structures \cite{Rosencher_Bois_EL_1989, book_Berger_1999}:

\begin{eqnarray}
  \chi^{(2)} (2\omega; \omega, \omega) \approx \frac{e^3 z_{12}z_{13}z_{23}\Delta n_0}{(\hbar \varepsilon \varepsilon_0 )^2 L_{cav}}
  \nonumber \\
 \times \frac{1}{(\omega_{21} - \omega + i\gamma/2)(\omega_{31} - 2\omega + i\gamma/2)} \label{chi2_SP} 
\end{eqnarray}

Here we introduced the position matrix elements, $z_{ij}$, such as the dipoles moments are expressed as $d_{ij}=ez_{ij}$ with $e$ the electron charge. We recognize the proportionality of the susceptibility function to the volume density of electrons, $\Delta n_0/L_{cav}$,  \cite{Rosencher_Bois_EL_1989, book_Berger_1999, boyd_nonlinear_2008}. The additional factor $\varepsilon \varepsilon_0$ arises from our choice to define susceptibilities with respect to electrical displacement field rather than the electric field as discussed above. 

\begin{figure*}
    \centering
    \includegraphics[width = \textwidth]{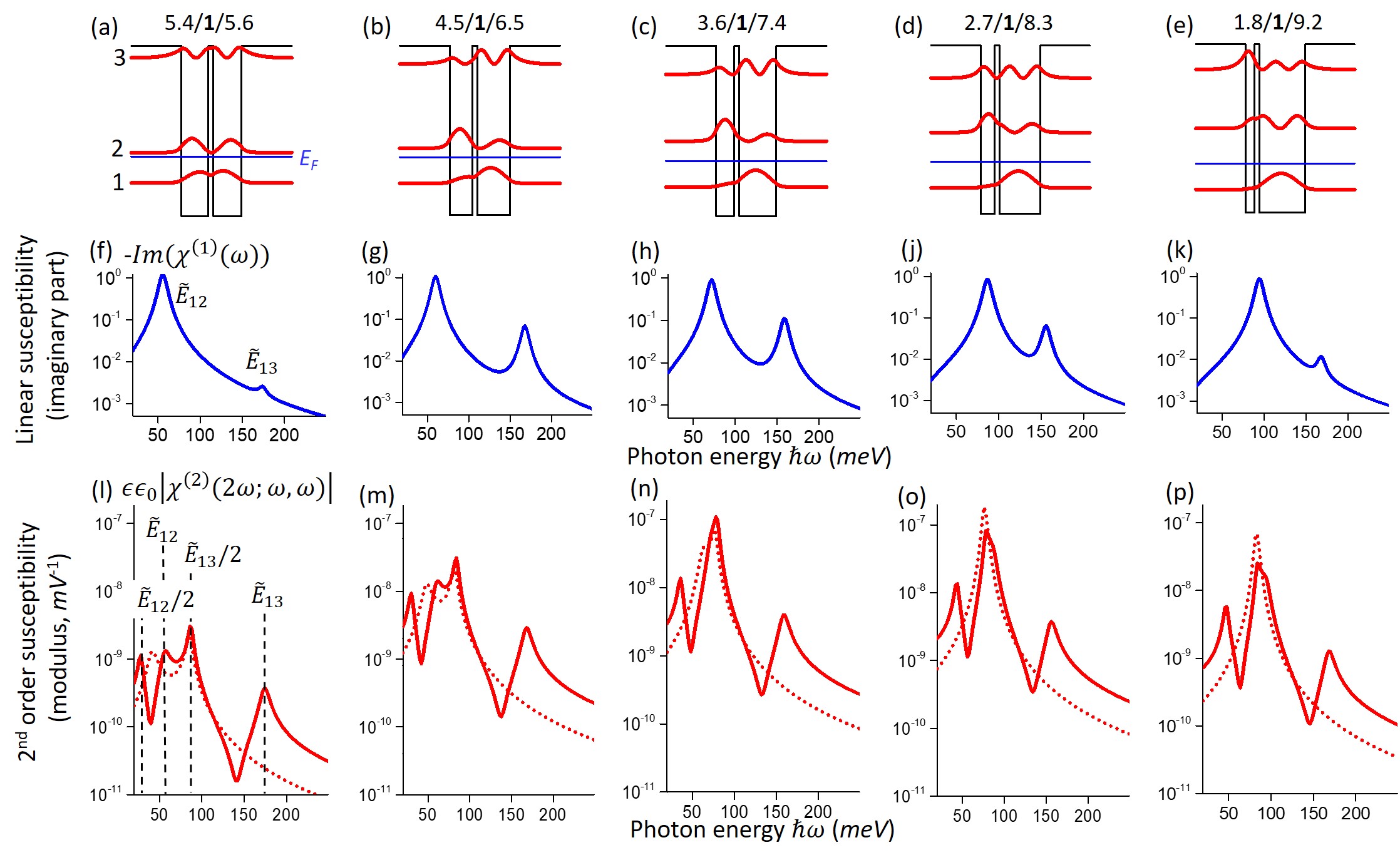}
    \caption{(a-e) A set of coupled double quantum well $\mathrm{GaAs/Al_{0.3}Ga_{0.7}As}$ heterostructures. The thickness of the wells and barriers has been indicated above each structure in nanometers, with the barrier thickness (bold) always equal to $1$ nm. For each potential there are three confined subbands, labeled 1-3.  The square modulus of the corresponding wavefunctions are represented by the red curves. The blue line indicates the position of the Fermi level $E_F$ which corresponds to an areal doping of $10^{12}$ $cm^{-2}$. The labels from (a) have not been reproduced for (b-c) as they are identical. (f-k) Imaginary part of the linear susceptibility  $-\Im(\chi^{(1)}(\omega))$ as a function of the photon energy $\hbar \omega$ for each quantum design from (a-e), as deduced from Eq.(\ref{chi1_3sub}). The two resonances correspond to the energies $\tilde{E}_{12}$ and $\tilde{E}_{13}$ of the collective modes of the three subband system. These labels have not been repeated in (g-k). The horizontal axes of all graphs share the same label.
    (l-m) Modulus of the second order susceptibility function $\varepsilon \varepsilon_0|\chi^{(2)} (2\omega; \omega, \omega)|$ in $m/V$. The full red curve correspond to the complete model (Eq. (\ref{Eq_chi2_3subfull})), the dotted curves correspond to the single-particle expression under the rotating wave approximation, (Eq. (\ref{chi2_SP} )). As indicated in (l) the four resonances of the second order susceptibility function correspond to $\tilde{E}_{12}/2$, $\tilde{E}_{12}$,   $\tilde{E}_{13}/2$ and $\tilde{E}_{13}$. The horizontal axes of all graphs share the same label.}
    \label{fig:chi3_var2sub}
\end{figure*}

We now come back to the full expression of the second order susceptibility, Eq. (\ref{Eq_chi2_3subfull}). There are much more degrees of freedom that we can explore with respect to the case of a two subband system. Here, in a non exhaustive way, we will evaluate the impact of the  electronic concentration and the quantum heterostructure design on the spectral dependence of the second order susceptibility. 

As a first example, we consider a structure composed of two coupled $\mathrm{GaAs}$ quantum wells with thicknesses $L_1 = 3.6$ nm and $L_2 = 7.4$ nm separated by a $L_b = 1$ nm $\mathrm{Al_{0.3}Ga_{0.7}As}$ barrier. As shown further, this design optimizes the peak value of the second-order susceptibility in the presence of collective effects. The results from our calculations are shown in Fig. \ref{fig:chi2_3sub}. For all results, we use identical broadening parameters $\hbar \gamma = 10$  meV for all transitions. 

Fig. \ref{fig:chi2_3sub}(a) shows the imaginary part of the linear susceptibility for this design. The areal electronic densities have been varied between  $10^{10}$ $cm^{-2}$ and $1.8\times 10^{12}$ $cm^{-2}$. Assuming Si dopants distributed along the total thickness of $11$ nm of the quantum wells, these values correspond to volume densities between $10^{16}$ $cm^{-3}$ and $1.6\times10^{18}$ $cm^{-3}$ which are achievable with AlGaAs material systems \cite{pisani_electronic_2023}. For simplicity, we do not consider electrostatic (Hartree) corrections owe to the Si donors to the heterostructure potential, and we assume that it keeps the same shape regardless of the doping. 

For a very low density, $10^{10}$ $cm^{-2}$, the imaginary part of the linear susceptibility is maximum at the single-particle transition energies $E_{12}=\hbar \omega_{21}$ and $E_{13}=\hbar \omega_{31}$, as expected. The oscillator strength of the transition $1\rightarrow 3$ is much lower; this transition is optically active only because of the asymmetry of the heterostructure potential. As the electronic density $\Delta n_0$ is increased,  there is a significant blueshift of the optical resonances owe to the dipole-dipole interactions. In Fig. \ref{fig:chi2_3sub}(b) we plot the corresponding curves for the modulus of the  nonlinear susceptibility $\varepsilon \varepsilon_0|\chi^{(2)} (2\omega; \omega, \omega)|$  multiplied by  $\varepsilon \varepsilon_0$. Typical peak values, on the order of $10^{-7}$ $m/V$ which are comparable with the values ($\sim 100$ $nm/V$) reported in the literature with metasurfaces \cite{yu_broadband_2024}.

We can clearly distinguish four resonances in the susceptibility function that evolve roughly as the collective resonances in the linear susceptibility. In order to gain understanding of this behavior, in Fig. \ref{fig:chi2_3sub}(c) we also compare our results (full lines) with the curves obtained from (Eq. (\ref{chi2_SP}), dotted lines) with the same amount of charges. For low doping, this comparison allows identifying the four resonances of the nonlinear susceptibility which are located at the energies $E_{12}/2$, $E_{12}$, $E_{13}/2$ and $E_{13}$. This can also be inferred from Eq. (\ref{chi2_SP_full}), as in that case, the quantities $1/\Delta_{ij} (\omega)$ are maximized at the single-particle transition frequencies $\omega_{ji}$. Eq. (\ref{chi2_SP}) predicts only the highest maxima at $E_{12}$ and $E_{13}/2$ due to the RWA employed for its derivation. 

At higher electronic densities, the maxima of the second order susceptibility are provided by the zeroes of the product $\mathrm{Det} (2\omega)\mathrm{Det}(\omega)$. The maxima thus appear for the renormalized  energies $\tilde{E}_{12}/2$, $\tilde{E}_{12}$, $\tilde{E}_{13}/2$ and $\tilde{E}_{13}$. We can observe that for the highest doping we have $\tilde{E}_{12} = \tilde{E}_{13}/2$, which allows optimization of the nonlinear response. The progressive merging between the resonances $\tilde{E}_{12}$ and $\tilde{E}_{13}/2$ as the electronic density increases is clearly visible in Fig. \ref{fig:chi2_3sub}(b). 
Quite interestingly, at very high doping our calculation provides a peak value of the nonlinear susceptibility that is almost a factor of 3 higher than what expected from the single-particle expression, Eq. (\ref{chi2_SP}). 

Next, we explore the impact of the heterostructure design on the behavior of the second-order susceptibility. It is well known that second-order nonlinear effects appear for asymmetric confining potentials \cite{boyd_nonlinear_2008, book_Sirtori_2000}.  In Fig. \ref{fig:chi3_var2sub} we consider a set of structures where the two thicknesses $L_1$ and $L_2$ are varied, while their sum is fixed, $L_1+L_2 = 11$ nm, Fig. \ref{fig:chi3_var2sub} (a-e). We start with the case of an almost symmetric potential, ($L_1=5.4$ nm, $L_1=5.6$ nm Fig.\ref{fig:chi3_var2sub} (a)), and the thickness $L_1$ is progressively reduced down to $L_1=1.8$ nm, in order to increase progressively the asymmetry of the double well potential. Note however that what matters is the shape of the electronic wavefunctions. Their square modulus have been shown as the red curves in Fig.\ref{fig:chi3_var2sub} (a-e). For vanishing  values of $L_1$ the first well becomes a small perturbation to the total hetero-structure potential, and wave-functions tend to the ones for a single quantum well of a thickness $L_2$. We then recover a centro-symmetric potential and the second order nonlinear susceptibility is expected to vanish in the limit $L_1 \rightarrow 0$. Therefore in the chosen set of double quantum wells we expect to observe an optimum for the nonlinear susceptibility $\chi^{(2)} (2\omega; \omega, \omega)$ as $L_1$ is varied. 

For all structures, the areal electronic density is taken to be $10^{12}$ $cm^{-2}$ such as the Fermi level lies below the second subband, and therefore for all structures there is a single occupied subband. In the second row of Fig. \ref{fig:chi3_var2sub}, panels (f-k) we plot the spectra for imaginary part of the first order susceptibility, $-\Im(\chi^{(1)}(\omega))$ as deduced from Eq. (\ref{chi1_3sub}). These plots reveal the energy positions of the collective resonances and their oscillator strengths. The visibility of the high energy resonance which originates from the plasmon of the $1 \rightarrow 3$ transition depends strongly on the symmetry of the wavefunctions. Indeed, it is proportional to the oscillator strength $f_{13}$ which vanishes for centro-symmetric systems, as the wavefunction from the first and the third subband have the same parity. Respectively, as seen from Fig.\ref{fig:chi3_var2sub} (g,h,j)  the second peak is important for the structures from  Fig.\ref{fig:chi3_var2sub} (b,c,d) that feature strong asymmetry. 

In Fig.\ref{fig:chi3_var2sub} (l-p) we plot the modulus of the second order susceptibility $\varepsilon \varepsilon_0 |\chi^{(2)} (2\omega; \omega, \omega)|$ as a function of the incident photon energy for the five structures depicted in Figs. \ref{fig:chi3_var2sub} (a-e) and an areal doping $10^{12}$ $cm^{-2}$. Like in Fig. \ref{fig:chi2_3sub}(c) the full curves correspond to the exact expression,  Eq. (\ref{Eq_chi2_3subfull}), whereas the dotted lines represent the single-particle result under RWA, Eq. (\ref{chi2_SP}). In the last expression only the two resonances at the energies $E_{12}$ and $E_{13}/2$ contribute to the susceptibility. As expected, the single-particle second order susceptibility increases for strong asymmetry, and the absolute maximum is increased for the structures from 
Fig.\ref{fig:chi3_var2sub} (d,e) which satisfy the doubly-resonant condition $E_{12} \approx E_{13}/2$. In particular the structure Fig.\ref{fig:chi3_var2sub}(d) optimizes the peak susceptibility as it combines strong asymmetry with the doubly-resonant condition \cite{book_Berger_1999, book_Sirtori_2000}. Instead, when dipole-dipole interactions are taken into account the optimum design becomes the structure Fig.\ref{fig:chi3_var2sub}(c) that was also examined in Fig. \ref{fig:chi2_3sub} for which the doubly-resonant condition is  now satisfied with the collective energies, $\tilde{E}_{12} \approx \tilde{E}_{13}/2$. 

These examples show that the optimization of the nonlinear susceptibilities $\chi^{(n)}(n\omega; \omega,...\omega)$ can begin with the analysis of the function $\mathrm{det} |M_{\alpha \beta} (\omega)|$ that provides the energies of the collective modes. Our approach from section \ref{General theory chi} indicates that the $n^{th}$ order susceptibility will be generally expressed from products the matrices $M^{-1}_{\alpha \beta} (\omega)$, $M^{-1}_{\alpha \beta} (2\omega)$,...$M^{-1}_{\alpha \beta} ((n-1)\omega)$, $M^{-1}_{\alpha \beta} (n\omega)$. Therefore the resonances of the function $\chi^{(n)}(n\omega; \omega,...\omega)$ are found at the collective energies and their $k^{th}$ multiples for all $k \leq n$. The multiple resonance condition can be used in order to select the optimal hetero-structure, and consequently the full calculation would provide the weights of the poles of the function $\chi^{(n)}(n\omega; \omega,...\omega)$ that quantify the peak values of the resonances.  However, as we will see in the next section the analysis of the function $\chi^{(n)}(n\omega; \omega,...\omega)$ alone is generally not enough to predict the optimum of the high-frequency generation for high electronic densities when polaritonic effects are present.

\begin{figure*}
    \centering
    \includegraphics[width = \textwidth]{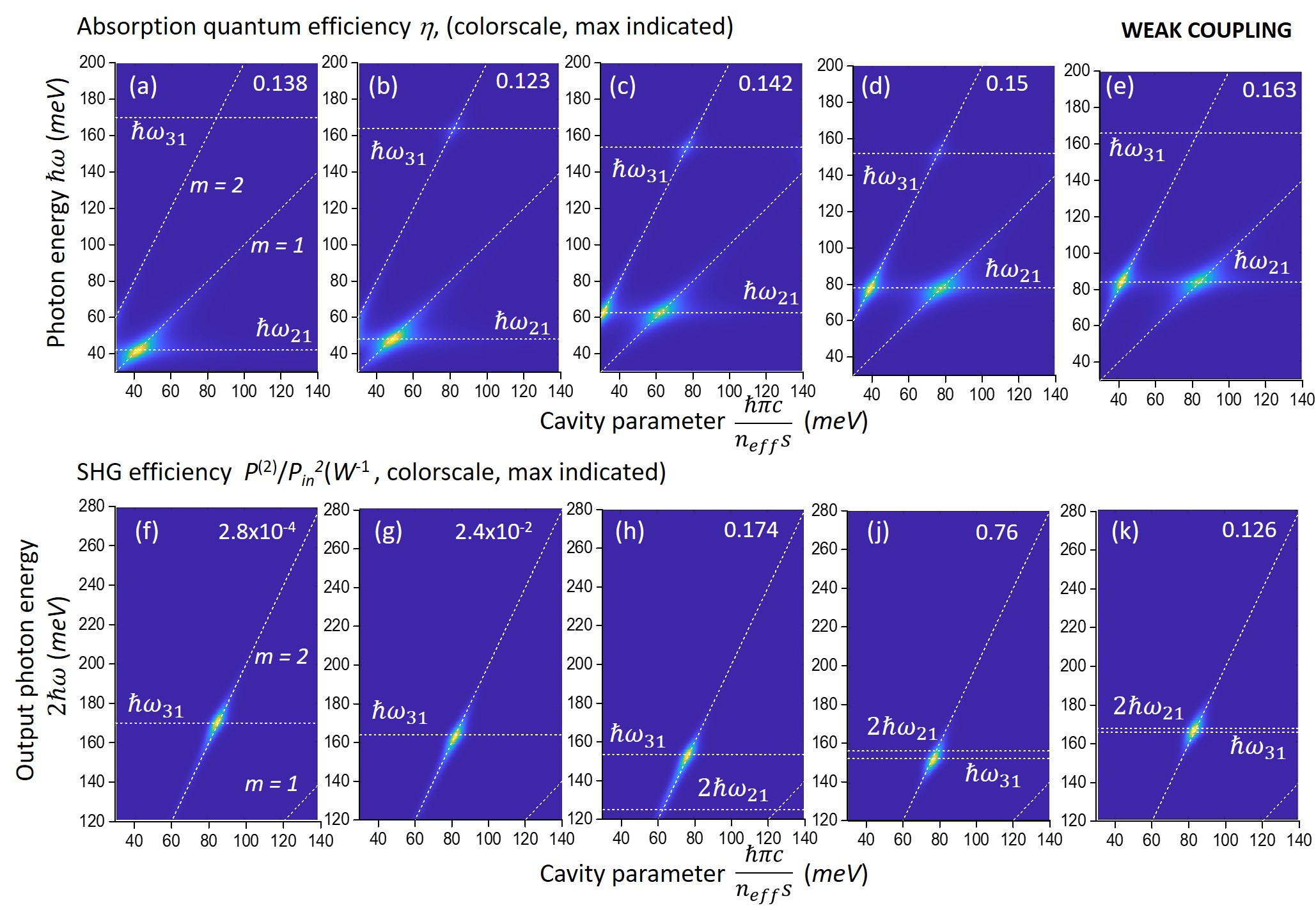}
    \caption{(a-e) Color plots for the absorption quantum efficiency $\eta(\omega)$ for each of the quantum well heterostructures shown in Fig.\ref{fig:chi3_var2sub} (a-e) and the case of low electronic concentration, $10^{10}$ $cm^{-2}$. The peak value of $\eta(\omega)$ is indicated in the top right corner of each panel. The two dotted diagonal lines correspond to the modes $m=1$ and $m=2$ as indicated in (a), this annotation is not repeated in (b-e). (f-k) Color plots of the SHG efficiency for each structure Fig.\ref{fig:chi3_var2sub} (a-e). The peak value of the SHG efficiency is indicated in the top right corner of each panel, in $W^{-1}$. The modes $m=1$ and $m=2$ are indicated in (f) and that annotation is not repeated in (g-k).}
    \label{fig:SHG_3sub1}
\end{figure*}

\begin{figure*}
    \centering
    \includegraphics[width = \textwidth]{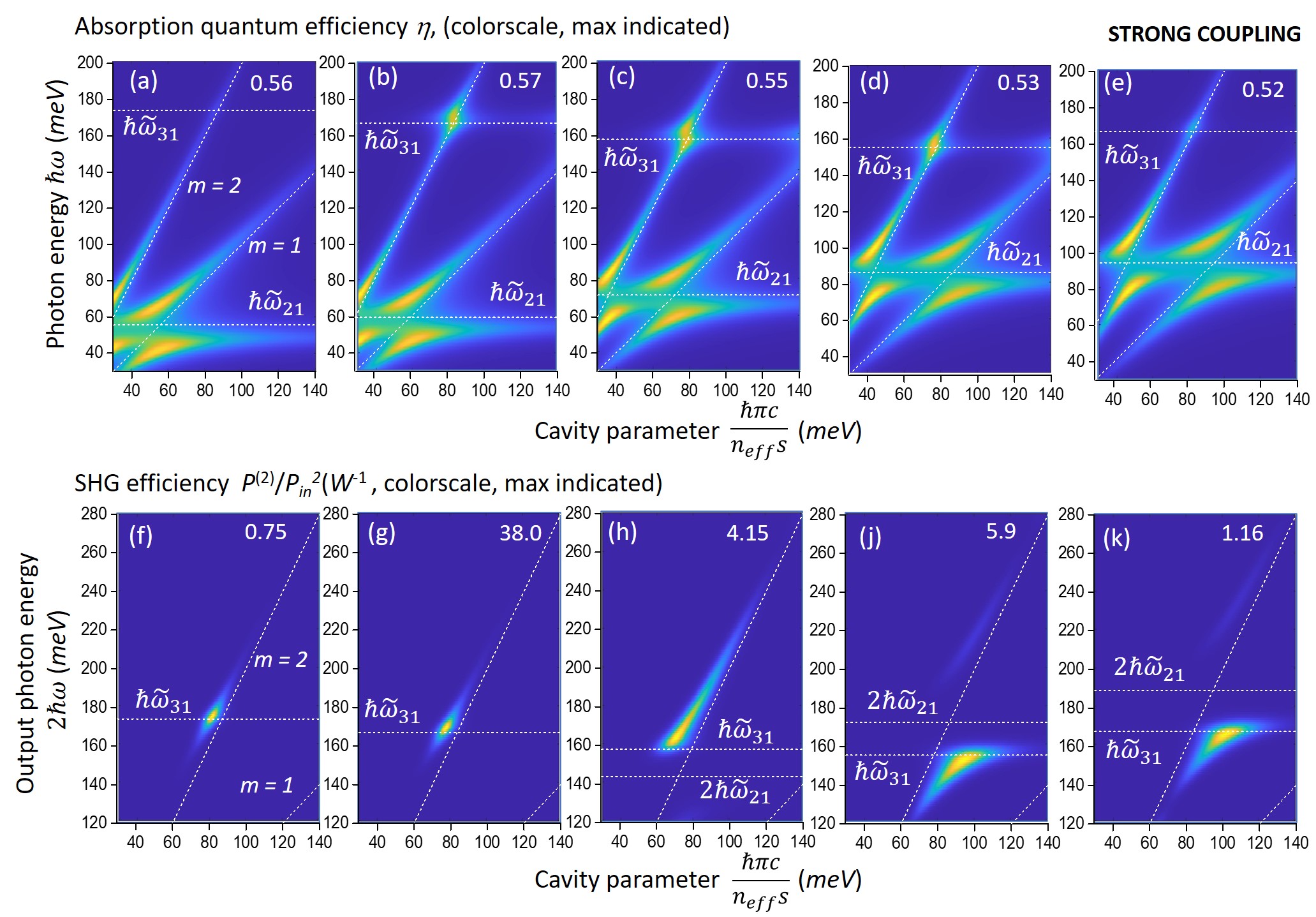}
    \caption{(a-e) Color plots for the absorption quantum efficiency $\eta(\omega)$ for each of the quantum well heterostructures shown in Fig.\ref{fig:chi3_var2sub} (a-e) for the case of high electronic concentration, $10^{12}$ $cm^{-2}$. The peak value of $\eta(\omega)$ is indicated in the top right corner of each panel. The two dotted diagonal lines correspond to the modes $m=1$ and $m=2$ as indicated in (a), this annotation is not repeated in (b-e). (f-k) Color plots of the SHG efficiency for each structure Fig.\ref{fig:chi3_var2sub} (a-e). The peak value of the SHG efficiency is indicated in the top right corner of each panel, in $W^{-1}$. The modes $m=1$ and $m=2$ are indicated in (f) and that annotation is not repeated in (g-k).}
    \label{fig:SHG_3sub2}
\end{figure*}

\subsubsection{Second harmonic generation}

We now consider SHG from the double quantum wells system described in last section.  We consider the $m=1$ ($\mathrm{TM}_{10}$) and $m=2$ $\mathrm{TM}_{20}$ resonances of the microcavity with $\omega_{c2} = 2\omega_{c1}$. Their eigenfunctions are:

\begin{equation}
    u_1 (x) = \sqrt{2} \cos (\pi x/s), \phantom{Q} u_2 (x) = \sqrt{2} \cos (2\pi x/s)
\end{equation}

 As the functions  $u_1(x)$ and $u_2(x)$ have different parities, all except two of the 6 coefficients $\kappa_{1[1^{k_1}2^{k_2}]}$ and:

\begin{equation}
\kappa_{1[1^{1}2^{1}]}  = \kappa_{2[1^{1}2^{0}]} =  \frac{1}{\sqrt{2}} 
\end{equation}

From Eq.(\ref{P_n_out}), $n=2$, the expression of the SHG power efficiency  becomes:

\begin{eqnarray}
\frac{P^{(2)} (2\omega)}{P_{in}^2} =  \frac{\varepsilon_0 \varepsilon}{V_{cav}} \frac{16 \omega^2|\chi^{(2)} (2\omega; \omega, \omega)|^2}{|1+i2\omega Y_p(2\omega)|^2 |1+i\omega Y_p(\omega)|^{4}} 
\nonumber \\ 
\times \Big| 2\Theta_1 (2 \omega) \Theta_1 (\omega) \Theta_2 (\omega) 
+ \Theta_2 (2 \omega) \Theta_1^2 (\omega) \Big|^2 \label{power_SGH_3sub}
\end{eqnarray}

As mentioned before in section \ref{sec HF generation} this expression does not include explicitly the angular dependence of the radiated SHG power. It is known that the modes $\mathrm{TM}_{10}$ and $\mathrm{TM}_{20}$ have different radiation patterns, and in particular the mode $\mathrm{TM}_{20}$ can not radiate along the normal of the microcavity array \cite{todorov_optical_2010}. We thus expect that the optimal direction for the pump into the $\mathrm{TM}_{10}$ mode is along the normal of the cavity array, while the SHG power will be distributed along the radiation pattern of the $\mathrm{TM}_{20}$ mode, with a maximum close to the grazing direction of the array. 

\begin{figure}
    \centering
    \includegraphics[width = \columnwidth]{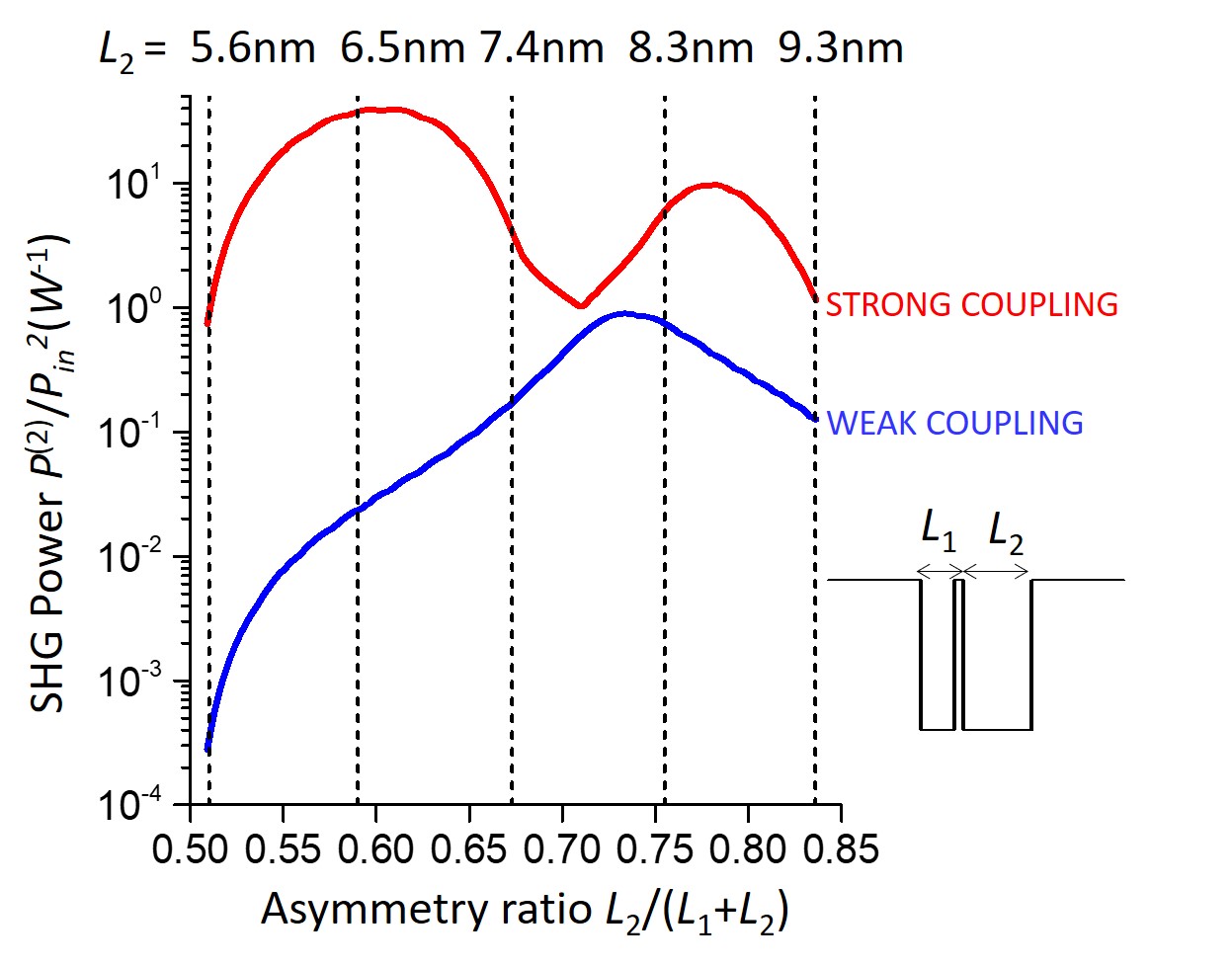}
    \caption{Absolute maximum of the SHG efficiency for a series of double quantum well designs where $L_1$ and $L_2$ are varied with $L_1+L_2=11$ nm and the barrier thickness is kept constant at $1$ $nm$. The dashed lines correspond to the five designs considered in Fig. \ref{fig:chi3_var2sub} (a-e). The weak coupling regime corresponds to an electronic density of $10^{10}$ $cm^{-2}$ and the strong coupling regime to a density $10^{12}$ $cm^{-2}$.}
    \label{fig:SHGeff}
\end{figure}

We have examined the SHG from the quantum well systems described in the previous section (Fig. \ref{fig:chi3_var2sub} (a-e)) both in the weak (Fig. \ref{fig:SHG_3sub1}) and strong coupling regime (Fig. \ref{fig:SHG_3sub2}).
Like in the previous part, we have considered a microcavity with a thickness $L_{cav}$ optimally filled with a repetition identical pairs of double quantum wells.  In Fig. \ref{fig:SHG_3sub1} we consider low electronic density, $10^{10}$ $cm^{-2}$. The upper panels (a-e) show the color plots of the absorption quantum efficiency $\eta (\omega)$ in the presence of $m=1$ and $m=2$ resonances of the cavity. The single-particle transitions energies $\hbar \omega_{21}$ and $\hbar \omega_{31}$ are also indicated. As for the case of the two subband system (Fig. \ref{fig:THG_2sub}(a)) the absorption is maximal when the cavity resonances are matched with the frequencies of the electronic transitions. In Fig. \ref{fig:SHG_3sub1}) (f-g) we show the corresponding color plots of  the SHG efficiency (Eq. (\ref{power_SGH_3sub})), with the vertical axis being the output photon energy $2\hbar \omega$. As a reference, we also provide the doubled photon energies of the two resonances of the nonlinear susceptibility: $2\times \hbar \omega_{21}$ and $2\times \hbar \omega_{31}/2$. These plots show that  SHG is maximal for cavities where the fundamental cavity mode $\omega_{c1}$ matches the resonance at $\omega_{31}/2$ of the susceptibility function $\chi^{(2)} (2\omega; \omega, \omega)$. Because of our choice $\omega_{c2} =  2\omega_{c1}$ this condition is also translated as $\omega_{c2} = 2\times  \omega_{31}/2$. The last condition means that SHG signal from frequency doubled photons $2 \times \omega_{31}/2$ is enhanced by the cavity effect when their frequency matches the second order resonance $m=2$. The SHG is optimized for the case Fig. \ref{fig:SHG_3sub1})(j) which corresponds to the structure Fig. \ref{fig:chi3_var2sub} (d). This structure is doubly resonant, $2\omega_{21} \approx \omega_{31}$  and it is sufficiently asymmetric such as the oscillator strength $f_{13}$ is important. In the single-particle picture the trend of the SHG can be entirely predicted form the properties of the susceptibility function, $\chi^{(2)} (2\omega; \omega, \omega)$, and the requirement that the frequency doubled photons match another (typically second order) microcavity resonance. 

The case of strong light-matter coupling is shown in Fig.\ref{fig:SHG_3sub2} where the electronic density is now $10^{12}$ $cm^{-2}$. The panels (a-e) from Fig.\ref{fig:SHG_3sub2} show that now the collective electronic resonance $\hbar \tilde{\omega}_{21}$ is in a strong coupling regime with the two modes $m=1$ and $m=2$. Instead, the resonance $\hbar \tilde{\omega}_{31}$ is only in a weak coupling regime, owe to its smaller oscillator strength; and onset of strong coupling between $\hbar \tilde{\omega}_{31}$ and the $m=2$ resonance can be observed in Fig.\ref{fig:SHG_3sub2}(c). 

The patterns that correspond to the SHG in the strong coupling regime are shown in Fig.\ref{fig:SHG_3sub2} (f-k). The photon energies  $\hbar \tilde{\omega}_{31}$ and $2\hbar \tilde{\omega}_{21}$ that correspond to the doubled resonances,$\hbar \tilde{\omega}_{31}/2$ and $\hbar \tilde{\omega}_{21}$ of the $\chi^{(2)} (2\omega; \omega, \omega)$ function are also indicated. From these plots we observe that the SHG power is maximally collected for frequency doubled photons with an energy close $\hbar \tilde{\omega}_{31}$. Because of the strong coupling phenomena, now the two photonic resonances that optimize the low frequency absorption are either the upper or lower polariton issued from the coupling between the $m=1$ mode and the $\hbar \tilde{\omega}_{21}$ resonance. Thus now the maximum of the susceptibility function $\chi^{(2)} (2\omega; \omega, \omega)$ must be matched with either of the two polariton states in order to obtain the highest SHG signal. For instance, in the panels Fig.\ref{fig:SHG_3sub2} (f,g,h) the resonance $\hbar \tilde{\omega}_{31}/2$ is matched with the upper polariton state, whereas in Fig.\ref{fig:SHG_3sub2} (j,k) the resonance $\hbar \tilde{\omega}_{31}/2$ is matched with the lower polariton state. Quite interestingly, in Fig.\ref{fig:SHG_3sub2} (j,k) we also observe signatures of the upper polariton state in the SHG power map; the latter arises from the frequency shoulder of the $\hbar \tilde{\omega}_{21}$ resonance that is at a higher energy than $\hbar \tilde{\omega}_{31}/2$ for the structures Fig. \ref{fig:chi3_var2sub} (d, e). 

It is also interesting to comment the absolute value of the SHG signal indicated in the panels Fig.\ref{fig:SHG_3sub2} (f-k). In particular, we observe that the structure from Fig. \ref{fig:chi3_var2sub} (c) that provided absolute maximum of the susceptibility function $\chi^{(2)} (2\omega; \omega, \omega)$ actually provides relatively small SHG efficiency with respect to structures Fig. \ref{fig:chi3_var2sub} (b,d). This is because for this structure the maximum of the function $\chi^{(2)} (2\omega; \omega, \omega)$ coincides with the frequency range in-between the two polariton states where there is low photonic density of states. Instead, the structure from Fig. \ref{fig:chi3_var2sub} (b) provides a maximum efficiency of about $P^{(2)} (2\omega)/P_{in}^2 \approx 40 W^{-1}$. This structure satisfies both conditions:

\begin{equation}
\omega_{c2} \approx  \tilde{\omega}_{31}, \phantom{Q} \omega_\pm \approx \tilde{\omega}_{31}/2, \label{CondBestSHG}
\end{equation}

with $\omega_\pm$ being the frequency of either the upper or lower polariton state. (Specifically for the structure Fig. \ref{fig:chi3_var2sub} (b) we have $\omega_+ \approx \tilde{\omega}_{31}/2$). Eq. (\ref{CondBestSHG}) appears as a more relevant design tool for optimal SHG in the strong and ultra-strong light-matter coupling regime. 

It is interesting to estimate the maximum SHG efficiency that can be obtained from the hetero-structure family described in Fig.\ref{fig:chi3_var2sub}. To that end we have performed simulations where we have continuously varied the thicknesses  $L_1$ and $L_2$ of the two quantum well while keeping constant the sum $L_1+L_2$. For each structure we have computed the colorplots of the SHG efficiency as those shown in Fig.\ref{fig:SHG_3sub2} (f-k) and we have numerically extracted its absolute maximum, both the case of weak and strong coupling (low and high  areal electronic density). In Fig. \ref{fig:SHGeff} we plot the absolute maximum of the SHG efficiency as a function of the parameter $L_2/(L_1+ L_2)$ that quantifies the asymmetry of the potential. As seen in Fig.\ref{fig:SHG_3sub2} (f-k) the output photon energy then varies in the interval between $150$ meV and $170$ meV and we can ignore its relative variations. In Fig. \ref{fig:SHGeff} there is a clear optimization of the SHG efficiency for the weak coupling regime, with a quantum design close to the doubly resonant case from Fig.\ref{fig:SHG_3sub2} (d). For the optimal structure, the peak value of the efficiency is $P^{(2)} (2\omega)/P_{in}^2 = 0.9 W^{-1}$, which yields an output SHG power on the order of $1\mu W$ for typical input powers $P_{in} = 10$ $mW$. (As a comparison, in Ref. \cite{yu_broadband_2024} a maximum efficiency $P^{(2)} (2\omega)/P_{in}^2 \approx 0.1 W^{-1}$ was reported). As expected, the SHG efficiency decreases significantly as the design becomes more symmetric, $L_2/(L_1+L_2) \rightarrow 0.5$. 

Instead, in the strong coupling regime there are two structures that optimize  the SHG process in a micro-cavity: a strongly asymmetric structure, $L_2/(L_1+L_2) = 0.8$ that leads to a secondary maximum of the plot with $P^{(2)} (2\omega)/P_{in}^2 = 10$ $W^{-1}$ and a structure with slight asymmetry, $L_2/(L_1+L_2) = 0.6$ for which we obtain a maximum efficiency $P^{(2)} (2\omega)/P_{in}^2 = 40 $ $W^{-1}$, which corresponds to a SHG power on the order of $40\mu W$. Based on the previous discussion the two maxima can be interpreted as respectively matching the  $\hbar \tilde{\omega}_{31}/2$ resonance   of the $\chi^{(2)} (2\omega; \omega, \omega)$ function respectively with the upper, $\omega_+$, and lower, $\omega_-$, polariton states of the system.  As seen from  Fig. \ref{fig:SHGeff} there is actually a broad range of weakly asymmetric designs that leads to an enhanced SHG. Instead, the optimal design in the weak coupling regime would display only a slight increase of the SHG efficiency, from $0.9$ $W^{-1}$ to $2$ $W^{-1}$ despite the two order of magnitude increase of the electronic density. These results show that the quantum design of microcavity-coupled SHG emitter must be completely reconsidered at high electronic densities where collective electronic effects and polaritonic effects become prominent: in particular doubly resonant designs studied so far in the literature and based on single-particle approximation are much less efficient than designs based in Eq. (\ref{CondBestSHG}).  

We recall that our calculations are valid for incident powers sufficiently low such as saturation effects can be neglected. Therefore we have considered typical maximal values $P_{in} \approx 10$ $mW$ based on the experimental results and our estimations from section  \ref{secBistability}. we expect that at very high powers the SHG efficiency drops significantly owe to saturation effects, as it was experimentally reported in Ref. \cite{yu_broadband_2024}. Furthermore, experimentally high incident powers lead to heating of the electron gas, which ultimately reduces the population difference between the electronic levels and leads to a lower absorption. Thus the relatively high SHG efficiencies estimated here $>40 W^{-1}$ are still compatible with energy conservation which requires that $P^{(2)} < P_{in}$. Actually the condition $P^{(2)} \approx P_{in}$ together with the estimation of the SHG efficiency provide a rough estimate on the saturation power of the system.  In particular for  $P^{(2)} =40 P_{in}^2$ we expect an onset of saturation effects a power $P_{in} = 25$ mW which sets a limit for the validity of any estimations without saturation effects.

\subsubsection{Multiphoton absorption}

\begin{figure}
    \centering
    \includegraphics[width = 5cm]{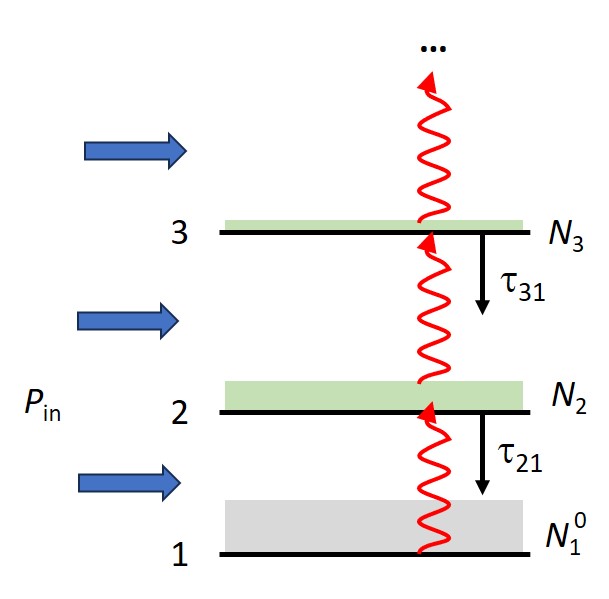}
    \caption{Illustration of the multi-photon absorption process, which induces populations $N_2, N_3,..$ on initially empty high energy electronic subbands. $N_1^0$ is the initial population of the lowest subband.}
    \label{fig:Multiphoton}
\end{figure}

\begin{figure*}
    \centering
    \includegraphics[width = \textwidth]{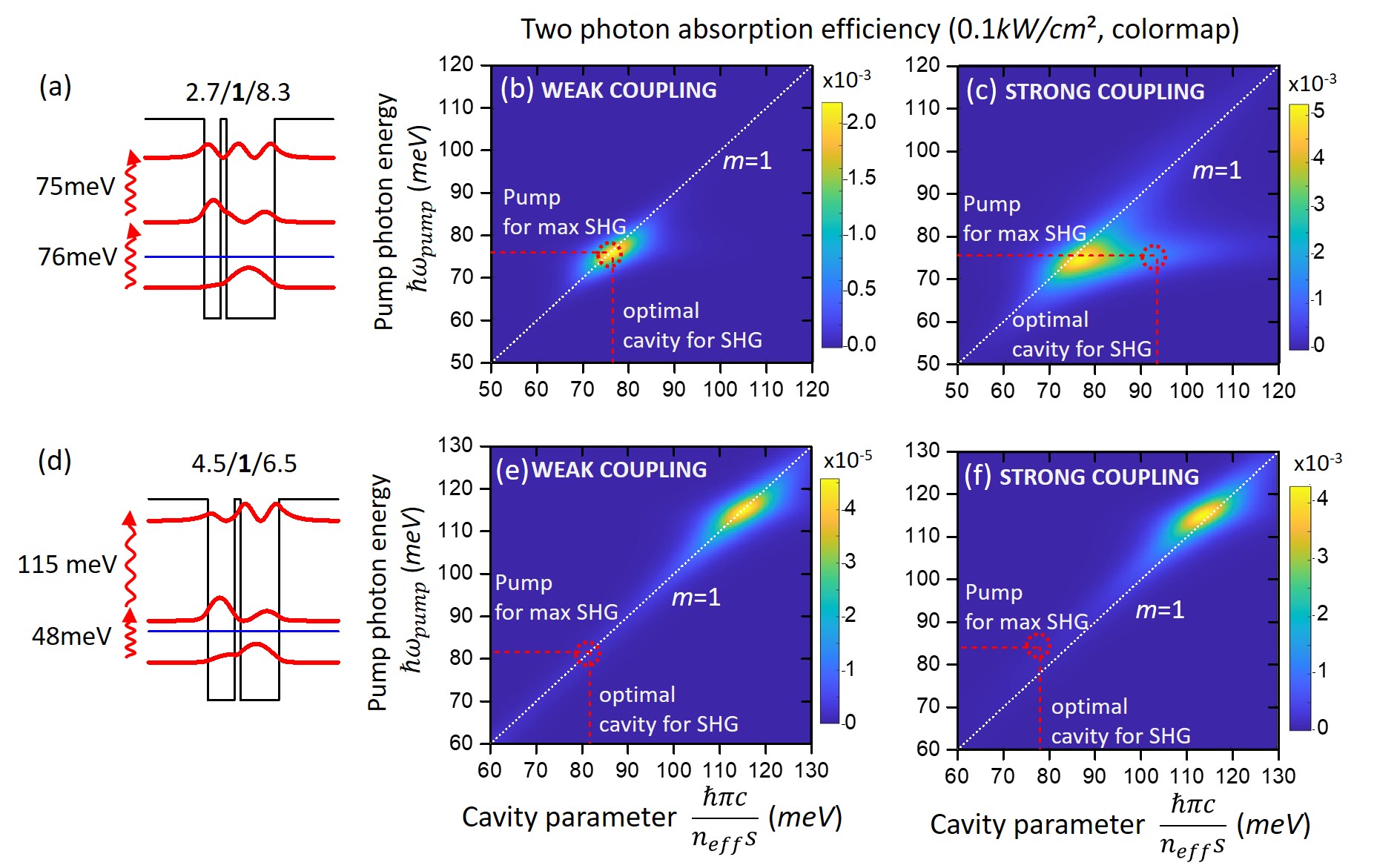}
    \caption{Two-photon absorption efficiency for the structures from Fig. \ref{fig:chi3_var2sub}(b,d) which have been reproduced in panels (a,b). (b,e) Two-photon absorption efficiency in the weak coupling regime, for an areal electronic concentration $\Delta n_0= 10^{10}$ $cm^{-2}$. (e,f) Two-photon absorption efficiency for the case where the $1\rightarrow 2$ transition is in strong coupling regime with the $m=1$ cavity resonance, for an electronic concentration  $\Delta n_0=10^{12}$ $cm^{-2}$.}
    \label{fig:Multiphoton_full}
\end{figure*}

In multi-subband  systems, we must also consider the phenomenon  of multi-photon absorption, which has a similar dependence on the incident power as the processes of high harmonic generation. In particular, this phenomenon is strong in systems with equally spaced subbands, such as those used for high harmonic generation. More generally, this phenomenon also leads to a correction of the approximate energy balance expressed in Eq.(\ref{approxEnBalance}).

This phenomenon is described in Fig. \ref{fig:Multiphoton}. For simplicity we consider a system  where initially only the first subband is populated. The incident pump is absorbed by the transition $1 \rightarrow 2$ and as a result a population $N_2$ is induced on the second subband, $N_2 = \eta_{1\rightarrow 2} (\omega) P_{in}\tau_{21}/(\hbar \omega_{21})$. Then the  probability of a sequential absorption $1\rightarrow 2 \rightarrow 3$ is merely provided by the value $\eta_{2\rightarrow 3} (\omega)$ of the absorption quantum efficiency between levels $2$ and $3$ computed for the light-induced population $N_2$.
Now, the light-induced population of the third subband being $N_3 =\eta_{2\rightarrow 3} (\omega) P_{in}$ the probability for a three photon absorption will be provided by the value of the absorption efficiency $\eta_{3\rightarrow 4}$ computed for the value of  $N_3$, etc. Clearly,  we can iterate this procedure to any number of subbands, thus obtaining the probability of the $n$-phonon sequential absorption which will be proportional to $P_{in}^{n-1}$. Since the absorption efficiencies $\eta_{n,n+1} (\omega)$ explicitly depend on the incident photon frequency $\omega$ and all other parameters of the system, there is no requirement for the subbands to be equally spaced, and our definition is applicable to any heterostructure potential. In the case where several subbands are populated, we can use our general theory described in section \ref{par: full set} to link the population of the first empty subband to the absoption quantum efficiency $\eta (\omega)$ defined in  Eq.(\ref{eta_lin}), and then the procedure described above can be applied. 

Let us consider the case with a single cavity resonance. We can then use Eq. (\ref{etasinglemode}) to determine the absorption coefficient $\eta_{1 \rightarrow 2} (\omega)$ and Eq. (\ref{N_2toPin}) which provides the light-induced population on the second subband $N_{2} =\eta_{1\rightarrow 2} (\omega) P_{in} \tau_{21}/\hbar \omega_{21}$. Consequently, the expression for the coefficient $\eta_{2\rightarrow 3} (\omega)$ becomes: 
    
\begin{equation}\label{eta23}
 \eta_{2\rightarrow 3} (\omega) \approx 4f_w^{23}f_{23}\frac{\gamma_{23}\Gamma_{r,m} \omega^2\omega_{cm}^2 \omega_{P23}^2}{|\Delta_{23}(\omega)|^2|\Delta_{cm}(\omega)|^2}   
\end{equation}

 As the population $N_2$ is considered low, here we assumed that the transition $2\rightarrow 3$ is in a weak coupling regime with the cavity mode. The denominator $\Pi_{23} (\omega, x)$ which describes the light-matter coupling with the cavity mode and the transition $2\rightarrow 3$ can then be factorized $\Pi_{23} (\omega, x) \approx \Delta_{23}(\omega)\Delta_{cm}(\omega)$. Since the plasma frequency of the $2\rightarrow 3$ transition is expressed as $\omega_{P23}^2 = e^2/(m^* \varepsilon \varepsilon_0 L_{eff, 23}S)N_2$ we arrive at the following expression of the efficiency of two-photon absorption, at the lowest order in the incident power:

 \begin{eqnarray}
  \eta_{2\rightarrow 3} (\omega) \approx 4f_w^{23}f_{23}\frac{\gamma_{23}\Gamma_{r,m} \omega^2\omega_{cm}^2 }{|\Delta_{23}(\omega)|^2|\Delta_{cm}(\omega)|^2} \nonumber \\
\times \eta_{1\rightarrow 2} (\omega) \frac{e^2}{m^* \varepsilon \varepsilon_0 L_{eff, 23}}\frac{P_{in} \tau_{21}}{S\hbar \omega_{21}} \label{2photAbs}
 \end{eqnarray}

 In the case of strong pump, we must infer $N_2$ using the general results from section \ref{parN20} (for instance, Eq. (\ref{EqDeltan12})) and revise accordingly the expression for the denominator in Eq. (\ref{eta23}). For a very large pump the transition $1\rightarrow 2$ is saturated, and the maximal value of the second order electronic population is $N_2 = N_{10}/2$. Typically, in this regime the second subband will be further depleted by sequential  absorption to higher energy subbands. Furthermore, the stimulated emission will start contributing to the shortening relaxation time $\tau_{21}$ \cite{book_QCL_faist, Micheletti_2021} leading to further reduction of $N_2$.

In doubly-resonant structures, such as the one depicted in Fig. \ref{fig:chi3_var2sub}(d) the two-photon absorption coexists with the process of SHG and can contribute to the depletion of the incident pump. In Fig. \ref{fig:Multiphoton_full} we have examined the two photon absorption efficiency in microcavity coupled double-quantum well structures from  Fig. \ref{fig:chi3_var2sub}(d) as well as the one from Fig. \ref{fig:chi3_var2sub}(b) which optimizes SHG. These structures are reproduced in Fig. \ref{fig:chi3_var2sub} (a,d). The outcome from Eq. (\ref{2photAbs}) has been plotted both for low, $10^{10}$ $cm^{-2}$, and high, $10^{12}$ $cm^{-2}$ electronic density, which leads to weak (Fig. \ref{fig:chi3_var2sub} (b,e)) and strong (Fig. \ref{fig:chi3_var2sub} (c,f)) coupling regime between the $1 \rightarrow 2$ transition and the $m=1$ cavity mode. Calculations are performed with an incident beam intensity $0.1$ $kW/cm^2$ so that saturation effects can be neglected. In the four panels Fig. \ref{fig:chi3_var2sub} (b,c,e,f) we have indicated the value of the cavity parameter $\hbar \pi c/n_{eff}s$ and the pump photon frequency $\hbar \omega_{pump}$ that optimizes the SHG process, as extracted from Fig. \ref{fig:SHG_3sub1} and \ref{fig:SHG_3sub2}.

In the case of a doubly resonant structure, and in the weak coupling regime Fig. \ref{fig:chi3_var2sub}(b) the two photon absorption is maximal for the same cavity and pump frequency as the one that optimizes the SHG process (Fig. \ref{fig:SHG_3sub1} (j)). This coincidence is lifted for the case of the strong coupling regime, Fig. \ref{fig:chi3_var2sub}(c), where the cavity optimizing the SHG process has to be chosen to match one of the polariton resonances (Fig. \ref{fig:SHG_3sub2} (j)). The two-photon absorption probability is now maximal close to the bare cavity frequency, which is coherent with the fact that the $2\rightarrow 3$ transition is in a weak coupling with the cavity. On the other hand, the polariton effect arising from the strong coupling between the  $1\rightarrow 2$ transition and the cavity is also visible, as there is a weak two-photon absorption probability along the lower polariton state, as seen from Fig. \ref{fig:chi3_var2sub}(c).  Instead, for the hetero-structure that optimizes the SHG in the strong coupling regime, the two-photon absorption takes place in a completely different frequency range and cavity parameters both for the weak and strong coupling regime (Fig. \ref{fig:chi3_var2sub}(e,f)). This is yet another example where the phenomenon of strong coupling and collective electronic effects can thus be used to engineer completely the various nonlinear process in the cavity-coupled quantum wells. 

\section{Conclusion}\label{Conclusion}

In summary, we provided a fully microscopic model for light-matter interactions and nonlinear phenomena in microcavity-coupled highly doped quantum well heterostructures, that takes fully into account the dipole-dipole interactions between electrons within the framework of the PZW formulation of quantum electrodynamics \cite{cohen-tannoudji_photons_1986}.
In the present work, we considered the case of a single frequency pump which leads to nonlinear susceptibilities of the form $\chi^{(n)} (n\omega, \omega, ....\omega)$. The latter are important for the description of THG and SHG in quantum wells that has been studied experimentally \cite{Capasso_1994, kim_giant_2020, sarma_strong_2021, sarma_all-dielectric_2022,Sarma_2022, chung_electrical_2023, yu_broadband_2024, Park2024}. Here, we examined these processes in the regime of high electronic densities, which in principle must increase the efficiency of the nonlinear generation process. We showed how the effect of high density can be leveraged in these structures by taking into account the collective nature of the electronic resonances.  Our model allows a full design of both the polaritonic resonances of the system, which set the optimal frequency of the input pump and the output radiation, as well as the maxima for the  high order susceptibility functions modified by collective electronic phenomena. In particular, our theory provides a unified approach to  describe both microcavity effects and local field effects (ENZ, \cite{Fomra_reviewENZ_2024}) that have been considered to enhance nonlinear response in quantum hetero-structures \cite{Fomra_reviewENZ_2024}. As an example, we considered possible designs of metamaterial-coupled quantum heterostructure for SHG and we showed that the doubly-resonant configuration must be revised in the ultra-strong coupling regime, where novel designs with much better performance can be provided.

At the present stage, our theory sets a framework for a systematic investigation of any quantum heterostructure with an arbitrary number of electronic levels in interaction with any number of photonic modes in the ultra-strong light-matter coupling regime. The core of our microscopic model is contained in the equation set described in section \ref{secHamiltonian_evolution}, that provides the Hamiltonian evolution of the microscopic variables. These equations must be coupled with the rate equations that describe the relaxation of population and coherences, which depend on the specifics configuration of electronic levels as well as the position of the Fermi level. Thus analytical or numerical solutions can be provided for any situation, beyond the examples considered in part \ref{partExamples}. In particular, the case of several occupied subbands \cite{delteil_charge-induced_2012} is very appealing and can be handled by our theory.

So far, our formalism was developed for the case of a single frequency pump. An immediate generalization can be made for the case where the driving pump is composed of several frequencies $\omega_1, \omega_2, ...$ in Eq.  (\ref{dAdt_av}). This will allow evaluating nonlinear susceptibilities such as $\chi^{(2)} (\omega_1 \pm \omega_2, \omega_1, \omega_2)$ that play important role for frequency conversion in intersubband devices, and namely for THz generation by frequency difference \cite{Belkin2007}. 

Concerning electron-electron interactions, we considered the dynamic dipole-dipole interactions where the polarization field (Eq. \ref{EqP}), that was previous used to describe the impact of collective effects on the lienar optical response \cite{todorov_intersubband_2012,delteil_charge-induced_2012}. In section \ref{secPolField} we discussed a complete expression for the polarization field, comprising both dynamic and static terms that are expressed from population operators. Developing the self-interacting Hamiltonian (Eq. (\ref{GeneralP2})) leads to  term that couple population and dipoles, for which other types of  nonlinear phenomena appear. In particular, we can show that these terms lead to the phenomenon of optical rectification \cite{Rosencher_1989, K_Unterrainer_1996} and a finite second order susceptibility $\chi^{(2)} (\omega_1 \pm \omega_2, \omega_1, \omega_2)$ in asymmetric quantum wells. These phenomena have never been fully explored for microcavity-coupled systems in the presence of strong collective effects and will be discussed elsewhere. Furthermore, it is an open question whether the dipole-density coupling terms could affect the electronic transport in devices that operate in USC \cite{pisani_electronic_2023}. 

While our  equation set described in section \ref{secHamiltonian_evolution} is fully quantum, here we developed a semi-classical theory which is based on the factorization scheme expressed in Eq. (\ref{factorization}). Going beyond this assumption would consist in evolving the system from a vacuum state, and studying the dynamics of the correlation functions $\expval{(a_m^\dagger - a_m) \hat{c}^\dagger_{\lambda i}\hat{c}_{\mu i} }$ which will allow describing spontaneous emission and parametric fluorescence in these systems \cite{book_OptResonance, grynberg2010introduction}. Our approach also allows expressing the coupling between various cavity modes through the matter nonlinearities. Then, yet another possibility is to eliminate the electronic degrees of freedom in favor of the photonic ones and introduce an effective photonic Hamiltonian \cite{IrvinePhysRevLett.96.057405, Chatterjee_PhysRevResearch.6.023288} in the context of USC.  An interesting open question is to which degree the USC can alter the quantum optical statistics of squeezed photon states. For instance, one can search for the signatures of  ground-state quantum correlations initially predicted for USC \cite{ciuti_quantum_2005} in the quantum statistics of the upconverted or down-converted microcavity photons. 

In summary, our work  lays the ground for new fundamental studies of the quantum optical phenomena in the THz and MIR frequency ranges  based on nonlinear optical conversion in the presence of micro-cavity and collective electronic effects. These studies provide a route for designing novel experiments based on the quantum nature of infrared photons and ultimately would contribute for the development of quantum technologies for these frequency ranges. Beyond this context, we are convinced that the PZW approach illustrated here for the case of electronic excitations in quantum well is a powerful tool to describe linear and nonlinear optical processes in condensed matter systems, and can be particularly relevant for two dimensional materials where collective interactions play important role \cite{Basov_science_Rev2D, Tame2013}.  

\begin{acknowledgments}
We acknowledge funding from the ERC-COG-863487 UNIQUE as well as  useful discussions with Francesco Pisani.
\end{acknowledgments}

\appendix
\section{Derivation of the polarization in the position lattice}

Using the definition (\ref{defci}) it is straightforward to prove the following commutation relations:

\begin{eqnarray}
\{ c_{\lambda i}^\dagger, c_{\mu j} \}  = \delta_{i,j}\delta_{\lambda,\mu}
\\
\{ c_{\lambda i}^\dagger, c_{\mu j}^\dagger \} = \{ c_{\lambda i}, c_{\mu j} \} = 0
\end{eqnarray}

For the first relation we used the fact that:

\begin{equation}
    \sum_\mathbf{k} e^{i\mathbf{k}(\mathbf{r}_i -\mathbf{r}_j )} = N_e \delta_{i,j}
\end{equation}

Once again, because of  Eq. (\ref{defci}) the following relation is satisfied:

\begin{eqnarray}\label{eqAppcc}
    \hat{c}_{\lambda i}^\dagger \hat{c}_{\mu i} = \frac{1}{N_e} \sum_{\mathbf{k}, \mathbf{q}} \hat{c}_{\lambda \mathbf{k} + \mathbf{q}}^\dagger \hat{c}_{\mu \mathbf{k}} e^{-i \mathbf{q} \mathbf{r}_i}
\end{eqnarray}

We can use Riemann sum rule to write:

\begin{eqnarray}
    \sum_i \delta\left(\mathbf{r} - \mathbf{r}_i\right)e^{i \mathbf{q} \mathbf{r}_i} &=& \frac{N_e}{S} \iint \delta\left(\mathbf{r} - \mathbf{r}'\right) e^{i \mathbf{q} \mathbf{r}'} d\mathbf{r}' \nonumber
    \\
    &=& \frac{N_e}{S} e^{i \mathbf{q} \mathbf{r}} 
\end{eqnarray}

Using this expression of the factor  $e^{i\mathbf{qr}}$ from the above equation together with Eq. (\ref{eqAppcc}) we obtain:

\begin{eqnarray}
    \frac{1}{S} \sum_{\mathbf{k}, \mathbf{q}} \hat{c}_{\lambda \mathbf{k} + \mathbf{q}}^\dagger \hat{c}_{\mu \mathbf{k}} e^{-i \mathbf{q} \mathbf{r}} = \sum_i \delta\left(\mathbf{r} - \mathbf{r}_i\right) \hat{c}_{\lambda i}^\dagger \hat{c}_{\mu i}
\end{eqnarray} 

This relation allows to transform Eq. (\ref{EqP}) into Eqs. (\ref{DP1}, \ref{DP2}).

\section{Amplitudes of the cavity modes and reflectivity} \label{AppendixB}

Eqs. (\ref{eqdDdt}), (\ref{eqdAdt}) and (\ref{eqIout}) are written as follows for the case of an incident monochromatic pump:

\begin{eqnarray}
     i\omega \tilde{D}_m &=& -\omega_{c m} \tilde{A}_m  \label{eqdDdt B} \\
    i\omega \tilde{A}_m &=& \omega_{c m} \tilde{D}_m - \Gamma_{nr , m} \tilde{A}_m - \sqrt{\Gamma_{r, m}} T  \nonumber\\
     &+&  2\sqrt{\Gamma_{r, m}} \tilde{I}_{in} 
     \label{eqdAdt B} \\
     \tilde{I}_{out} &=& -\tilde{I}_{in} + T \label{IoutB}
     \\
     T &=& \sum_m \sqrt{\Gamma_{r, m}} \tilde{A}_m \label{defT}
\end{eqnarray}

Eliminating the amplitudes $\tilde{D}_m$ and expressing $\tilde{A}_m$ as a function of $I_{in}$ and the quantity $T$ defined in Eq. (\ref{defT}) we obtain:

\begin{align}
    \Delta_{cnr, m} (\omega) \tilde{A}_m + i\omega \sqrt{\Gamma_{r, m}} T 
    = 2i\omega \sqrt{\Gamma_{r, m}} \tilde{I}_{in} \label{dAm B}
    \\
    \Delta_{cnr,m}(\omega) = \omega_{cm}^2 - \omega^2 + i\omega \Gamma_{nr , m}
\end{align}

By expressing $\tilde{A}_m$ as a function of $T$ and $I_{in}$ we obtain an expression for $T$ as a function of the incident field:

\begin{equation}
    T = \frac{2i\omega Y(\omega)}{1 + i\omega Y(\omega)} I_{in} \label{solutionT}
\end{equation}

The quantity $Y(\omega)$ is defined in Eq. (\ref{defY}):

\begin{align}
Y(\omega) = \sum_m \frac{\Gamma_{\text{r}, m}}{\Delta_{\text{cnr}, m}(\omega)}.
\end{align}

This result is then substituted back into equation (\ref{dAm B}) to yield the  expression for the mode amplitudes:

\begin{equation}
     \tilde{A}_m = 2\tilde{I}_{in} \frac{i\omega \sqrt{\Gamma_{r , m}}}{\Delta_{cnr,m}(\omega)} \frac{1}{1+i\omega Y(\omega)}
\end{equation}

The reflectivity is defined as $R = |\tilde{I}_{out}|^2/|\tilde{I}_{in}|^2$. By using Eqs. (\ref{IoutB}) and (\ref{solutionT}) we obtain:

\begin{equation}
    R (\omega) = \Bigg|\frac{1-i\omega Y(\omega)}{1+i\omega Y(\omega)}\Bigg|^2
\end{equation}

which clearly defines the reflectivity as a positive quantity. Expanding for the real and imaginary part of the function $Y(\omega)$ we have:

\begin{equation}
 R (\omega) = \frac{\omega^2 \mathrm{Re}{Y(\omega)}^2+[1+\omega \mathrm{Im}Y(\omega)]^2}{\omega^2 \mathrm{Re}{Y(\omega)}^2+[1-\omega \mathrm{Im}Y(\omega)]^2}  \label{intermediateR}
\end{equation}

Since we have 

\begin{eqnarray}
 \mathrm{Im}Y(\omega) =  \sum_m \frac{\Gamma_{\text{r}, m} \mathrm{Im}\Delta_{\text{cnr}, m}(\omega)^* }{|\Delta_{\text{cnr}, m}(\omega)|^2} 
 \nonumber \\
 = - \omega \sum_m  \frac{\Gamma_{\text{nr}, m} \Gamma_{\text{r}, m}}{|\Delta_{\text{cnr}, m} (\omega)|^2} < 0, \label{ImY}
\end{eqnarray}

the reflectivity (\ref{intermediateR}) is always bounded,  $R(\omega) \leq1$ for arbitrary values of the linewidth broadenings $\Gamma_{r,m}$ and $\Gamma_{nr,m}$. Finally, using the identity $(1+x)^2 = (1-x)^2+4x$ the reflectivity (\ref{intermediateR}) can be written as:

\begin{align}
 R (\omega) = 1 + \frac{ 4\omega \mathrm{Im}Y(\omega) }{|1 + i\omega Y(\omega)|^2}
\end{align}

which leads to Eq. (\ref{Reflvoidcav}) from the main text with the help of  Eq. (\ref{ImY}).

\bibliography{Biblio}

\end{document}